\documentclass[12pt]{iopart}

\usepackage[utf8]{inputenc} 
\usepackage[T1]{fontenc}    
\usepackage{amsmath}
\usepackage{url}
\usepackage{booktabs}       
\usepackage{amsfonts}       
\usepackage{nicefrac}       
\usepackage{microtype}      
\usepackage{graphicx}
\usepackage{subfig}
\usepackage{xcolor}         
\usepackage{bm}
\usepackage[]{algorithm2e}

\DeclareMathOperator*{\argmax}{arg\,max}
\DeclareMathOperator{\diag}{diag}
\DeclareMathOperator{\Diag}{Diag}

\begin{document}

\title[Spectrally Adaptive Common Spatial Patterns]{Spectrally Adaptive Common Spatial Patterns}

\author{Mahta Mousavi,  Eric Lybrand,  Shuangquan Feng, Shuai Tang, Rayan Saab, Virginia de Sa}

\ead{mahta@ucsd.edu, desa@ucsd.edu}
\vspace{10pt}

\begin{abstract}
The method of Common Spatial Patterns (CSP) is widely used for feature extraction of electroencephalography (EEG) data, such as in motor imagery brain-computer interface (BCI) systems. It is a data-driven method estimating a set of spatial filters so that the power of the filtered EEG signal is maximized for one motor imagery class and minimized for the other. This method, however, is prone to overfitting and is known to suffer from poor generalization especially with limited calibration data. Additionally, due to the high heterogeneity in brain data and the non-stationarity of brain activity, CSP is usually trained for each user separately resulting in long calibration sessions or frequent re-calibrations that are tiring for the user. In this work, we propose a novel algorithm called \textit{Spectrally Adaptive Common Spatial Patterns} (SACSP) that improves CSP by learning a temporal/spectral filter for each spatial filter so that the spatial filters are concentrated on the most relevant temporal frequencies for each user. We show the efficacy of SACSP in providing better generalizability and higher classification accuracy from calibration to online control compared to existing methods. Furthermore, we show that SACSP provides neurophysiologically relevant information about the temporal frequencies of the filtered signals. Our results highlight the differences in the motor imagery signal among BCI users as well as spectral differences in the signals generated for each class, and show the importance of learning robust user-specific features in a data-driven manner. 
\end{abstract}

\vspace{2pc}

\section{Introduction}

The method of common spatial patterns (CSP) \cite{ramoser2000optimal} is among the most commonly used methods for feature extraction in electroencephalography (EEG) data \cite{nicolas2012brain, wierzgala2018most}. One of its main applications is to learn relevant features in motor imagery brain-computer interface (MI-BCI) systems \cite{pfurtscheller2001motor}. To operate an MI-BCI,  users imagine movement of different parts of their body -- for instance, their right or left hand -- without actually moving them and the goal of the BCI is to infer the user's intended imagery class from the EEG data \cite{pfurtscheller2006mu}.  Imagined movement, similar to actual movement, results in a decrease in power in the mu (7-13 Hz) and often beta (13-30Hz) frequency bands over the contralateral sensorimotor cortex. However, as movement imagination is self-generated with varying success and there are differences in cortical folding patterns across people, there is significant heterogeneity in motor imagery signals between BCI users, including differences in the precise temporal frequencies that are modulated. Furthermore, as EEG data is recorded non-invasively through electrodes that are located on the scalp, several tissue layers away from the neural activity, the recorded EEG is spatially blurred and can be very similar between the different motor imagery classes while differing significantly among people
\cite{allison2010could}.

CSP is a data-driven supervised approach that learns a set of spatial filters to maximally discriminate the two classes.  That is, the power of the projected EEG data through the selected spatial filters is maximally different for the two imagery classes \cite{blankertz2008optimizing}. After learning the feature space as the logarithm of the power through the selected filters, a classifier -- commonly a linear one such as those associated with linear discriminant analysis (LDA) \cite{blankertz2008optimizing} -- is trained on the selected features.

Due to the difficulty of the task and the above mentioned differences between people, there is currently no universal motor-imagery BCI classifier, and at least a short calibration session (usually supervised) for every user is required to achieve  reliable control. In the case of a binary BCI, during the calibration session, the user is instructed to imagine the movement of a body part (e.g., the right or the left hand) and is provided with no feedback or sham feedback \cite{nicolas2012brain}. The recorded EEG data during this calibration session -- calibration data -- is then used to train the classifier (CSP filters together with a linear classifier) and deployed for real-time online control of the BCI. Movement imagination of different body parts is usually translated to real-world application, for instance switching a light on/off, moving a wheelchair or a robotic arm, etc. 

CSP extracts the class-correlated spatial features and acts as a supervised dimensionality reduction technique that also reduces the inherent noise in the EEG data and counters the spatial blurring caused by measuring electrical activity through bone and several layers of tissue \cite{blankertz2008optimizing}. The corresponding EEG data is usually temporally filtered to the 7--30 Hz frequency band (which contains the relevant motor imagery information - a decrease in power in the alpha and beta bands \cite{pfurtscheller2006mu, blankertz2008optimizing}) before applying CSP. However, CSP suffers from overfitting and poor generalizability from calibration to online control \cite{shenoy2006towards, mousavi2021motor}.

Variants of CSP have been proposed to learn the spatial patterns together with temporal filters, using various optimization approaches (e.g., \cite{lemm2005spatio,zhao2009multilinear,noh2013canonical,dornhege2006combined, willems2009body, tomioka2006spectrally, wu2014probabilistic, wu2008classifying, higashi2012simultaneous, novi2007sub, ang2008filter, ang2012filter,aghaei2015separable, qi2015rstfc, li2016unified}). Common spatio-spectral patterns (CSSP) \cite{lemm2005spatio} finds a simple temporal finite-impulse response (FIR) filter comprising only a temporal shift. Therefore, CSSP is also limited in its frequency selectivity. The authors proposed to use cross-validation to find the optimal temporal shift which requires longer calibration data for its robust estimation. The canonical correlation approach to common spatial patterns (CCACSP) \cite{noh2013canonical} algorithm combines the canonical correlation analysis (CCA) and CSP approaches and its cost function considers a temporal filter with a forward and backward one sample temporal shift. It performs better than CSSP with limited calibration data since at lower sampling rates (around 100 Hz), a one sample shift is optimal for the majority of the users \cite{mousavi2021motor}. Another method called the common sparse-spectral spatial pattern (CSSSP) \cite{dornhege2006combined} finds a general finite impulse response (FIR) filter through optimizing a more sophisticated cost function. However, CSSSP assumes that the two motor imagery classes share the same temporal filter which may not be true \cite{willems2009body}. 

Spectrally-weighted CSP (spec-CSP) attempts to find a set of spatial and spectral (frequency transformed temporal) filters through alternating optimization of the cost function \cite{tomioka2006spectrally}. Iterative spatio-spectral patterns learning (ISSPL) \cite{wu2008classifying} is a similar approach that uses a maximal margin optimization for the spectral filters as opposed to a Fisher's criteria in spec-CSP. They both have high computation costs as well as multiple hyper-parameters to optimize through cross-validation, which limits their generalizability under a limited amount of calibration data as is usually the case for EEG data. Furthermore, one of the ISSPL optimization steps has no closed-form solution which adds to the computational cost. Discriminative filter bank CSP (DFBCSP) \cite{higashi2012simultaneous} finds spatio-temporal filters via a modified CSP cost function in an alternating fashion similar to spec-CSP and ISSPL. DFBCSP, however, assumes orthogonality among the temporal filters and solves two generalized eigenvalue decomposition problems, one for the spatial filters that is the same as the one in CSP and another for the temporal filters. However, since the optimal temporal filters corresponding to different spatial filters may overlap considerably, the orthogonality assumption is often not realistic. 

Instead of learning temporal/spectral filters via various optimization approaches, another line of CSP variants finds the relevant sub-bands through a filter bank approach \cite{novi2007sub, ang2008filter, ang2012filter,aghaei2015separable}. These methods require a priori assumptions on the length and location of the sub-bands and require significant amounts of calibration data to find a robust feature space to be transferred from calibration data to online BCI control \cite{lotte2018review}. 

Another line of work implements the method of CSP through deep neural networks \cite{lawhern2018eegnet, mousavi2019temporally, al2021deep}. EEGNet is \cite{lawhern2018eegnet} one popular approach that facilitates training of the features (e.g., CSP filters) together with the classifier in an end-to-end fashion. However, these methods have a large number of parameters to train that would require a large number of training samples limiting their performance in conditions when the amount of calibration (or training) data is small. 

In this work, we propose a novel approach to learn spatial and spectral filters in motor imagery data by solving an optimization problem that generalizes the cost function in CCACSP. We use an alternating direction method \cite{Boyd2011DistributedOA} to optimize our proposed cost function. We apply our method to motor imagery data that we recorded from 12 participants and show improved classification accuracy as well as more interpretable spatial patterns and spectral filters compared to the closely related methods of CSP, CCACSP, spec-CSP and EEGNet.

\section{Methods}
In this section, we will first describe our proposed SACSP method and then briefly summarize its relation to CSP, CCASP and spec-CSP methods. Finally, we will describe our experiment and data collection procedure as well as the data processing pipeline. The algorithms were implemented in Python and ran on an Intel Xeon e5-2680 v2 server running Ubuntu version 18.04.

\subsection{Spectrally adaptive CSP (SACSP)}
Due to the subtlety of an underlying motor imagery signal, its detection requires finding a subspace on which the power (or variance since EEG signal is usually bandpassed) of the projected signal is maximized for one class while minimized for the other class. To do so, CSP estimates the spatial covariance for the two classes and uses a Raleigh quotient cost function to find the optimal directions for such projection \cite{ramoser2000optimal, blankertz2008optimizing}. In SACSP however, we define a spectrally-weighted spatial covariance instead, to allow for learning weights on the motor imagery related frequencies in the estimation of the covariance. 

Let $X_{i}\in \mathbb{R}^{n\times t}$ represent the bandpassed EEG data epoch over $n$ channels with $t$ time samples that belongs to class $i\in\{1,2\}$, e.g. the two motor imagery classes. Since EEG data is bandpassed, $\mathbb{E}[X_1] = \mathbb{E}[X_2] = \boldsymbol{0} $ where $\boldsymbol{0}$ is a vector of size ${n\times 1}$ and $\mathbb{E}$ refers to the expectation with respect to the time variable $t$. Let $C$ be a circulant matrix. We define the spectrally-weighted spatial covariance matrices for classes 1 and 2 as follows: 
  \begin{align}
     \Gamma_{S1} = \mathbb{E} [X_1 C X_1^T], ~~~~ \Gamma_{S2} = \mathbb{E} [X_2 C X_2^T], 
\end{align}
where $(\cdot)^T$ indicates the transpose of a matrix/vector. 
Since $C$ is circulant, we can write it as $C = F \diag(h) F^H$ where $h$ is a vector, $diag(h)$ is a matrix whose diagonal elements are the elements of the vector $h$, $F$ is the $t\times t$ Fourier matrix and $(\cdot)^H$ indicates the complex conjugate transpose of a matrix. Replacing this in the above we get: 
  \begin{align}
    \Gamma_{S1} = \mathbb{E} [X_1  F\diag(h) F^H X_1^H] = \mathbb{E} [(X_1F) \diag(h) (X_1F)^H],\\ 
    \Gamma_{S2} = \mathbb{E} [X_2  F\diag(h) F^H X_2^H] = \mathbb{E} [(X_2F) \diag(h) (X_2F)^H].
     \end{align}
Let $\hat{X_1} = X_1F$ and $\hat{X_2} = X_2F$ be the Fourier transform of an epoch along its temporal dimension; we would like to find the optimal parameters:
  \begin{align}
    \argmax_{w, h} \quad \frac{w^T\mathbb{E}[\hat{X_1}\diag(h)\hat{X}_1^H]w}{w^T(\Sigma_1+\Sigma_2)w}  \hspace{1cm} \text{ subject to} \hspace{1cm}\|h\|_2\leq 1 , \label{opt-prob1}
     \end{align}
     
\begin{align}
    \argmax_{v, l}  \quad
 \frac{v^T\mathbb{E}[\hat{X_2}\diag(l)\hat{X}_2^H]v}{v^T(\Sigma_1+\Sigma_2)v} \hspace{1cm} \text{ subject to} \hspace{1cm} \|l\|_2\leq 1 , \label{opt-prob2}
\end{align}
where $\Sigma_i = \mathbb{E}[X_iX_i^T]$ is the spatial covariance of each class. The above optimization problems are non-convex, thus it is generally difficult to find their global solutions. On the other hand, upon fixing one of the parameters, the resulting optimization problem (in the other parameter) becomes much easier. So we will approach \eqref{opt-prob1} and \eqref{opt-prob2} via an alternating maximization procedure. Indeed, for fixed $h$ and $l$, the optimal $w$ and $v$ are the leading (generalized) eigenvectors associated with the generalized eigenvalue problems:
\begin{align}
    \mathbb{E}[\hat{X_1}\diag(h)\hat{X}_1^H] w = \lambda_w (\Sigma_1+\Sigma_2)w, 
\label{findw}
\end{align}

\begin{align}
    \mathbb{E}[\hat{X_2}\diag(l)\hat{X}_2^H] v = \lambda_v (\Sigma_1+\Sigma_2)v.    \label{findv}
\end{align}

On the other hand if $w$ is fixed, solving for $h$ reduces to finding
    \begin{align}
 & \argmax_{h} \quad w^T\mathbb{E}[\hat{X_1}\diag(h)\hat{X}_1^H]w \nonumber \\
 = & \argmax_{h} \quad \mathbb{E}[tr(w^T\hat{X_1}\diag(h)\hat{X}_1^Hw)] \nonumber \\
 = & \argmax_{h} \quad tr(\diag(h) \mathbb{E}[\hat{X}_1^Hww^T\hat{X_1}]) \nonumber \\
 =& \argmax_{h} \quad \langle h,\Diag(\mathbb{E}[\hat{X}_1^Hww^T\hat{X_1}])\rangle,
 \label{findh}
  \end{align}
where $Diag(.)$ is an operator that extracts the main diagonal of a square matrix into a vector. Equation \ref{findh} subject to the $\ell_2$-norm constraint $||h||_2\leq 1$ leads to the following solution: 
 \begin{equation}
     h(k) = (\mathbb{E}[\hat{X}_1^Hww^T\hat{X_1}])_{kk}/\|\Diag(\mathbb{E}[\hat{X}_1^Hww^T\hat{X_1}])\|_2, ~~k\in\{1,2,..,t\}, 
     \label{fixed-w}
 \end{equation}
where $kk$ indicates the $k-\text{th}$ diagonal element of $(\mathbb{E}[\hat{X}_1^Hww^T\hat{X_1}])$ and $h(k)$ indicates the $k-\text{th}$ element of vector $h$. 
Similarly, the solution for $l$ will be: 
 \begin{equation}
     l(k) = (\mathbb{E}[\hat{X}_2^Hvv^T\hat{X_2}])_{kk}/\|\Diag(\mathbb{E}[\hat{X}_2^Hvv^T\hat{X_2}])\|_2, ~~k\in\{1,2,..,t\},
     \label{fixed-v}
 \end{equation}
 where $kk$ indicates the $k-\text{th}$ diagonal element of $(\mathbb{E}[\hat{X}_2^Hvv^T\hat{X_2}])$ and $l(k)$ indicates the $k-\text{th}$ element of vector $l$.

Putting the above together, we propose Alg. \ref{SACSP-algo}.

\paragraph{\underline{Practical Notes:}} In practice, including when implementing Algorithm \ref{paradigm} below, the expected values of the various quantities above are estimated by taking sample averages over epochs.  

Furthermore, since EEG data is often rank deficient (due to for instance re-referencing to the common average), we recommend projecting the EEG data to a full-rank subspace by finding $L$ and $\Psi$ such that $\Sigma_1+\Sigma_2 = L\Psi L^T$. Then we find the projection operator as $Q = \Psi^{-\frac{1}{2}}L^T$ and project the EEG data as $Y_1 = QX_1$ and $Y_2=QX_2$. We apply algorithm \ref{SACSP-algo} and eventually project the spatial filters back to the original space by multiplying $Q^T$ to the left side of the selected spatial filters. 

Since SACSP requires that the epochs be transformed to the frequency domain using the Fourier matrix, for motor imagery data, we recommend downsampling to 100 Hz and use the number of frequency samples as 100. If the length of the EEG data is more than 1 second, then we recommend estimating $\mathbb{E}[\hat{X_1}\diag(h)\hat{X}_1^H]$, $\mathbb{E}[\hat{X_1}\diag(l)\hat{X}_1^H]$, $\mathbb{E}[\hat{X}_1^Hww^T\hat{X_1}]$ and $\mathbb{E}[\hat{X}_2^Hvv^T\hat{X_2}]$ for every 1 second of the data and then averaging them.

 \begin{algorithm}
 \KwData{$R$ = number of spectral/spatial filters for each class, $M$ = number of initializations of the spectral filters, $\epsilon>0$}
 \KwResult{$w_r$, $v_r$, $h_r$ and $l_r$ for $r\in\{1,2,..,R\}$}
 initialization: $h^m/||h^m||_2$ and $l^m/||l^m||_2$ for $m\in\{1,2,..,M\}$\;
 \For{$m$ = $1$ to $M$}{
select $w_1$ to $w_R$ corresponding to the $R$ top eigenvectors from Eq. \ref{findw}\;
 select $v_1$ to $v_R$ corresponding to the $R$ top eigenvectors from Eq. \ref{findv} \;
 \For{$r$ = $1$ to $R$}{
 \While{the cost function for class 1 (Eq. \ref{opt-prob1}) increases more than $\epsilon$}{
  update the spectral filter $h^m_r$ from Eq. \ref{fixed-w}  \;
  update the spatial filter, $w_r$, by selecting the eigenvector corresponding to the largest $\lambda_w$ from from Eq. \ref{findw}\;
 }
  \While{the cost function for class 2 (Eq. \ref{opt-prob2}) increases more than $\epsilon$}{
  update the spectral filter $l^m_r$ from Eq. \ref{fixed-v} \;
  update the spatial filter, $v_r$, by selecting the eigenvector corresponding to the largest $\lambda_v$ from Eq. \ref{findv}\;
 }
 }
 }
 Among the $M\times R$ spatial and spectral filter pairs for each class, select $R$ that maximizes the cost function on the training data for the corresponding class (Eq. \ref{opt-prob1} for class 1 and Eq. \ref{opt-prob2} for class 2) $\Rightarrow$ $w_r, l_r, v_r$ and $h_r$ for $r\in\{1,2,..,R\}$ \;
 
 \caption{Spectrally adaptive common spatial patterns (SACSP).
 }
 \label{SACSP-algo}
\end{algorithm}

\paragraph{\underline{Classification:}}
Similar to CSP, SACSP also requires a user-selected number of filters $R$ for each class. After finding the optimal $2\times R$ spatial and spectral filter pairs $\{(f_{z^r}, f_{q^r} ), r\in\{1,...,2R\}\} = \{(w_r,h_r),r\in\{1,...,R\}\} \cup \{(v_r,l_r),r\in\{1,...,R\}\}$, then $2\times R$ features were extracted by filtering each epoch, $x$ as follows: 
\begin{equation}
    \log(f_{z^r}^T~(\hat{x}~\diag(f_{q^r})\hat{x}^H) ~f_{z^r}), 
\end{equation}
where $\hat{x} = xF$. Next, linear discriminant analysis (LDA) \cite{scikit-learn} with automatic shrinkage using the Ledoit-Wolf lemma \cite{ledoit2004honey} was trained as the classifier.

\paragraph{\underline{Choice of parameters in SACSP:}}
We chose the top 3 filters for each class, i.e., $R=3$ as this is usually the choice for CSP and other similar methods \cite{blankertz2008optimizing,lotte2018review}. The iterations were set to stop when the improvement in the cost function was smaller than a threshold which was set to $\epsilon = 10^{-6}$; however, our investigations showed that SACSP is not highly dependent on this parameter. Of course if the results for a specific dataset vary significantly for different values of $\epsilon$, one can always set it through cross-validation. 
Furthermore, the spectral filters $h$ and $l$ were initialized with 3 initializations each(i.e., $M=3$): 1) $h^1(k) = l^1(k) = 1, \forall k\in\{0,1,...,t-1\}$, 2) $h^2(k) = l^2(k) = 1$ only in 7--15 Hz to include the mu frequency band, and 3) $h^3(k) = l^3(k) = 1$ only in 15--30 Hz to include the beta frequency band.

\paragraph{\underline{Comparison with other methods:}}
We compared our proposed method with CSP and two closely related methods in the literature, namely spec-CSP \cite{tomioka2006spectrally} and CCACSP \cite{noh2013canonical} as well as an end-to-end neural network based implementation of the CSP method, namely EEGNet \cite{lawhern2018eegnet}. Note that SACSP reduces to CSP if $h(k) = l(k) = 1, \forall k\in\{0,1,...,t-1\}$ and reduces to CCACSP if $h(k) = l(k) =  \cos(\frac{2\pi k}{t}), \forall k\in\{0,1,...,t-1\}$.

For CSP, spec-CSP and CCACSP, we chose the top 3 filters for each class. For spec-CSP, per the authors' recommendation in the original paper \cite{tomioka2006spectrally}, the number of iteration steps was set to 10 (with one spatial and spectral filter update in each step), hyperparameters $p^\prime \in\{0, 0.5, 1\}$ and $q^\prime \in\{0, 0.5, 1, 1.5, 2\}$ were optimized using cross-validation on the calibration data and the 7--30 Hz frequency band was selected as apriori knowledge on the motor imagery signal similar to \cite{tomioka2006spectrally}. For EEGNet, we used the default parameters as per the authors' recommendation in \cite{lawhern2018eegnet}.

\subsection{Experiment}
After the study was approved by the institutional review board of UC San Diego, data were recorded from 12 participants (7 females, 1 left-handed, average age = 20.4 $\pm$ 1.0) who were naive to BCI systems. Each participant signed an informed consent form prior to participating in one session of motor imagery. They were compensated with experimental course credit or at the rate of \$20/hour based on their preference. During the experiment which took roughly 2.5 hours, participants were comfortably sitting in an arm chair about one meter away from the screen (Dell 19” CRT monitor). They were asked to minimize their movements to minimize the occurrence of eye movements and muscular artifacts in the EEG recordings. Furthermore, participants could take as much break between the blocks as needed.

Each session had two main parts: in the first part, participants were instructed (over 30 trials) on how to perform movement imagination of their right or left hand \cite{mousavi2017towards}. Note that this part of the study was only designed to provide some exposure to the otherwise naive participants and the recorded EEG data in this part was not used in the analysis that is presented next. 

\begin{figure}
    \centering
        \includegraphics[trim=10 0 550 0, clip,width=\textwidth]{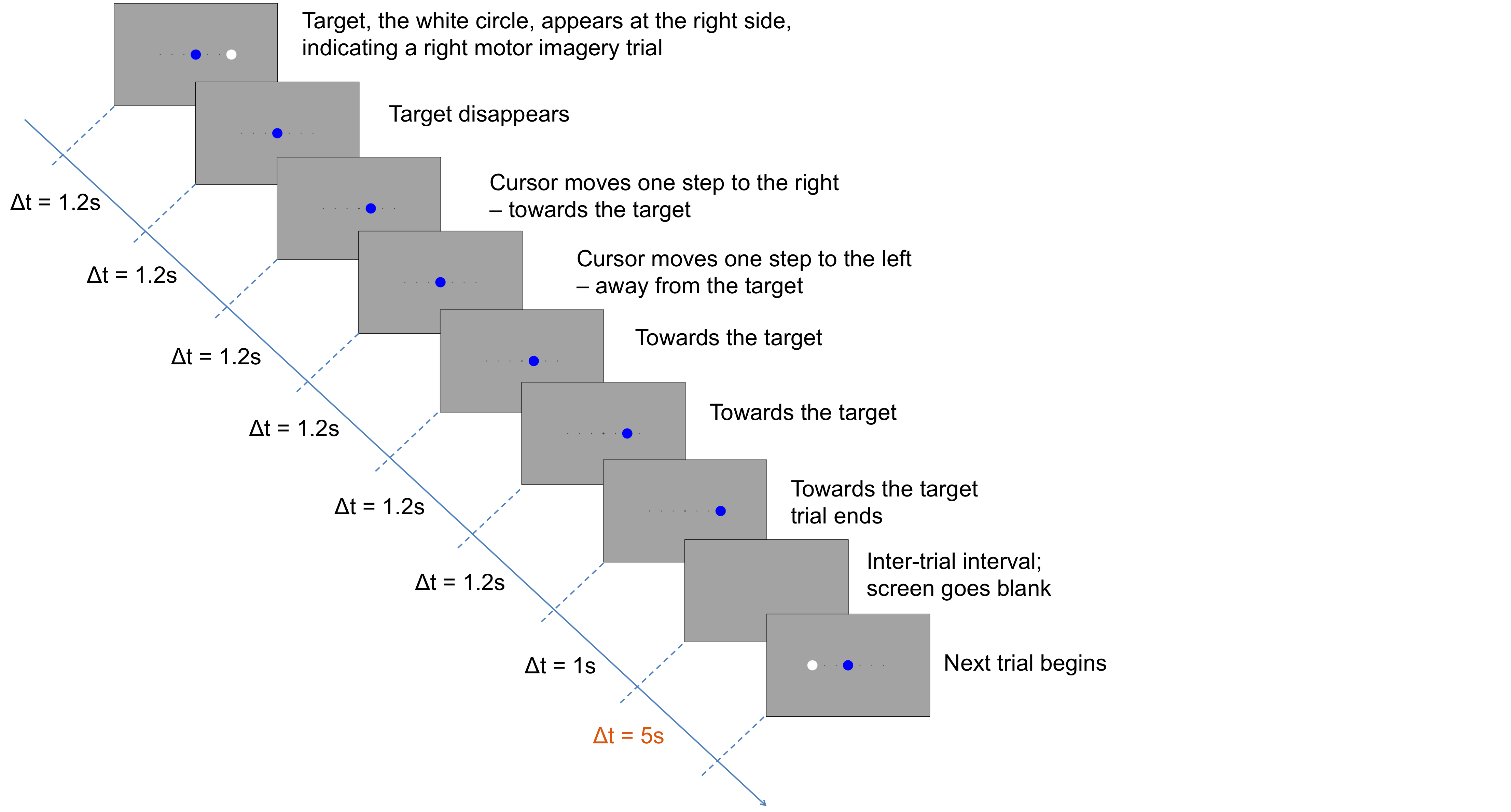}
    \caption{An example of a trial \cite{mousavi2020hybrid}: Each trial began with the cursor (blue circle) at the center and the target (white circle) at either side of the screen, exactly three steps from the cursor. Trial ended when the cursor hit the target location or the corresponding location on the other side or after 12 cursor steps, whichever came first. Participants had 5 seconds to rest before the next trial began. Note that the background of the frames was dark gray in the experiment but is depicted lighter here for easier visualization of the details. } 
    \label{paradigm}
\end{figure}

For the second part of the experiment, participants were instructed to use movement imagination of their right or left hand to move a cursor on the screen. Each trial began with a cursor (blue circle about 2~cm in diameter) in the center and a target (white circle about 2~cm in diameter) at either side of the screen and three steps away from the cursor. The cursor moved one step per 1.2 seconds towards or away from the target until it reached the target location or the corresponding location on the other side of the screen. Each participant participated in 9 blocks comprising 20 trials each. During the first three blocks, the cursor moved according to a pre-determined pseudo-randomly generated sequence of movements but the participants were not aware of this and were led to believe that they were in control of the cursor movements. The recorded EEG data from the first three blocks were used for calibration. CSP was trained for feature extraction where the top three CSP filters for each class were selected. The logarithm of the power of the CSP-filtered EEG data were selected as features and shrinkage-LDA \cite{ledoit2004honey} was trained for classification. This classifier was used for the motor imagery signal in the online control of the cursor in the latter 6 blocks which had a similar structure as the calibration data except that the participant was in control of the cursor movement. In half of the blocks, the motor imagery classifier was combined with another classifying error-related brain activity based on the last cursor movement. An example of one trial in our experiment is presented in figure \ref{paradigm}.
For more details about the experiment set-up please refer to \cite{mousavi2020hybrid}.

In an offline analysis, we used the recorded data from the calibration blocks to train CSP, CCACSP, spec-CSP, EEGNet and SACSP separately. Each method was then tested on the data that were recorded during the online control blocks. Note that we trained and tested the methods on the cursor movements (or steps) and the results are reported accordingly in the next section.

\paragraph{\underline{Data collection:}}
EEG data were recorded with a 64-channel BrainAmp system (Brain Products GmbH) at 5000 Hz sampling rate and downsampled to 100 Hz for online control and later for the offline analysis. Electrodes were located according to the international 10-20 system \cite{klem1999ten}, the impedance of the electrodes was set to be below 20 $K\Omega$. The online reference and ground electrodes were located at FCz and AFz, respectively. 

We used the python-based Simulation and Neuroscience Application Platform (SNAP) \cite{snap} to design the stimuli and Lab Streaming Layer (LSL) \cite{lsl} as an interface between the computer and the EEG system as well as to record the EEG data. For data processing and classification in online control and later for offline analysis of the data, we used Numpy \cite{numpy}, Scipy \cite{scipy} and Scikit-learn \cite{scikit-learn}. MATLAB \cite{MATLAB18} and EEGLAB \cite{Delorme2004} were used for epoching the EEG data and for plotting the results. 

\paragraph{\underline{Pre-processing:}}
All 64 channels were used in this study. Data from calibration and online blocks were downsampled to 100 Hz and epoched 0-1 second with respect to each cursor movement excluding the last cursor movement. This is because participants stopped motor imagery when the trial ended with the last cursor movement. After downsampling and epoching, data were re-referenced to the common average and bandpass filtered in 7--30~Hz to include the mu (about 7--13~Hz) and beta (about 13--30~Hz) frequency bands \cite{pfurtscheller2006mu, mousavi2017improving}. The bandpass filter was designed as a 6th order Butterworth filter and was applied to each epoch with the \texttt{filtfilt} function from Scipy. This function applies the filter once forward and once backward and results in a zero-phase filtering.

Since in calibration blocks, all participants were shown the same pre-determined sequence of cursor movements, the number of calibration steps in each class did not vary across participants. However, the number of steps in each class in online blocks varied across participants depending on their online performance. We balanced the right and left classes (including movements towards and away from the target in each class) for both calibration and online blocks to alleviate any potential bias in learning features and for easier interpretation of the results. After balancing the number of right and left motor imagery steps, there were a total of 136 steps in the calibration data and on average across participants, a total of  272.7 $\pm$ 38.4 steps in the online blocks. Balancing the classes was done 10 times to allow steps to be in the training (from calibration data) or test (from online data) sets.  

\paragraph{\underline{Visualization:}}
We visualized the found spatial patterns and spectral filters to neurophysiologically interpret the methods under study. 
Let $A$ be a matrix comprising the sorted eigenvectors from equations \ref{findw} or \ref{findv}. The spatial patterns for the corresponding filters are defined as the columns of the matrix $(A^{-1})^T$ \cite{blankertz2008optimizing}. We call the patterns corresponding to filters $w_i$ and $v_i$, $Pw_i$ and $Pv_i$, respectively. 

Normalized spectral filters were plotted against frequency (Hz) as line plots using MATLAB. Spatial patterns were estimated for each method accordingly and \texttt{topoplot} from EEGLAB was used to plot them as heatmaps on a 2-D scalp.

\section{Results}
\paragraph{\underline{Classification accuracy:}}
The classification accuracy of the CSP, CCACSP, spec-CSP, EEGNet and SACSP when the feature extraction and classifiers were trained on the calibration data and tested on the online data, are presented in table \ref{accuraciesWS}. The first number in each entry for participants 1 to 12 (P1-P12), represents the mean classification accuracy over 10 instances of train-test data and the second number reports the standard deviation. In the last row, the average performance across participants is reported where in each entry, the first number indicates the average for the corresponding column and the second number indicates the standard error of the mean. Across participants, the performance of SACSP is significantly better compared to CSP, CCACSP, spec-CSP and EEGNet (Wilcoxon signed rank test, $p<0.01$). 

We also looked at the classification accuracy when each method was trained on calibration data and tested on calibration data through cross-validation (the `calibCV' case) as well as when trained on online data and tested on online data through cross-validation (the `onlineCV' case). Note that both of these conditions are different from the practical scenario of interest in this work, where the methods are trained on the calibration data and tested on the online data (the `test' case). Our results show that for CSP, calibCV and onlineCV are not different across participants (average 0.73 and 0.71, Wilcoxon signed-rank test, $p=0.42$) -- results for each participant are presented in supplementary material section 1. However, as table \ref{accuraciesWS} shows, there is a large drop in performance of CSP in the test condition (across participants significantly lower that calibCV and onlineCV, Wilcoxon signed rank test, $p<0.001$). Our results show that spec-CSP, CCACSP, and SACSP were all able to close this gap from calibration to online performance with SACSP performing significantly better than all other methods (Wilcoxon signed rank test, $p<0.01$).

\begin{figure}
  \centering
  \includegraphics[width=0.55\textwidth]{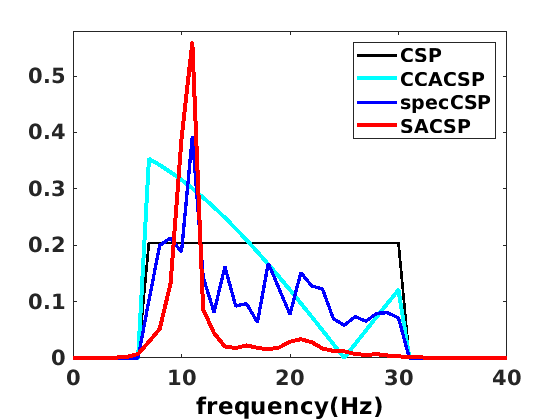}
  \caption{Comparison of the normalized spectral filters in P9. SACSP and spec-CSP lines show the top spectral filter for the right hand motor imagery class. Note that CSP and CCACSP have fixed spectral filters by design. }
  \label{spectral}
\end{figure}

\begin{table}
  \caption{Motor imagery (right/left hand) classification accuracy. For P1-P12, the first number in each entry reports the average accuracy across 10 instances of the data and the second number reports the standard deviation. In the last row, the first number in each entry reports the average classification accuracy across participants for the corresponding column and the second number reports the standard error of the mean.  }
  \centering
  \begin{tabular}{cccccc}
    \toprule
    PID  & CSP &  spec-CSP & CCACSP & EEGNet & SACSP  \\ 
    \midrule
P1  &  0.55/0.035  & 0.61/0.039  & 0.74/0.033  & 0.61/0.032  & 0.81/0.022  \\ 
P2  &  0.56/0.028  & 0.63/0.091  & 0.69/0.012  & 0.56/0.051  & 0.69/0.030  \\ 
P3  &  0.63/0.038  & 0.61/0.037  & 0.61/0.057  & 0.60/0.074  & 0.72/0.096 \\ 
P4  &  0.72/0.019  & 0.70/0.051  & 0.71/0.020  & 0.60/0.028  & 0.73/0.009  \\ 
P5  &  0.59/0.118  & 0.59/0.105  & 0.56/0.071  & 0.56/0.058  & 0.62/0.117  \\ 
P6  &  0.50/0.014  & 0.52/0.032  & 0.54/0.017  & 0.52/0.016  & 0.52/0.013  \\ 
P7  &  0.62/0.027  & 0.65/0.035  & 0.68/0.019  & 0.58/0.044  & 0.70/0.040  \\ 
P8  &  0.57/0.023  & 0.59/0.034  & 0.71/0.023  & 0.56/0.050  & 0.72/0.036  \\ 
P9  &  0.68/0.048  & 0.67/0.033  & 0.73/0.019  & 0.69/0.023  & 0.76/0.023  \\ 
P10  &  0.49/0.021  & 0.49/0.024  & 0.49/0.017  & 0.52/0.027  & 0.50/0.019  \\ 
P11  &  0.60/0.019  & 0.64/0.028  & 0.64/0.034  & 0.61/0.023  & 0.67/0.027  \\ 
P12  &  0.47/0.032  & 0.45/0.025  & 0.49/0.024  & 0.49/0.026  & 0.51/0.010  \\  
\midrule
Average  &  0.58/0.022  & 0.60/0.022  & 0.63/0.027  & 0.58/0.015 & {\textbf{0.66/0.030}} \\
    \bottomrule
  \end{tabular}
  \label{accuraciesWS}
\end{table}

\begin{figure}
  \subfloat[P1 -- CSP spatial patterns.]{
	   \centering
	   \includegraphics[trim={0cm 5cm 0cm 4.2cm},clip,width=0.9\textwidth]{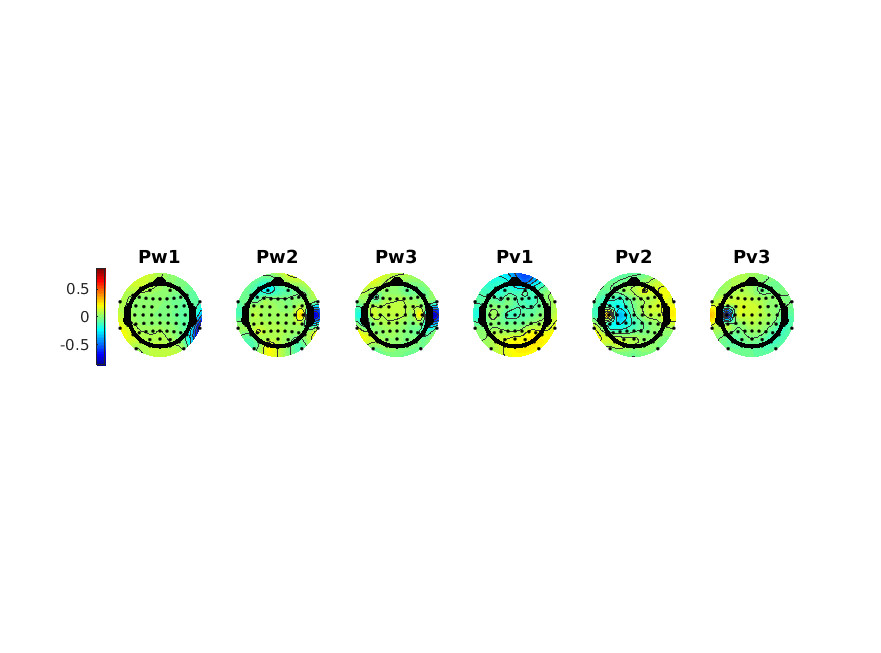}}
 \hfill 	
  \subfloat[P1 -- CCACSP spatial patterns.]{
	   \centering
	   \includegraphics[trim={0cm 5cm 0cm 4.2cm},clip,width=0.9\textwidth]{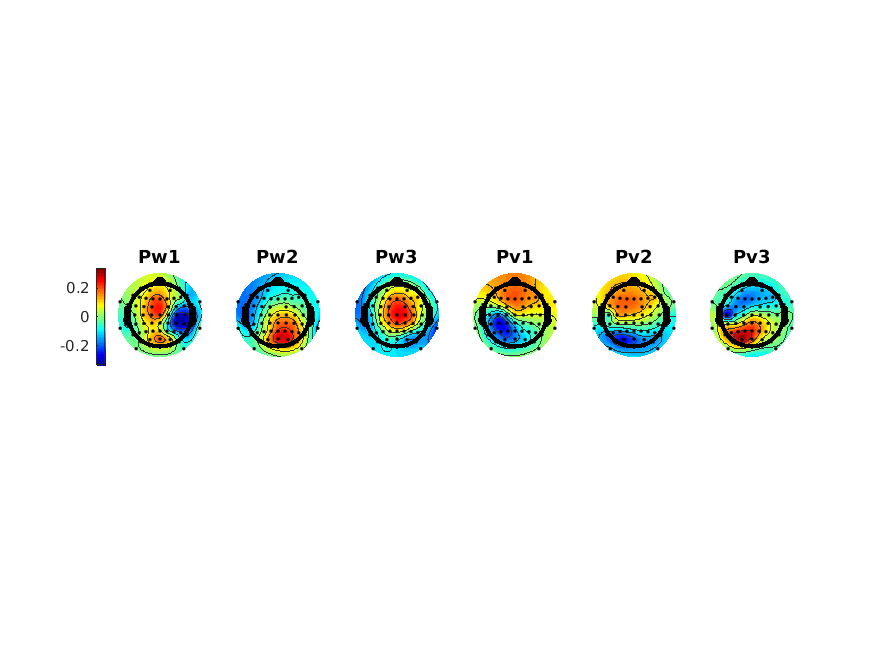}}
 \hfill	
  \subfloat[P1 -- spec-CSP spatial patterns.]{
	   \centering
	   \includegraphics[trim={0cm 5cm 0cm 4.2cm},clip,width=0.9\textwidth]{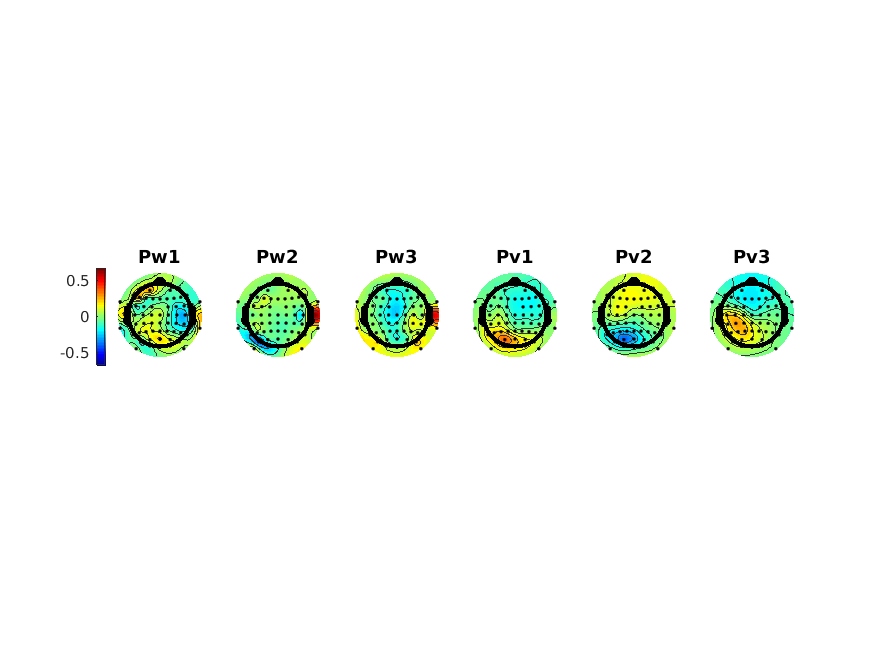}}
 \hfill 	
  \subfloat[P1 -- spec-CSP spectral filters.]{
	   \centering
	   \includegraphics[trim={1cm 0.5cm 1cm 13cm},clip,width=0.9\textwidth]{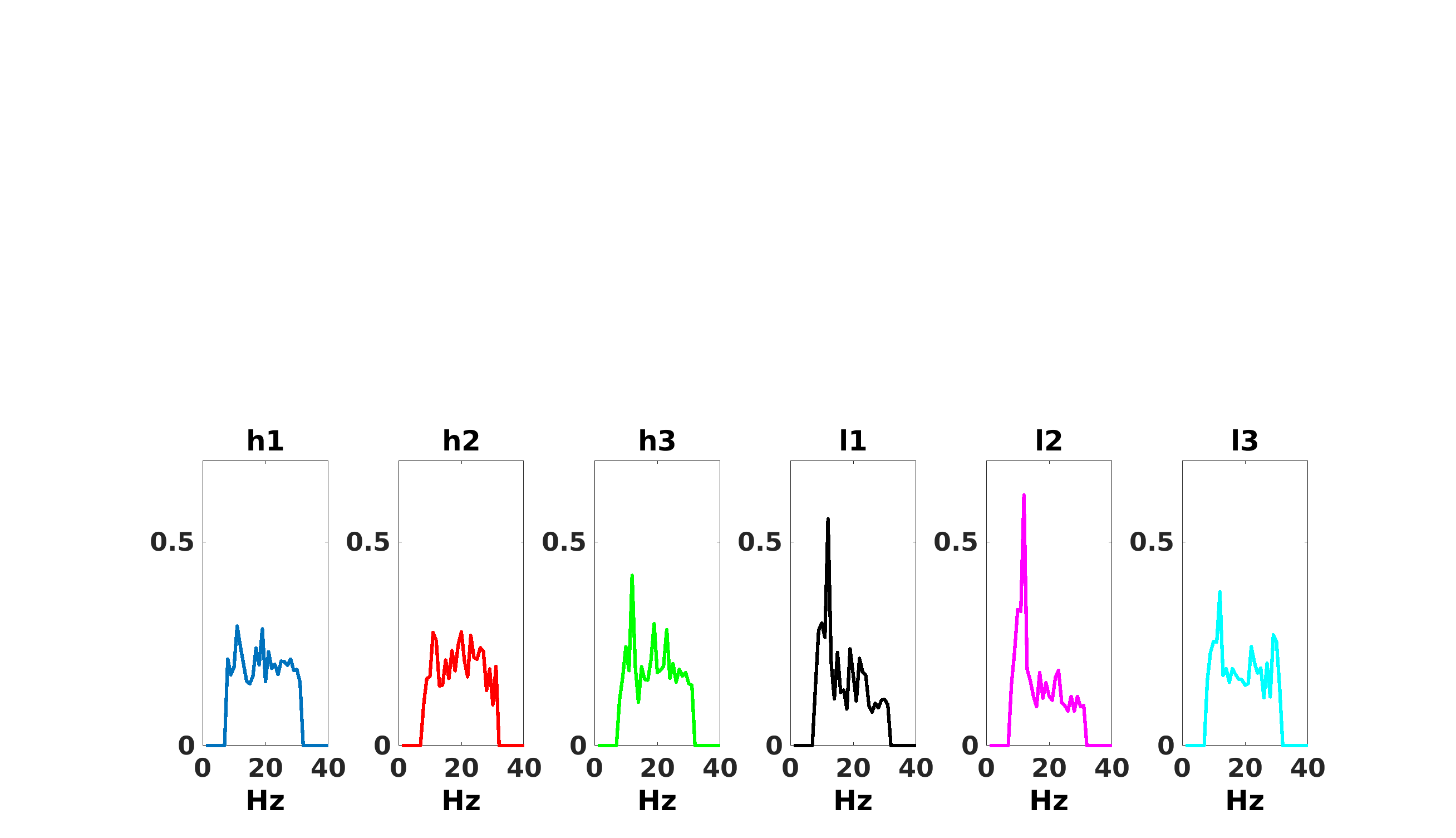}}
 \hfill	
 \subfloat[P1 -- SACSP spatial patterns.]{
	   \centering
	   \includegraphics[trim={0cm 5cm 0cm 4.2cm},clip,width=0.9\textwidth]{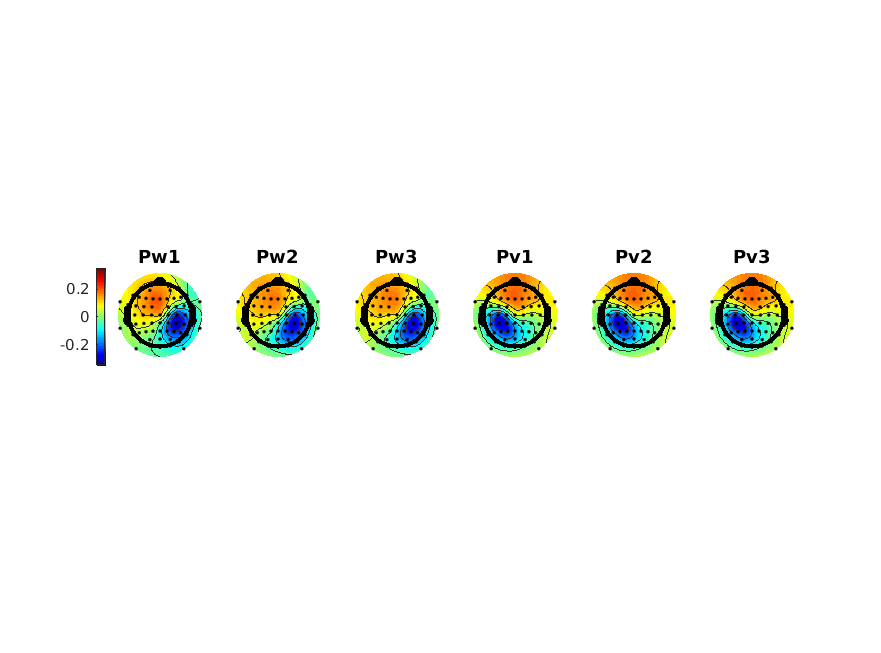}}
 \hfill 	
  \subfloat[P1 -- SACSP spectral filters.]{
	   \centering
	   \includegraphics[trim={1cm 0.5cm 1cm 13cm},clip,width=0.9\textwidth]{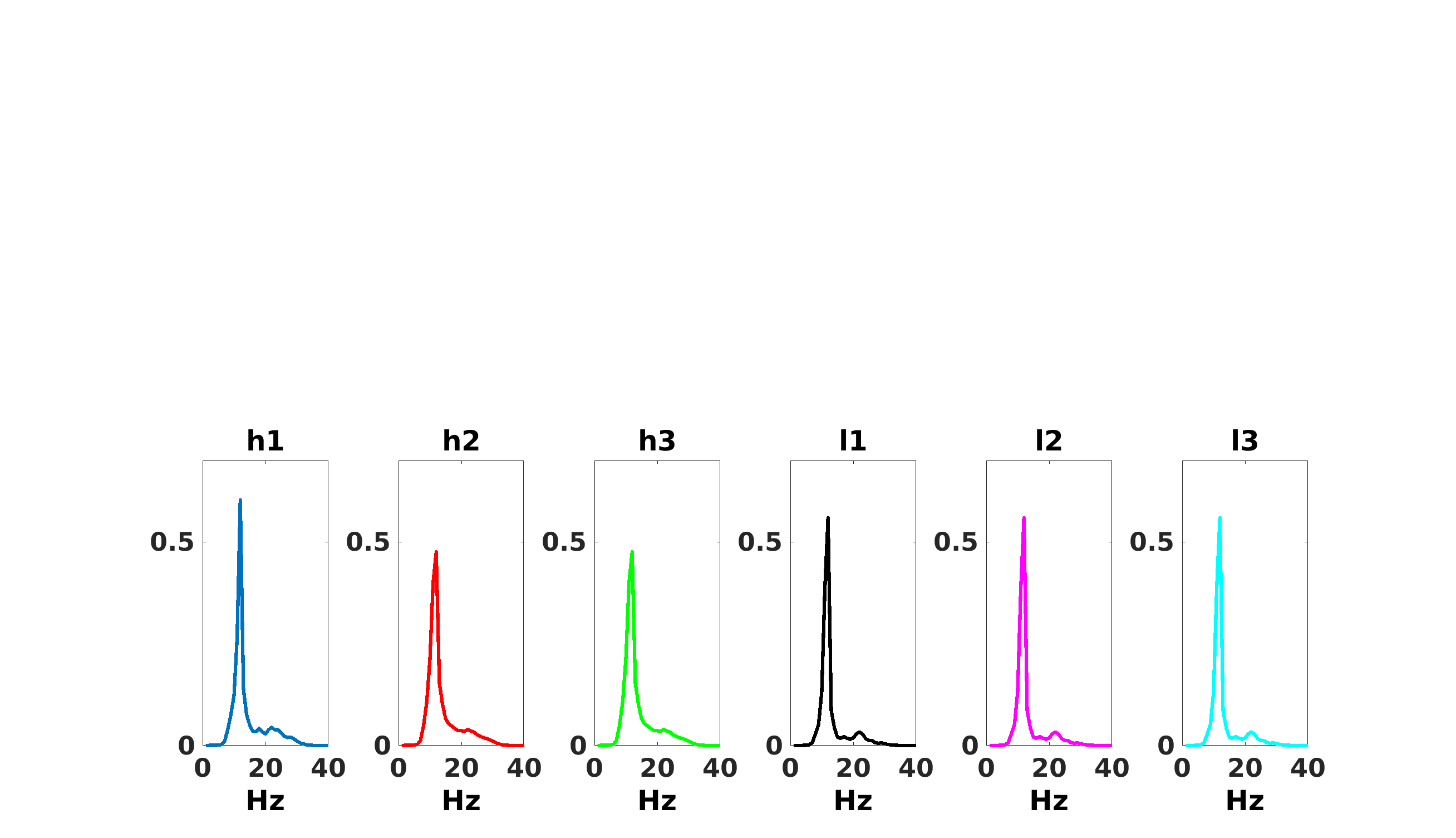}}
\caption{Spatial patterns and spectral filters trained on the calibration data for P1. Compared to other methods, SACSP results in spatial patterns that are typical for right and left hand motor imagery while providing neurophysiologically relevant frequency selectivity in spectral filters. }
\label{SACSP1}
\end{figure}

\paragraph{\underline{Visualization of spatial patterns and spectral filters:}}
Figure \ref{SACSP1} compares the spatial patterns and spectral filters of the SACSP with CSP, CCACSP, and spec-CSP. Note that these are the filters and patterns that were trained on the calibration data and used later for classification of the online data. Due to limited space, we present the results for the rest of the participants in the supplementary material section 2. The last two rows show the results for SACSP. The filters and patterns for each class ($Pw_1$, $Pw_2$ and $Pw_3$, $h_1$, $h_2$ and $h_3$) and ($Pv_1$, $Pv_2$ and $Pv_3$, $l_1$, $l_2$ and $l_3$) converged to similar values. This is because our algorithm, different from the others presented here, does not result in spatial filters/patterns that are necessarily orthogonal to each other and convergence to the same solution from different initializations in fact shows that SACSP converged to its optimal solution over a large range in initializations. Furthermore, we tested SACSP with random initialization and received the same filters and patterns.    

Comparing the spectral filters for spec-CSP and SACSP, we can see that the spec-CSP spectral filters fail at selectively extracting the underlying features in an interpretable way. SACSP shows smoother and less noisy filters with a peak at/around the mu band for both the right and left hand motor imagery classes and a small beta peak for the left hand motor imagery class. Another example comparing spectral filters for P9 is presented in figure \ref{spectral}. 

Comparing the spatial patterns, SACSP shows patterns that are typical for right and left hand motor imagery. $Pv_3$ in spec-CSP can be regarded as a similar pattern to $Pv_3$ in SACSP but the rest of the patterns in spec-CSP and the ones in CSP are neurophysiologically irrelevant for motor imagery. $Pv_1$ in CCACSP is a motor imagery pattern and $Pv_3$ could also be interpreted as such. However, CCACP fails at finding appropriate patterns for the other class.

\section{Discussion}
In this work, we proposed SACSP, a novel feature extraction method for motor imagery data, and showed its superiority in classification accuracy when trained on calibration data and tested on online data as well as neurophysiologically plausible spatial patterns and spectral filters. Note that the classification accuracy was reported for one second of motor imagery data and still more than half of the participants performed at a classification accuracy of $0.7$ or higher with the proposed algorithm. The majority of existing work in the literature focuses on cross-validation accuracy (not calibration to online transfer) and over longer lengths of motor imagery data. 

The reported SACSP spatial patterns were very typical of the ones from motor imagery data \cite{blankertz2008optimizing}. However, CSP, CCACSP and spec-CSP fully or partially failed at learning them. Furthermore, the spectral filters were also as expected, revealing expected information in mu and beta frequency bands \cite{pfurtscheller2001motor, pfurtscheller2006mu}. 

One important and valuable observation from an engineering and scientific perspective, is the difference among the spectral filters learned for the right and left hand motor imagery classes and the ones learned for different participants. Note that SACSP, different from sub-band and filterbank variations of CSP, learns the spectral filters directly instead of a priori assuming multiple frequency bands and searching for class-correlated information in those bands \cite{novi2007sub, ang2008filter, ang2012filter,aghaei2015separable}. 

All compared methods except for CSP itself performed better than EEGNet. While neural network and deep learning methods for EEG and BCI applications are becoming evermore popular especially in training end-to-end networks to combine feature extraction and classification \cite{craik2019deep}, they generally require a large amount of training data. In our case, since the calibration data were very limited as is the case for many BCI applications, a deep learning approach such as EEGNet would not be an appropriate choice as reflected by our results. 

SACSP has a few hyperparameters to choose: 1) the number of spectral/spatial filters, 2) initialization for the spectral filters, and 3) the threshold for the stopping of the iterations. We selected the number of spatial filters for each class to be $3$ as is usually selected for CSP in the literature \cite{blankertz2008optimizing, lotte2018review}. In the case of CSP, three orthogonal filters were selected and the LDA classifier weighs them according to their relevance. However, SACSP initialized the spatial filters in each class to be orthogonal to each other but through iterations, they would not necessarily remain orthogonal. In fact, our results showed that often the top three selected spatial filters that maximized the cost function for each class, converged to the same solution. This is an interesting observation and highlights the fact that SACSP converged with different initial conditions and paths to convergence.

We also investigated SACSP with different initializations of the spectral filters including random initialization; but neither the classification accuracy nor the spatial patterns/spectral filters were different upon convergence. Moreover, we varied $\epsilon$ from $10^{-8}$ to $10^{-3}$, but the number of steps to convergence remained the same varying from 1 to 15 for different initializations and for different participants/classes. On a Linux machine, this took a few seconds for each participant to converge as opposed to spec-CSP that took a few minutes for each participant. Therefore, the main parameter to choose in SACSP is the number of spatial/spectral filters and $3$ seems to be a good choice given that it often converged to the same solution and increasing the number would not be of benefit. 

As mentioned earlier, one major downside of the CSP method is its poor generalizability from calibration to online control \cite{shenoy2006towards}. Various approaches can be taken to solve this issue. The first approach, as discussed in the Introduction section, is to design neurophysiologically plausible cost functions to extract robust spectral/temporal or spatial information from EEG data. 
A second approach is to make use of available EEG data from other users to learn a classifier and then fine-tune it to the new user either through a supervised or unsupervised approach \cite{jin2018adaptive, azab2019weighted,chen2022multi}. For a review, please see \cite{lotte2018review}. A third approach is to use other available classifiable information (such as the user response to the BCI erroneous output \cite{chavarriaga2014errare}) in combination with the motor imagery signal in a hybrid BCI fashion \cite{MahtaThesis, pfurtscheller2010hybrid, mousavi2020hybrid}. The proposed method of SACSP belongs to the first category of approaches above. As future work, the performance of SACSP when transferred from one participant to another should be further studied. Nevertheless, it can be easily combined with other classifiers in a hybrid fashion replacing CSP to further improve the BCI performance.

\section*{Acknowledgments}
This work was supported by NSF IIS 1219200, IIS 1817226, SMA 1041755, and IIS 1528214, FISP G2171, G3155, NIH 5T32MH020002-18, and UC San Diego Mary Anne Fox dissertation year fellowship. 

\section*{References}
\bibliographystyle{unsrt}  
\bibliography{references} 

\begin{thebibliography}{10}

\bibitem{ramoser2000optimal}
Herbert Ramoser, Johannes Muller-Gerking, and Gert Pfurtscheller.
\newblock Optimal spatial filtering of single trial {EEG} during imagined hand
  movement.
\newblock {\em IEEE Transactions on Rehabilitation Engineering}, 8(4):441--446,
  2000.

\bibitem{nicolas2012brain}
Luis~Fernando Nicolas-Alonso and Jaime Gomez-Gil.
\newblock Brain computer interfaces, a review.
\newblock {\em Sensors}, 12(2):1211--1279, 2012.

\bibitem{wierzgala2018most}
Piotr Wierzga{\l}a, Dariusz Zapa{\l}a, Grzegorz~M Wojcik, and Jolanta Masiak.
\newblock Most popular signal processing methods in motor-imagery {BCI}: a
  review and meta-analysis.
\newblock {\em Frontiers in neuroinformatics}, 12:78, 2018.

\bibitem{pfurtscheller2001motor}
Gert Pfurtscheller and Christa Neuper.
\newblock Motor imagery and direct brain-computer communication.
\newblock {\em Proceedings of the IEEE}, 89(7):1123--1134, 2001.

\bibitem{pfurtscheller2006mu}
Gert Pfurtscheller, Clemens Brunner, Alois Schl{\"o}gl, and FH~Lopes Da~Silva.
\newblock Mu rhythm (de) synchronization and {EEG} single-trial classification
  of different motor imagery tasks.
\newblock {\em NeuroImage}, 31(1):153--159, 2006.

\bibitem{allison2010could}
Brendan~Z Allison and Christa Neuper.
\newblock Could anyone use a {BCI}?
\newblock In {\em Brain-computer interfaces}, pages 35--54. Springer, 2010.

\bibitem{blankertz2008optimizing}
Benjamin Blankertz, Ryota Tomioka, Steven Lemm, Motoaki Kawanabe, and K-R
  Muller.
\newblock Optimizing spatial filters for robust {EEG} single-trial analysis.
\newblock {\em IEEE Signal Processing Magazine}, 25(1):41--56, 2008.

\bibitem{shenoy2006towards}
Pradeep Shenoy, Matthias Krauledat, Benjamin Blankertz, Rajesh~PN Rao, and
  Klaus-Robert M{\"u}ller.
\newblock Towards adaptive classification for {BCI}.
\newblock {\em Journal of Neural Engineering}, 3(1):R13, 2006.

\bibitem{mousavi2021motor}
Mahta Mousavi and Virginia~R de~Sa.
\newblock Motor imagery performance from calibration to online control in
  {EEG}-based brain-computer interfaces.
\newblock In {\em 2021 10th International IEEE/EMBS Conference on Neural
  Engineering (NER)}, pages 491--494. IEEE, 2021.

\bibitem{lemm2005spatio}
Steven Lemm, Benjamin Blankertz, Gabriel Curio, and K-R Muller.
\newblock Spatio-spectral filters for improving the classification of single
  trial {EEG}.
\newblock {\em IEEE transactions on biomedical engineering}, 52(9):1541--1548,
  2005.

\bibitem{zhao2009multilinear}
Qibin Zhao, Liqing Zhang, and Andrzej Cichocki.
\newblock Multilinear generalization of common spatial pattern.
\newblock In {\em 2009 IEEE International Conference on Acoustics, Speech and
  Signal Processing}, pages 525--528. IEEE, 2009.

\bibitem{noh2013canonical}
Eunho Noh and Virginia~R De~Sa.
\newblock Canonical correlation approach to common spatial patterns.
\newblock In {\em 2013 6th International IEEE/EMBS Conference on Neural
  Engineering (NER)}, pages 669--672. IEEE, 2013.

\bibitem{dornhege2006combined}
Guido Dornhege, Benjamin Blankertz, Matthias Krauledat, Florian Losch, Gabriel
  Curio, and K-R Muller.
\newblock Combined optimization of spatial and temporal filters for improving
  brain-computer interfacing.
\newblock {\em IEEE transactions on biomedical engineering}, 53(11):2274--2281,
  2006.

\bibitem{willems2009body}
Roel~M Willems, Ivan Toni, Peter Hagoort, and Daniel Casasanto.
\newblock Body-specific motor imagery of hand actions: neural evidence from
  right-and left-handers.
\newblock {\em Frontiers in Human Neuroscience}, 3:39, 2009.

\bibitem{tomioka2006spectrally}
Ryota Tomioka, Guido Dornhege, Guido Nolte, Benjamin Blankertz, Kazuyuki
  Aihara, and Klaus-Robert M{\"u}ller.
\newblock Spectrally weighted common spatial pattern algorithm for single trial
  {EEG} classification.
\newblock {\em Dept. Math. Eng., Univ. Tokyo, Tokyo, Japan, Tech. Rep}, 40,
  2006.

\bibitem{wu2014probabilistic}
Wei Wu, Zhe Chen, Xiaorong Gao, Yuanqing Li, Emery~N Brown, and Shangkai Gao.
\newblock Probabilistic common spatial patterns for multichannel {EEG}
  analysis.
\newblock {\em IEEE transactions on pattern analysis and machine intelligence},
  37(3):639--653, 2014.

\bibitem{wu2008classifying}
Wei Wu, Xiaorong Gao, Bo~Hong, and Shangkai Gao.
\newblock Classifying single-trial {EEG} during motor imagery by iterative
  spatio-spectral patterns learning ({ISSPL}).
\newblock {\em IEEE Transactions on Biomedical Engineering}, 55(6):1733--1743,
  2008.

\bibitem{higashi2012simultaneous}
Hiroshi Higashi and Toshihisa Tanaka.
\newblock Simultaneous design of {FIR} filter banks and spatial patterns for
  {EEG} signal classification.
\newblock {\em IEEE transactions on biomedical engineering}, 60(4):1100--1110,
  2012.

\bibitem{novi2007sub}
Quadrianto Novi, Cuntai Guan, Tran~Huy Dat, and Ping Xue.
\newblock Sub-band common spatial pattern ({SBCSP}) for brain-computer
  interface.
\newblock In {\em 2007 3rd International IEEE/EMBS Conference on Neural
  Engineering}, pages 204--207. IEEE, 2007.

\bibitem{ang2008filter}
Kai~Keng Ang, Zheng~Yang Chin, Haihong Zhang, and Cuntai Guan.
\newblock Filter bank common spatial pattern ({FBCSP}) in brain-computer
  interface.
\newblock In {\em Neural Networks, 2008. IJCNN 2008.(IEEE World Congress on
  Computational Intelligence). IEEE International Joint Conference on}, pages
  2390--2397. IEEE, 2008.

\bibitem{ang2012filter}
Kai~Keng Ang, Zheng~Yang Chin, Chuanchu Wang, Cuntai Guan, and Haihong Zhang.
\newblock Filter bank common spatial pattern algorithm on {BCI} competition
  {IV} datasets 2a and 2b.
\newblock {\em Frontiers in Neuroscience}, 6:39, 2012.

\bibitem{aghaei2015separable}
Amirhossein~S Aghaei, Mohammad~Shahin Mahanta, and Konstantinos~N Plataniotis.
\newblock Separable common spatio-spectral patterns for motor imagery {BCI}
  systems.
\newblock {\em IEEE Transactions on Biomedical Engineering}, 63(1):15--29,
  2015.

\bibitem{qi2015rstfc}
Feifei Qi, Yuanqing Li, and Wei Wu.
\newblock {RSTFC}: A novel algorithm for spatio-temporal filtering and
  classification of single-trial {EEG}.
\newblock {\em IEEE transactions on neural networks and learning systems},
  26(12):3070--3082, 2015.

\bibitem{li2016unified}
Xinyang Li, Cuntai Guan, Haihong Zhang, and Kai~Keng Ang.
\newblock A unified fisher’s ratio learning method for spatial filter
  optimization.
\newblock {\em IEEE transactions on neural networks and learning systems},
  28(11):2727--2737, 2016.

\bibitem{lotte2018review}
Fabien Lotte, Laurent Bougrain, Andrzej Cichocki, Maureen Clerc, Marco Congedo,
  Alain Rakotomamonjy, and Florian Yger.
\newblock A review of classification algorithms for {EEG}-based brain--computer
  interfaces: a 10 year update.
\newblock {\em Journal of Neural Engineering}, 15(3):031005, 2018.

\bibitem{lawhern2018eegnet}
Vernon~J Lawhern, Amelia~J Solon, Nicholas~R Waytowich, Stephen~M Gordon,
  Chou~P Hung, and Brent~J Lance.
\newblock {EEGN}et: a compact convolutional neural network for {EEG}-based
  brain--computer interfaces.
\newblock {\em Journal of neural engineering}, 15(5):056013, 2018.

\bibitem{mousavi2019temporally}
Mahta Mousavi and Virginia~R de~Sa.
\newblock Temporally adaptive common spatial patterns with deep convolutional
  neural networks.
\newblock In {\em 2019 41st Annual International Conference of the IEEE
  Engineering in Medicine and Biology Society (EMBC)}, pages 4533--4536. IEEE,
  2019.

\bibitem{al2021deep}
Ali Al-Saegh, Shefa~A Dawwd, and Jassim~M Abdul-Jabbar.
\newblock Deep learning for motor imagery {EEG}-based classification: A review.
\newblock {\em Biomedical Signal Processing and Control}, 63:102172, 2021.

\bibitem{Boyd2011DistributedOA}
Stephen~P. Boyd, N.~Parikh, E.~Chu, B.~Peleato, and Jonathan Eckstein.
\newblock Distributed optimization and statistical learning via the alternating
  direction method of multipliers.
\newblock {\em Found. Trends Mach. Learn.}, 3:1--122, 2011.

\bibitem{scikit-learn}
Fabian Pedregosa, Ga{\"e}l Varoquaux, Alexandre Gramfort, Vincent Michel,
  Bertrand Thirion, Olivier Grisel, Mathieu Blondel, Peter Prettenhofer, Ron
  Weiss, Vincent Dubourg, et~al.
\newblock Scikit-learn: Machine learning in {P}ython.
\newblock {\em Journal of Machine Learning Research}, 12(Oct):2825--2830, 2011.

\bibitem{ledoit2004honey}
Olivier Ledoit and Michael Wolf.
\newblock Honey, {I} shrunk the sample covariance matrix.
\newblock {\em The Journal of Portfolio Management}, 30(4):110--119, 2004.

\bibitem{mousavi2017towards}
Mahta Mousavi and Virginia~R de~Sa.
\newblock Towards elaborated feedback for training motor imagery brain computer
  interfaces.
\newblock In {\em Proceedings of the 7th Graz Brain-Computer Interface
  Conference 2017}, pages 332--337, 2017.

\bibitem{mousavi2020hybrid}
Mahta Mousavi, Laurens~Ruben Krol, and Virginia de~Sa.
\newblock Hybrid brain-computer interface with motor imagery and error-related
  brain activity.
\newblock {\em Journal of Neural Engineering}, 2020.

\bibitem{klem1999ten}
George~H Klem, Hans~Otto L{\"u}ders, HH~Jasper, C~Elger, et~al.
\newblock The ten-twenty electrode system of the {International Federation}.
\newblock {\em Electroencephalography and Clinical Neurophysiology},
  52(3):3--6, 1999.

\bibitem{snap}
UC~San~Diego Swartz Center~for Computational~Neuroscience.
\newblock Simulation and {N}euroscience {A}pplication {P}latform ({SNAP}).
\newblock \url{https://github.com/sccn/SNAP}.

\bibitem{lsl}
UC~San~Diego Swartz Center~for Computational~Neuroscience.
\newblock Lab {S}treaming {L}ayer ({LSL}).
\newblock \url{https://github.com/sccn/labstreaminglayer}.

\bibitem{numpy}
Travis~E Oliphant.
\newblock {\em A guide to NumPy}, volume~1.
\newblock Trelgol Publishing USA, 2006.

\bibitem{scipy}
Eric Jones, Travis Oliphant, Pearu Peterson, et~al.
\newblock {SciPy}: Open source scientific tools for {Python}.
\newblock 2001.

\bibitem{MATLAB18}
MATLAB and Statistics Toolbox~Release 2018b.
\newblock The MathWorks Inc., Natick, Massachusetts, United States, 2018.

\bibitem{Delorme2004}
Arnaud Delorme and Scott Makeig.
\newblock {EEGLAB}: an open source toolbox for analysis of single-trial {EEG}
  dynamics including independent component analysis.
\newblock {\em Journal of Neuroscience Methods}, 134(1):9--21, 2004.

\bibitem{mousavi2017improving}
Mahta Mousavi, Adam~S Koerner, Qiong Zhang, Eunho Noh, and Virginia~R de~Sa.
\newblock Improving motor imagery {BCI} with user response to feedback.
\newblock {\em Brain-Computer Interfaces}, 4(1-2):74--86, 2017.

\bibitem{craik2019deep}
Alexander Craik, Yongtian He, and Jose~L Contreras-Vidal.
\newblock Deep learning for electroencephalogram ({EEG}) classification tasks:
  a review.
\newblock {\em Journal of neural engineering}, 16(3):031001, 2019.

\bibitem{jin2018adaptive}
Yiming Jin, Mahta Mousavi, and Virginia~R de~Sa.
\newblock Adaptive {CSP} with subspace alignment for subject-to-subject
  transfer in motor imagery brain-computer interfaces.
\newblock In {\em 2018 6th International Conference on Brain-Computer Interface
  (BCI)}, pages 1--4. IEEE, 2018.

\bibitem{azab2019weighted}
Ahmed~M Azab, Lyudmila Mihaylova, Kai~Keng Ang, and Mahnaz Arvaneh.
\newblock Weighted transfer learning for improving motor imagery-based
  brain--computer interface.
\newblock {\em IEEE Transactions on Neural Systems and Rehabilitation
  Engineering}, 27(7):1352--1359, 2019.

\bibitem{chen2022multi}
Zhining Chen, Mahta Mousavi, and Virginia~R de~Sa.
\newblock Multi-subject unsupervised transfer with weighted subspace alignment
  for common spatial patterns.
\newblock In {\em 2022 10th International Conference on Brain-Computer
  Interface (BCI)}, pages 1--4. IEEE, 2022.

\bibitem{chavarriaga2014errare}
Ricardo Chavarriaga, Aleksander Sobolewski, and Jos{\'e} del~R Mill{\'a}n.
\newblock Errare machinale est: the use of error-related potentials in
  brain-machine interfaces.
\newblock {\em Frontiers in Neuroscience}, 8:208, 2014.

\bibitem{MahtaThesis}
Mahta Mousavi.
\newblock {\em Novel Machine Learning and Design Methods to Improve EEG-based
  Motor Imagery Brain-Computer Interfaces}.
\newblock PhD thesis, UC San Diego, 2019.

\bibitem{pfurtscheller2010hybrid}
Gert Pfurtscheller, Brendan~Z Allison, G{\"u}nther Bauernfeind, Clemens
  Brunner, Teodoro Solis~Escalante, Reinhold Scherer, Thorsten~O Zander, Gernot
  Mueller-Putz, Christa Neuper, and Niels Birbaumer.
\newblock The hybrid {BCI}.
\newblock {\em Frontiers in Neuroscience}, 4:3, 2010.

\end{thebibliography}

\end{document}


\title[Spectrally Adaptive Common Spatial Patterns]{Spectrally Adaptive Common Spatial Patterns-- Supplementary material}

\author{Mahta Mousavi,  Eric Lybrand,  Shuangquan Feng, Shuai Tang, Rayan Saab, Virginia de Sa}

\ead{mahta@ucsd.edu, desa@ucsd.edu}
\vspace{10pt}

\section{Results -- cross-validation accuracy}
Table \ref{CV-csp} shows the classification accuracy of the CSP method when trained and tested on calibration data through cross-validation (called `calibCV' and depicted in blue color), when trained and tested on online data through cross-validation (called `onlineCV and depicted in red color) and when trained on calibration data and tested on the online data as reported in the main article (called `test' and depicted in bold font). Similarly, table \ref{CV-sacsp} shows the classification accuracy of the SACSP method when trained and tested on calibration data through cross-validation (called `calibCV' and depicted in blue color), when trained and tested on online data through cross-validation (called `onlineCV and depicted in red color) and when trained on calibration data and tested on the online data as reported in the main article (called `test' and depicted in bold font). 

Note that the `test' condition (reported in bold font here) is the natural scenario that is relevant for practical applications, i.e., trained on calibration data and tested on online data. This is also reported in the main text in Table 1. As for the CV rates, note that CSP performs significantly better in calibCV and onlineCV conditions (red and blue fonts) compared to the test condition in bold font (Wilcoxon signed rank test $p<0.001$). This suggests that CSP is overfitting to the data in calibration and online blocks. On the other hand, SACSP does not perform different in the onlineCV (red font) and the test condition in black bold font (Wilcoxon signed-rank test $p=0.97$) suggesting that SACSP is not suffering from overfitting as CSP does.

\begin{table}
  \centering
  \caption{Comparison of motor imagery (right/left hand) classification accuracy with \underline{CSP} when trained and tested on calibration data ({\color{blue}{CSP-calibCV}}), trained and tested on online data ({\color{red}{CSP-onlineCV}}) and trained on calibration and tested on online data (\textbf{CSP-test}). For P1-P12, the first number in each entry reports the average accuracy across 10 instances of the data and the second number reports the standard deviation of the mean. In the last row, the first number in each entry reports the average classification accuracy across participants for the corresponding column and the second number reports the standard error of the mean.  }
  \begin{tabular}{cccc}
    \toprule
    PID  & \color{blue}{CSP-calibCV} &  \color{red}{CSP-onlineCV} & \textbf{CSP-test}  \\ 
    \midrule
P1  &  \color{blue}{0.73/0.019}  & \color{red}{0.84/0.023}  &\textbf{0.55/0.035} \\
P2  &  \color{blue}{0.74/0.021}  & \color{red}{0.70/0.022}  & \textbf{0.56/0.028}  \\
P3  &  \color{blue}{0.75/0.024}  & \color{red}{0.78/0.050}  & \textbf{0.63/0.038} \\
P4  &  \color{blue}{0.87/0.027}  & \color{red}{0.72/0.017}  & \textbf{0.72/0.019} \\
P5  &  \color{blue}{0.79/0.051}  & \color{red}{0.81/0.026}  & \textbf{0.59/0.118} \\
P6  &  \color{blue}{0.60/0.046}  & \color{red}{0.57/0.022}  & \textbf{0.50/0.014} \\
P7  &  \color{blue}{0.63/0.062}  & \color{red}{0.77/0.036}  & \textbf{0.62/0.027} \\
P8  &  \color{blue}{0.67/0.039}  & \color{red}{0.70/0.044}  & \textbf{0.57/0.023} \\
P9  &  \color{blue}{0.81/0.040}  & \color{red}{0.73/0.017}  & \textbf{0.68/0.048} \\ 
P10  &  \color{blue}{0.73/0.039}  & \color{red}{0.65/0.023}  & \textbf{0.49/0.021} \\
P11  &  \color{blue}{0.81/0.032}  & \color{red}{0.64/0.050}  & \textbf{0.60/0.019} \\
P12  &  \color{blue}{0.67/0.038}  & \color{red}{0.59/0.019}  & \textbf{0.47/0.032} \\
\midrule
Average  &  \color{blue}{0.73/0.023}  & \color{red}{0.71/0.024}  & \textbf{0.58/0.022}\\
    \bottomrule
  \end{tabular}
  \label{CV-csp}
\end{table}

\begin{table}
  \centering
  \caption{Comparison of motor imagery (right/left hand) classification accuracy with \underline{SACSP} when trained and tested on calibration data ({\color{blue}{SACSP-calibCV}}), trained and tested on online data ({\color{red}{SACSP-onlineCV}}) and trained on calibration and tested on online data (\textbf{SACSP-test}). For P1-P12, the first number in each entry reports the average accuracy across 10 instances of the data and the second number reports the standard deviation of the mean. In the last row, the first number in each entry reports the average classification accuracy across participants for the corresponding column and the second number reports the standard error of the mean.  }
  \begin{tabular}{cccc}
    \toprule
    PID  & \color{blue}{SACSP-calibCV} & \color{red}{SACSP-onilneCV} & \textbf{SACSP-test}  \\ 
    \midrule
P1  &  \color{blue}{0.79/0.020}  & \color{red}{0.82/0.023}  & \textbf{0.81/0.022}  \\
P2  &  \color{blue}{0.68/0.018}  & \color{red}{0.67/0.021}  & \textbf{0.69/0.030} \\
P3  &  \color{blue}{0.77/0.046}  & \color{red}{0.79/0.027}  & \textbf{0.72/0.096}  \\
P4  &  \color{blue}{0.89/0.013}  & \color{red}{0.71/0.017}  & \textbf{0.73/0.009} \\
P5  &  \color{blue}{0.65/0.103}  & \color{red}{0.52/0.040}  & \textbf{0.62/0.117}  \\
P6  &  \color{blue}{0.56/0.022}  & \color{red}{0.55/0.022}  & \textbf{0.52/0.013} \\
P7  &  \color{blue}{0.68/0.036}  & \color{red}{0.79/0.015}  & \textbf{0.70/0.040}  \\
P8  &  \color{blue}{0.72/0.030}  & \color{red}{0.71/0.032}  & \textbf{0.72/0.036} \\
P9  &  \color{blue}{0.85/0.022}  & \color{red}{0.69/0.018}  & \textbf{0.76/0.023}  \\ 
P10  &  \color{blue}{0.75/0.030} & \color{red}{0.57/0.035}  & \textbf{0.50/0.019}  \\
P11  &  \color{blue}{0.76/0.023}  & \color{red}{0.63/0.032}  & \textbf{0.67/0.027}  \\
P12  & \color{blue}{0.67/0.061}  & \color{red}{0.51/0.018}  & \textbf{0.51/0.010} \\
\midrule
Average  &  \color{blue}{0.73/0.026}  & \color{red}{0.66/0.031}  & \textbf{0.66/0.030}\\
    \bottomrule
  \end{tabular}
  \label{CV-sacsp}
\end{table}

\clearpage

\section{Results -- spatial patterns and spectral filters}
\begin{figure}
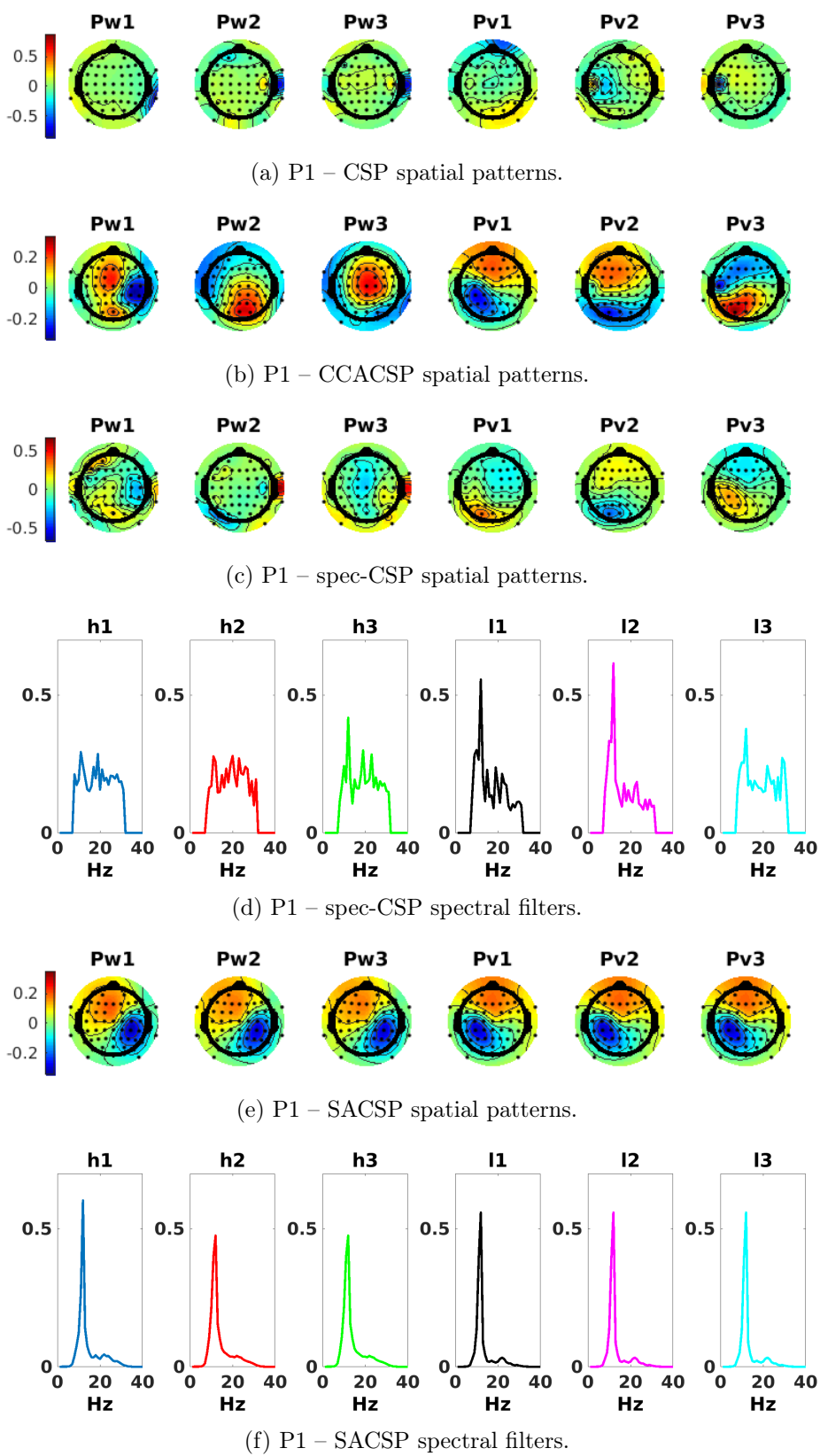

  \subfloat[P1 -- CSP spatial patterns.]{
	   \centering
	   \includegraphics[trim={0cm 4.7cm 0cm 4.2cm},clip,width=0.9\textwidth]{figs/S1_CSP_test_patts.png}}
 \hfill 	
  \subfloat[P1 -- CCACSP spatial patterns.]{
	   \centering
	   \includegraphics[trim={0cm 4.7cm 0cm 4.2cm},clip,width=0.9\textwidth]{figs/S1_CCACSP_test_patts.png}}
 \hfill	
  \subfloat[P1 -- spec-CSP spatial patterns.]{
	   \centering
	   \includegraphics[trim={0cm 4.7cm 0cm 4.2cm},clip,width=0.9\textwidth]{figs/S1_specCSP_test_patts.png}}
 \hfill 	
  \subfloat[P1 -- spec-CSP spectral filters.]{
	   \centering
	   \includegraphics[trim={1cm 0.5cm 1cm 13cm},clip,width=0.9\textwidth]{figs/S1_specCSP_test_h.png}}
 \hfill	
 \subfloat[P1 -- SACSP spatial patterns.]{
	   \centering
	   \includegraphics[trim={0cm 4.7cm 0cm 4.2cm},clip,width=0.9\textwidth]{figs/S1_SACSP_test_patts.png}}
 \hfill 	
  \subfloat[P1 -- SACSP spectral filters.]{
	   \centering
	   \includegraphics[trim={1cm 0.5cm 1cm 13cm},clip,width=0.9\textwidth]{figs/S1_SACSP_test_h.png}}
\caption{Spatial patterns and spectral filters trained on the calibration data for P1. }
\label{SACSP1}
\end{figure}

\begin{figure}
  \subfloat[P2 -- CSP spatial patterns.]{
	   \centering
	   \includegraphics[trim={0cm 4.7cm 0cm 4.2cm},clip,width=0.9\textwidth]{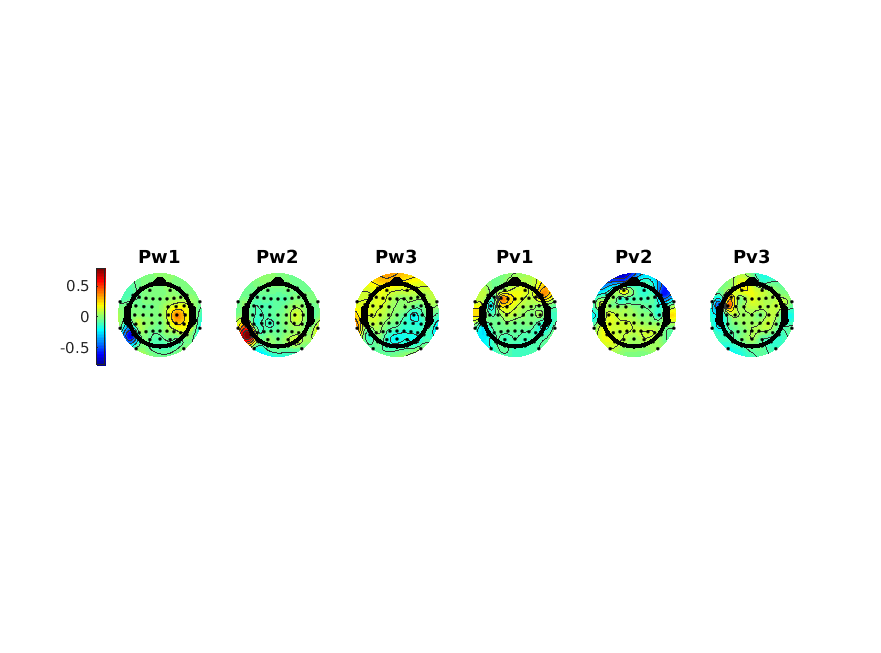}}
 \hfill 	
  \subfloat[P2 -- CCACSP spatial patterns.]{
	   \centering
	   \includegraphics[trim={0cm 4.7cm 0cm 4.2cm},clip,width=0.9\textwidth]{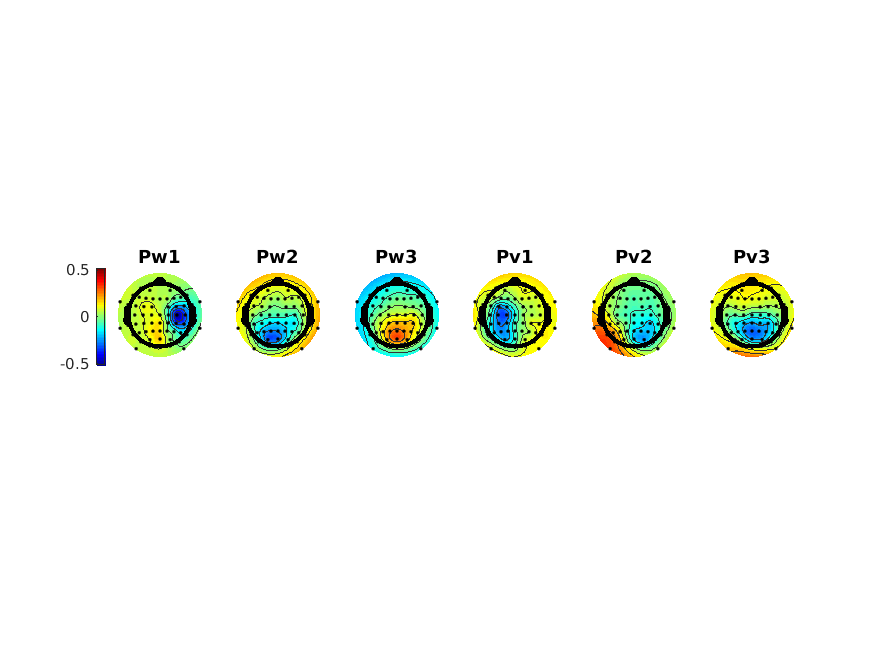}}
 \hfill	
  \subfloat[P2 -- spec-CSP spatial patterns.]{
	   \centering
	   \includegraphics[trim={0cm 4.7cm 0cm 4.2cm},clip,width=0.9\textwidth]{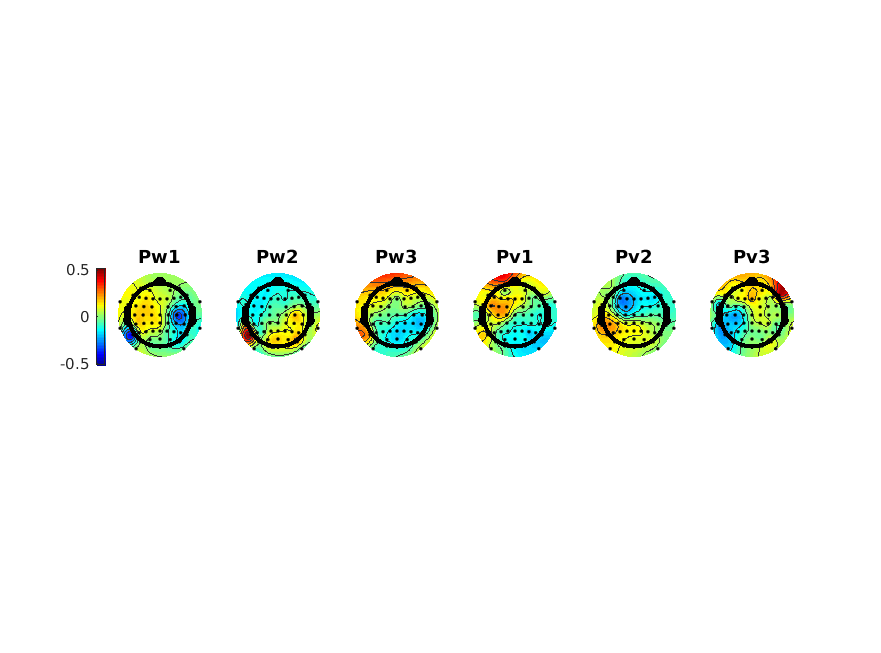}}
 \hfill 	
  \subfloat[P2 -- spec-CSP spectral filters.]{
	   \centering
	   \includegraphics[trim={1cm 0.5cm 1cm 13cm},clip,width=0.9\textwidth]{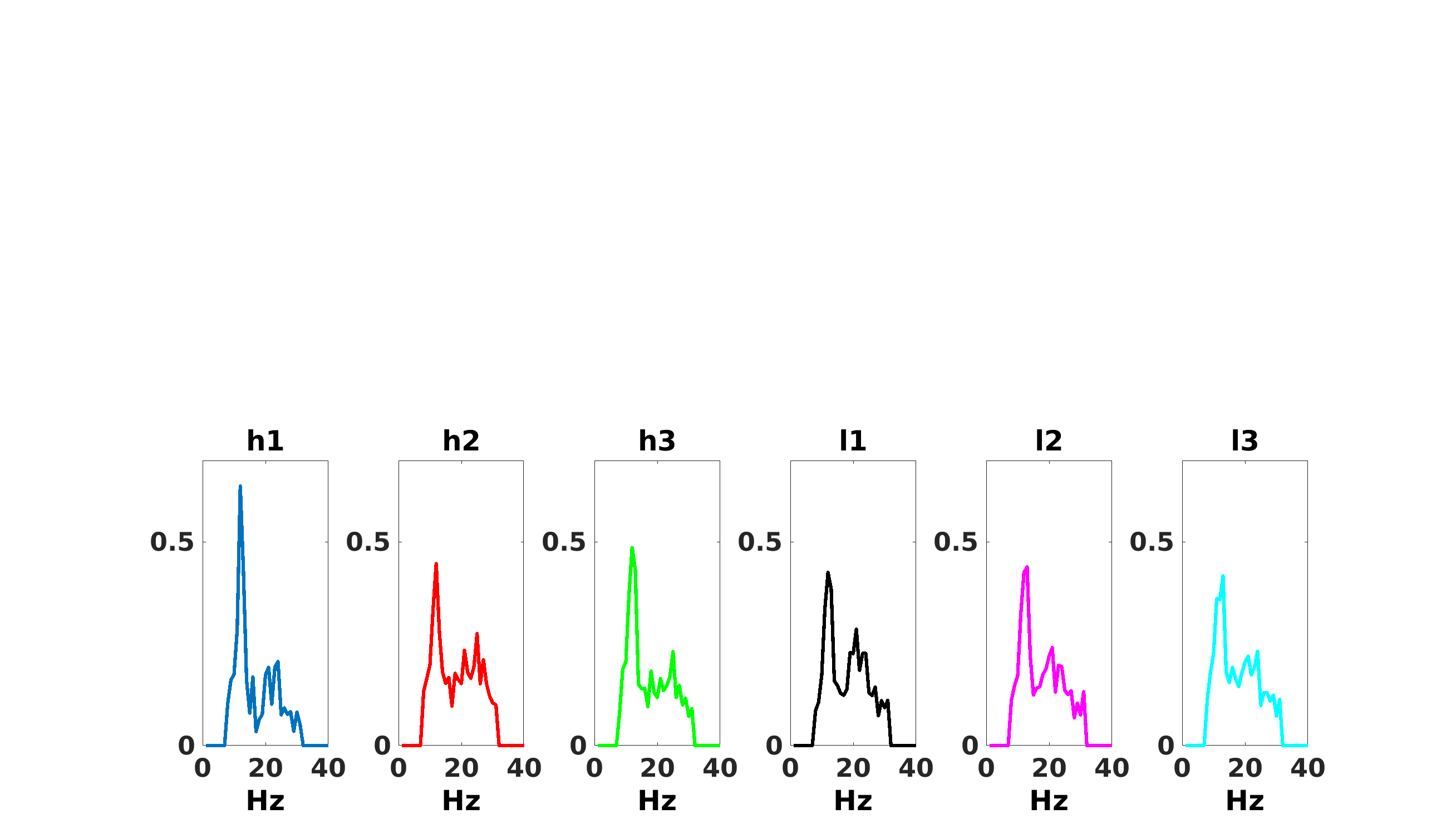}}
 \hfill	
 \subfloat[P2 -- SACSP spatial patterns.]{
	   \centering
	   \includegraphics[trim={0cm 4.7cm 0cm 4.2cm},clip,width=0.9\textwidth]{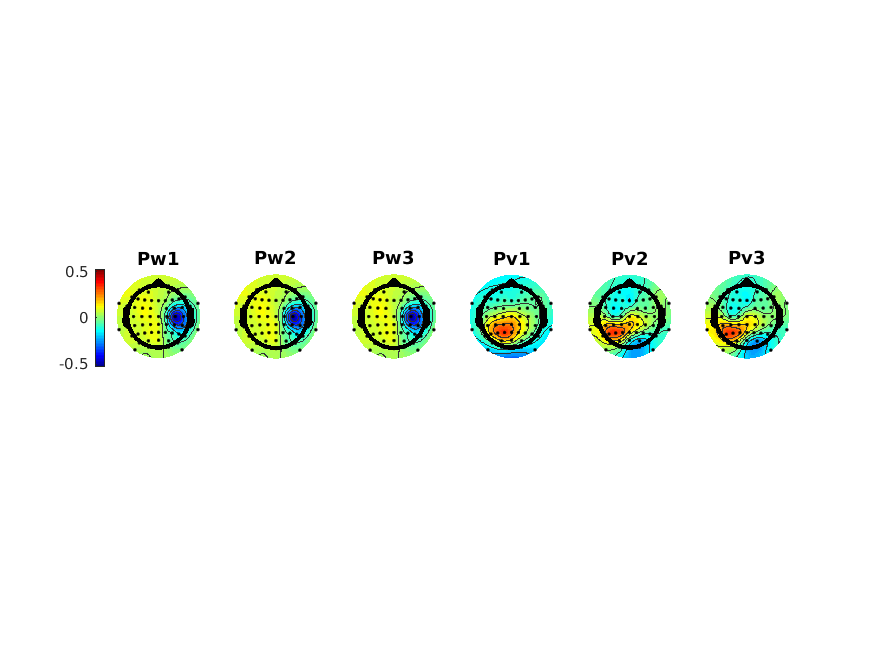}}
 \hfill 	
  \subfloat[P2 -- SACSP spectral filters.]{
	   \centering
	   \includegraphics[trim={1cm 0.5cm 1cm 13cm},clip,width=0.9\textwidth]{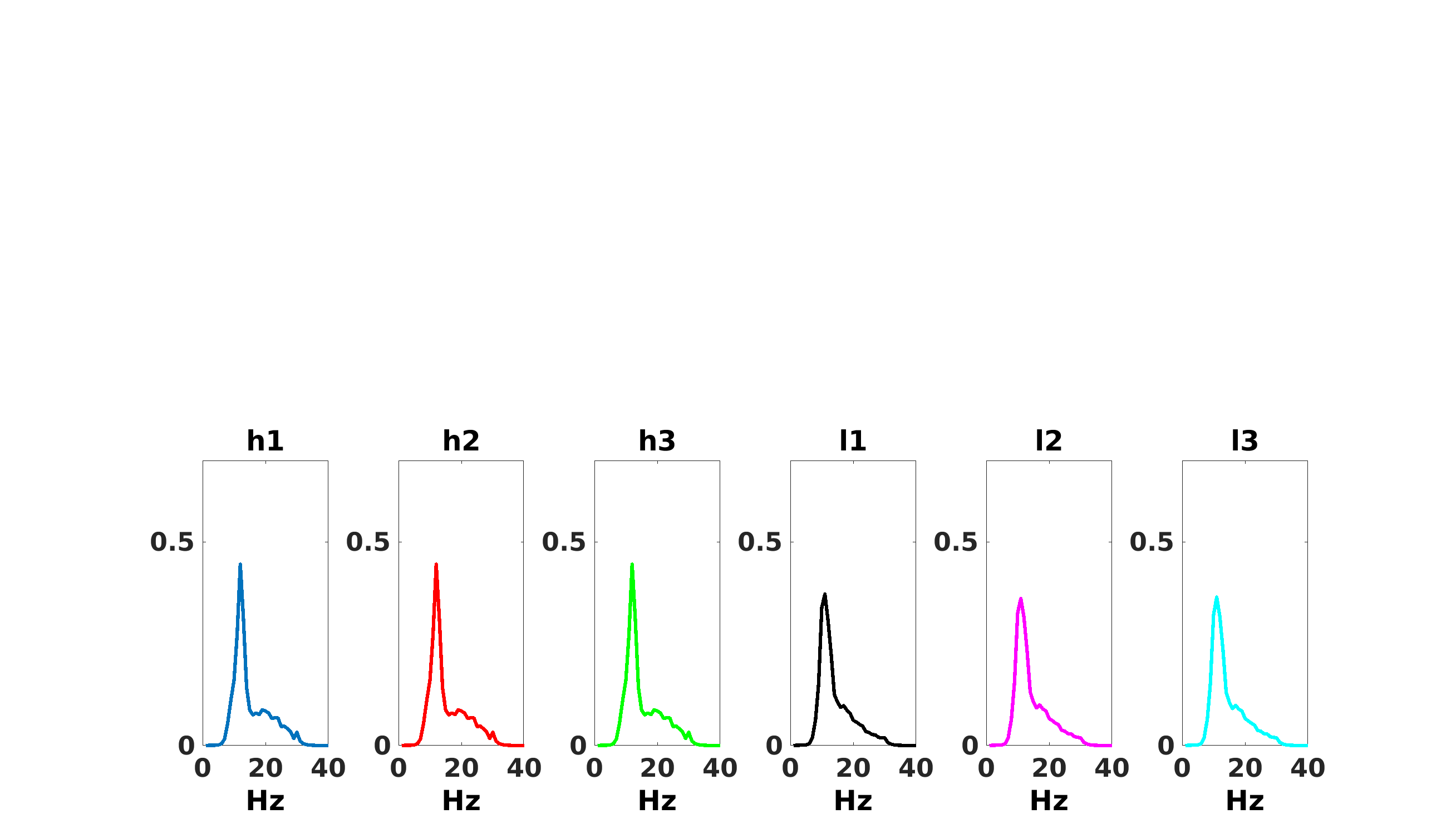}}
\caption{Spatial patterns and spectral filters trained on the calibration data for P2. }
\label{SACSP2}
\end{figure}

\begin{figure}
  \subfloat[P3 -- CSP spatial patterns.]{
	   \centering
	   \includegraphics[trim={0cm 4.7cm 0cm 4.2cm},clip,width=0.9\textwidth]{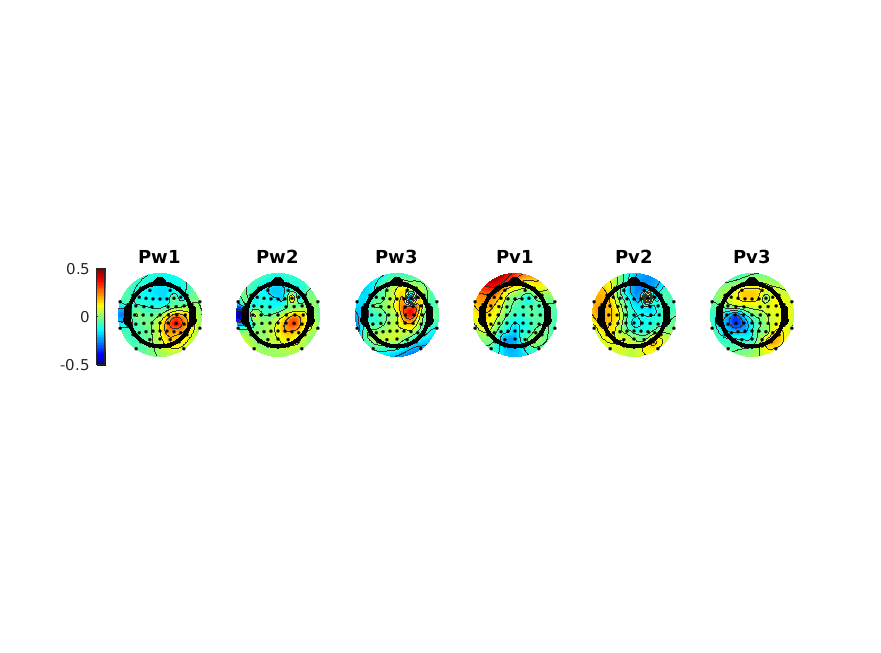}}
 \hfill 	
  \subfloat[P3 -- CCACSP spatial patterns.]{
	   \centering
	   \includegraphics[trim={0cm 4.7cm 0cm 4.2cm},clip,width=0.9\textwidth]{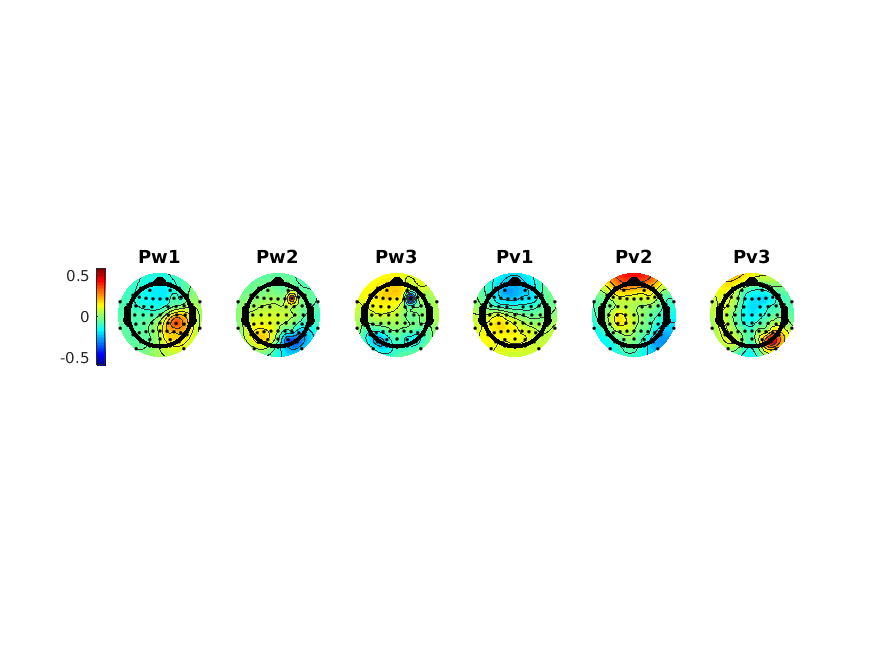}}
 \hfill	
  \subfloat[P3 -- specCSP spatial patterns.]{
	   \centering
	   \includegraphics[trim={0cm 4.7cm 0cm 4.2cm},clip,width=0.9\textwidth]{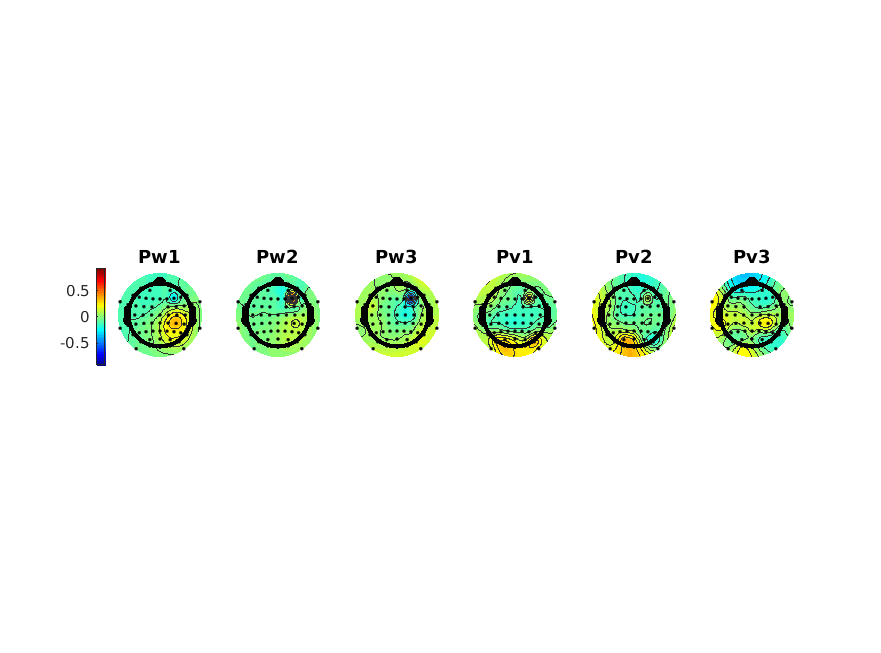}}
 \hfill 	
  \subfloat[P3 -- spec-CSP spectral filters.]{
	   \centering
	   \includegraphics[trim={1cm 0.5cm 1cm 13cm},clip,width=0.9\textwidth]{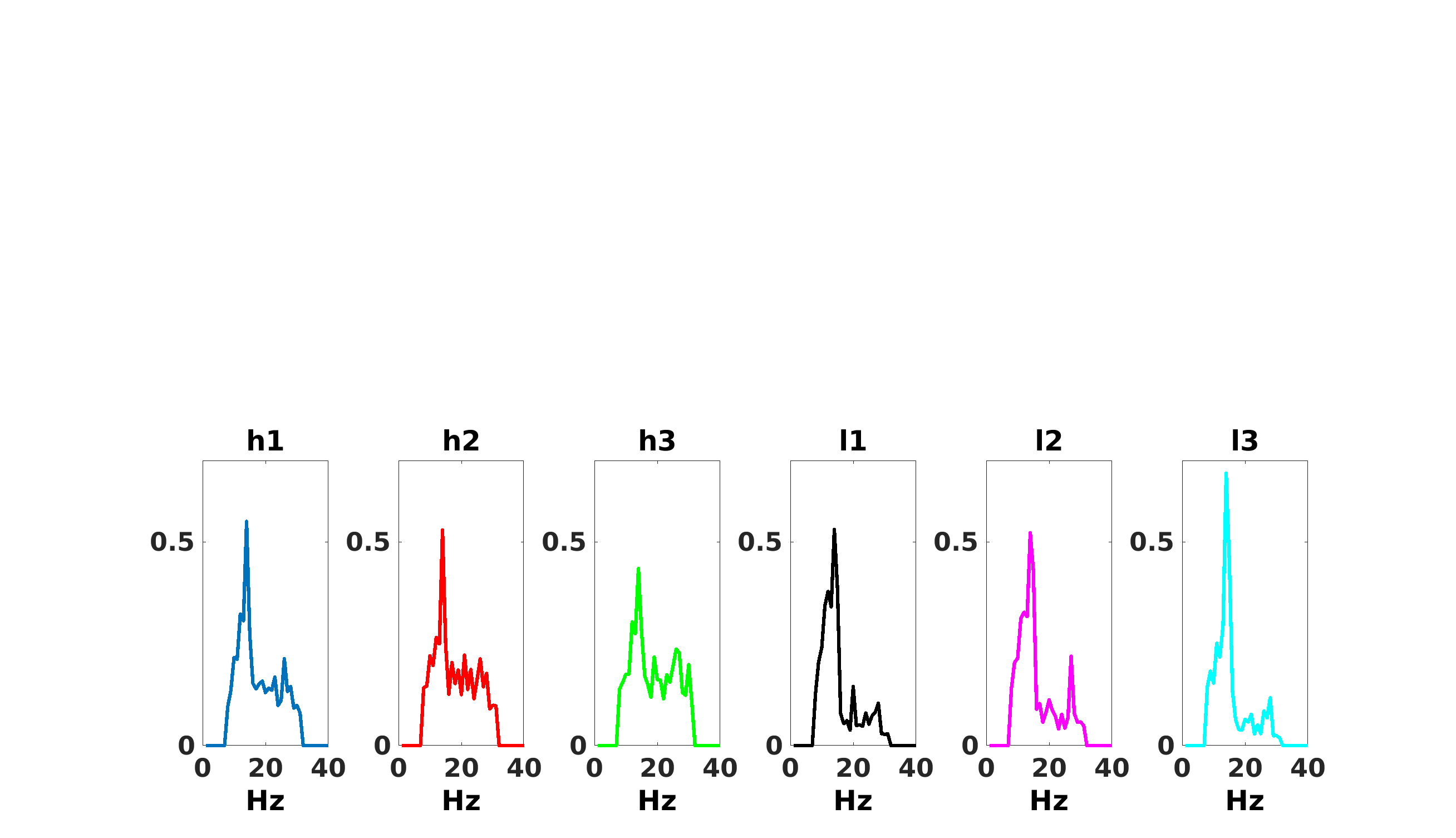}}
 \hfill	
 \subfloat[P3 -- SACSP spatial patterns.]{
	   \centering
	   \includegraphics[trim={0cm 4.7cm 0cm 4.2cm},clip,width=0.9\textwidth]{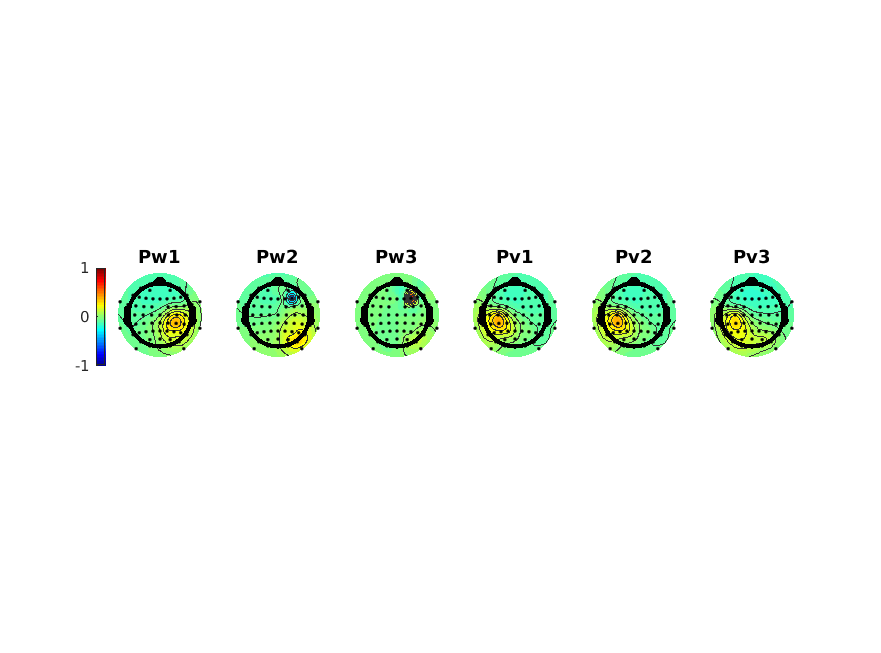}}
 \hfill 	
  \subfloat[P3 -- SACSP spectral filters.]{
	   \centering
	   \includegraphics[trim={1cm 0.5cm 1cm 13cm},clip,width=0.9\textwidth]{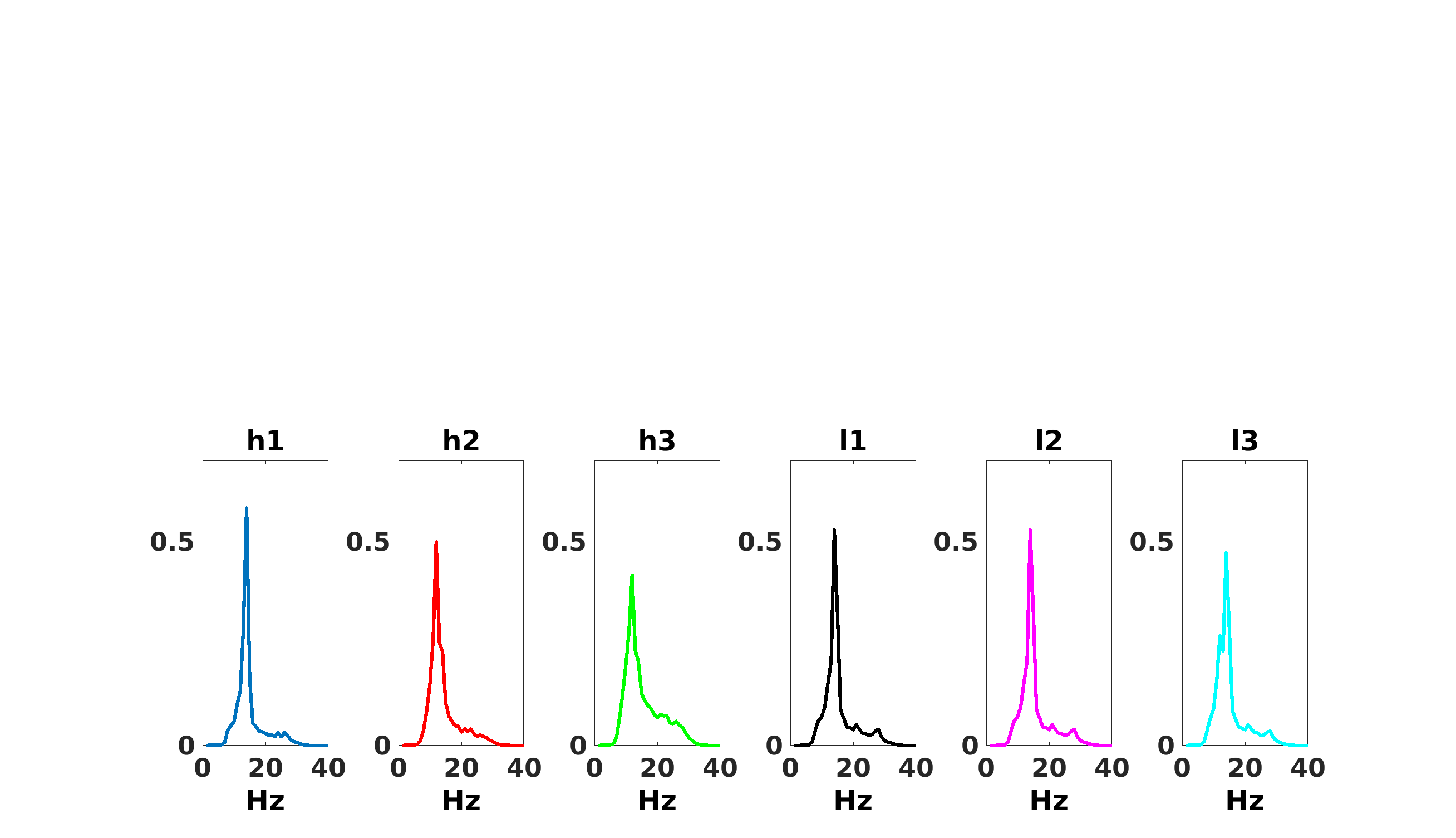}}
\caption{Spatial patterns and spectral filters trained on the calibration data for P3. }
\label{SACSP3}
\end{figure}

\begin{figure}
  \subfloat[P4 -- CSP spatial patterns.]{
	   \centering
	   \includegraphics[trim={0cm 4.7cm 0cm 4.2cm},clip,width=0.9\textwidth]{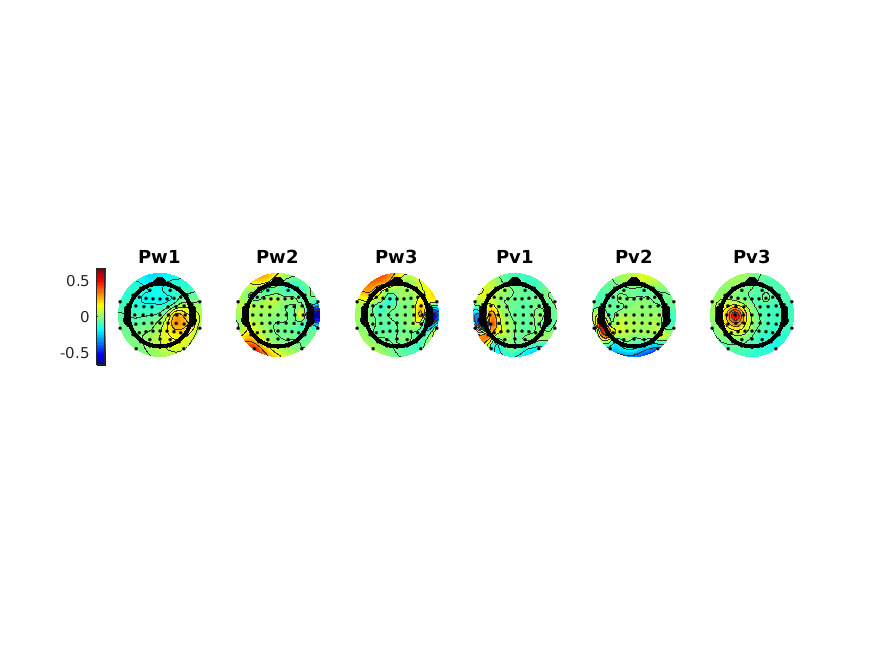}}
 \hfill 	
  \subfloat[P4 -- CCACSP spatial patterns.]{
	   \centering
	   \includegraphics[trim={0cm 4.7cm 0cm 4.2cm},clip,width=0.9\textwidth]{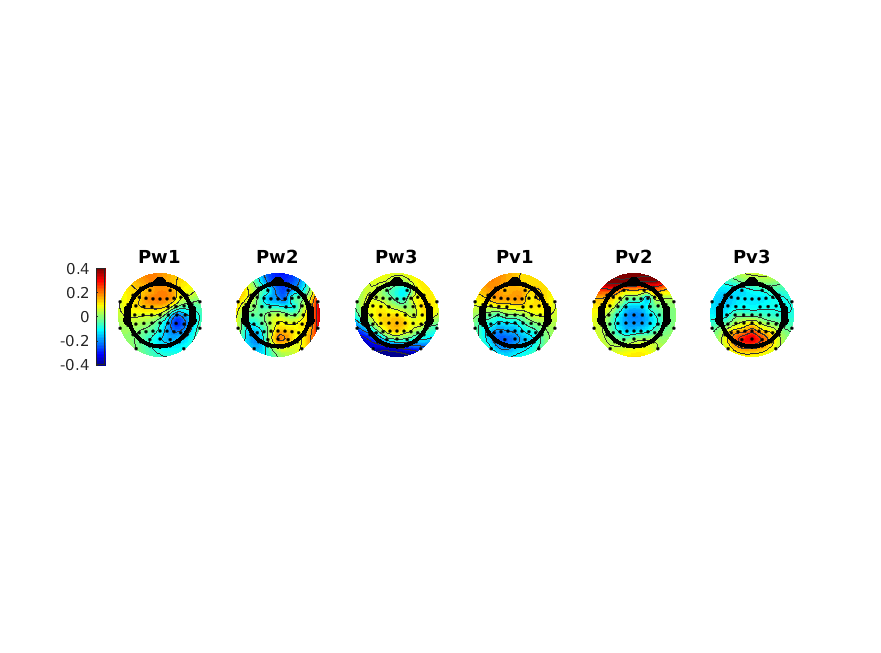}}
 \hfill	
  \subfloat[P4 -- spec-CSP spatial patterns.]{
	   \centering
	   \includegraphics[trim={0cm 4.7cm 0cm 4.2cm},clip,width=0.9\textwidth]{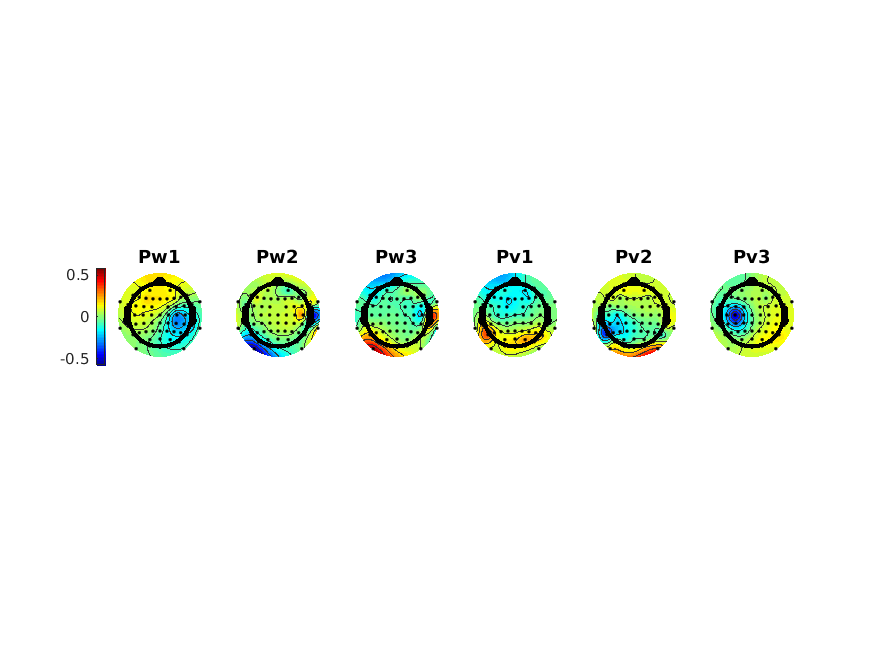}}
 \hfill 	
  \subfloat[P4 -- spec-CSP spectral filters.]{
	   \centering
	   \includegraphics[trim={1cm 0.5cm 1cm 13cm},clip,width=0.9\textwidth]{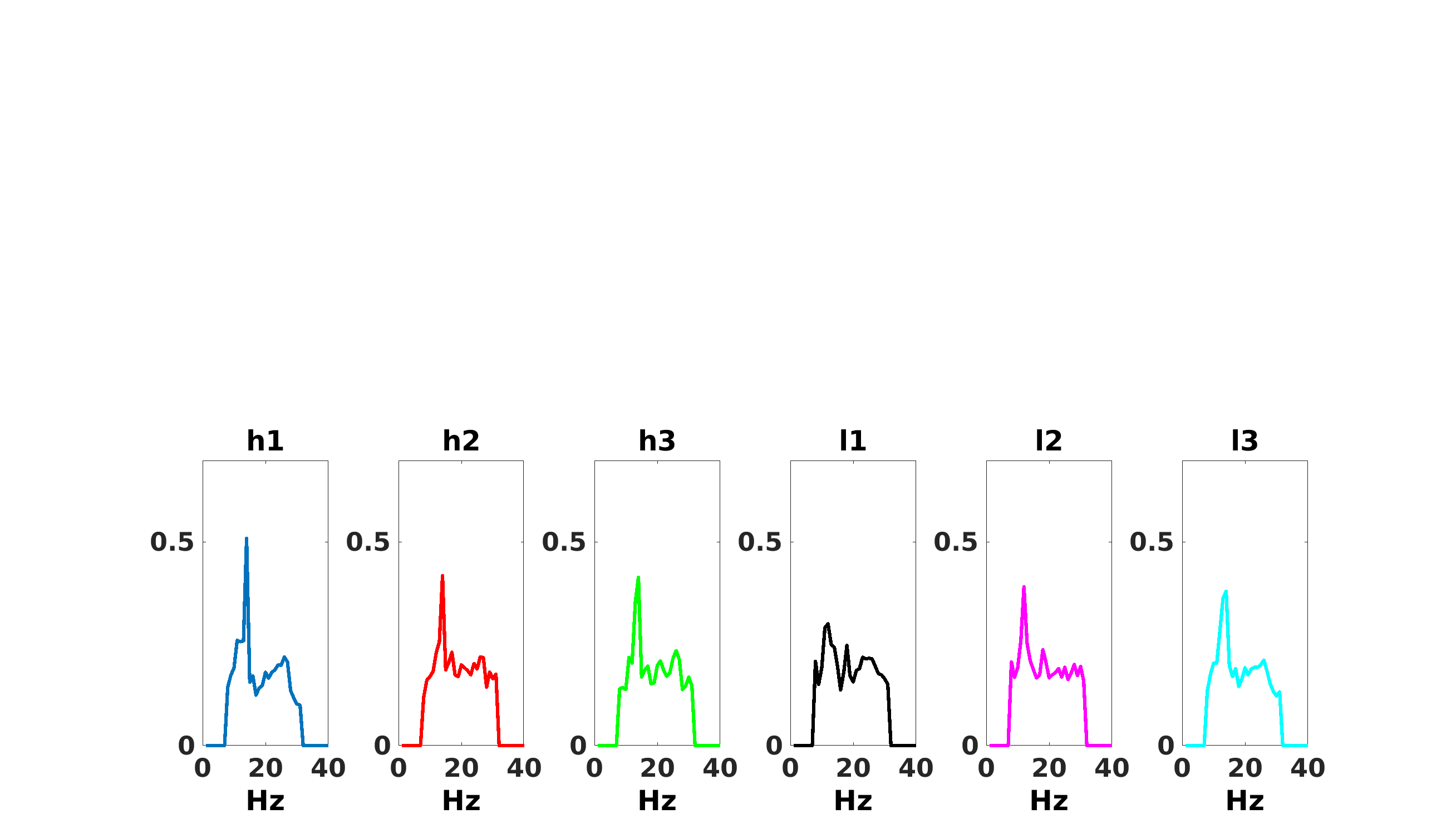}}
 \hfill	
 \subfloat[P4 -- SACSP spatial patterns.]{
	   \centering
	   \includegraphics[trim={0cm 4.7cm 0cm 4.2cm},clip,width=0.9\textwidth]{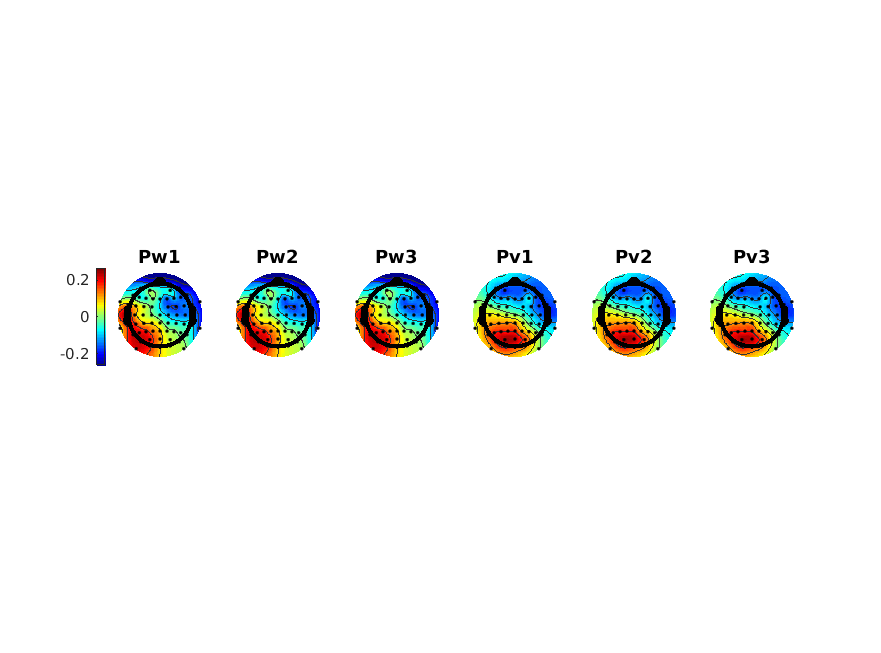}}
 \hfill 	
  \subfloat[P4 -- SACSP spectral filters.]{
	   \centering
	   \includegraphics[trim={1cm 0.5cm 1cm 13cm},clip,width=0.9\textwidth]{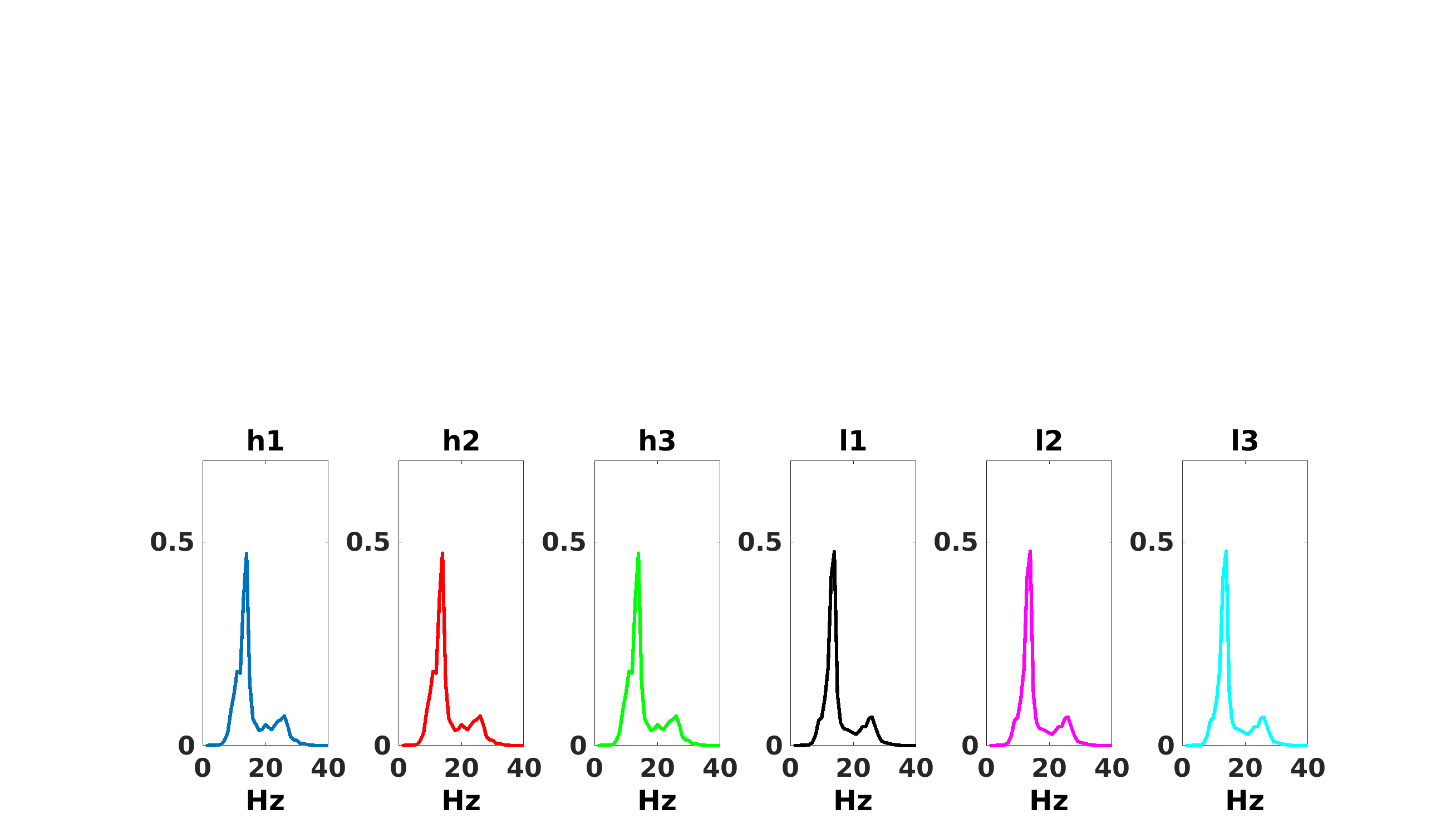}}
\caption{Spatial patterns and spectral filters trained on the calibration data for P4. }
\label{SACSP4}
\end{figure}

\begin{figure}
  \subfloat[P5 -- CSP spatial patterns.]{
	   \centering
	   \includegraphics[trim={0cm 4.7cm 0cm 4.2cm},clip,width=0.9\textwidth]{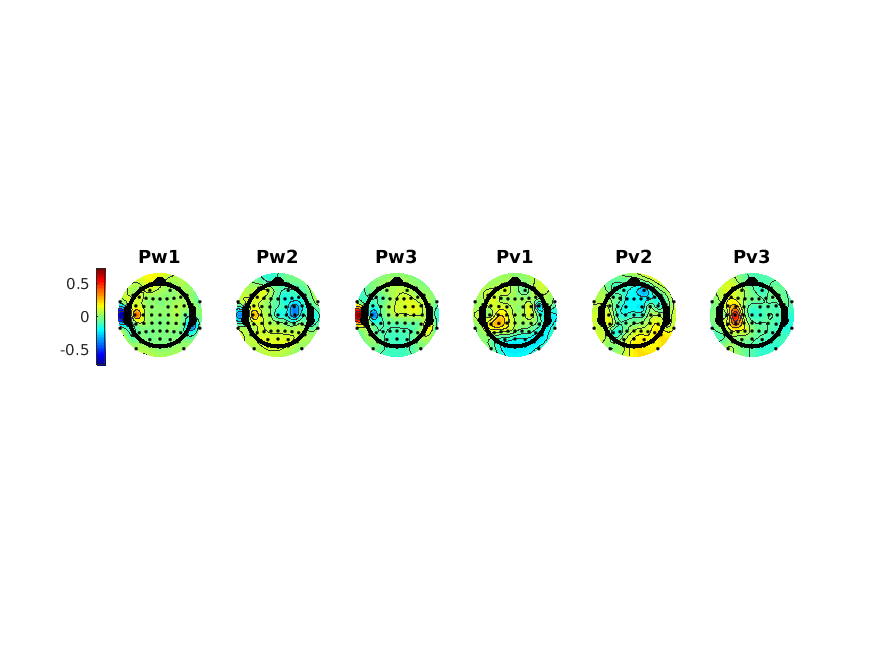}}
 \hfill 	
  \subfloat[P5 -- CCACSP spatial patterns.]{
	   \centering
	   \includegraphics[trim={0cm 4.7cm 0cm 4.2cm},clip,width=0.9\textwidth]{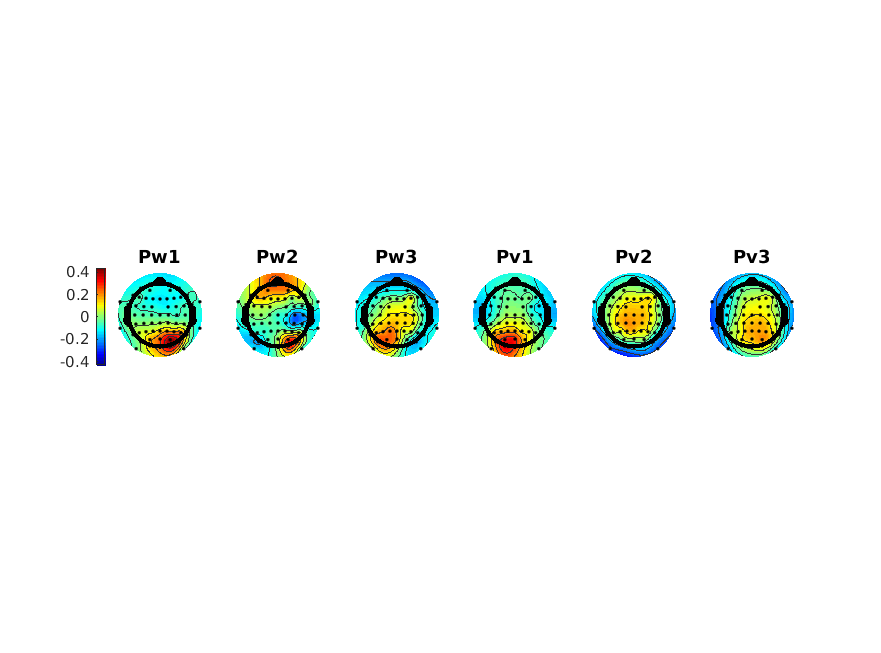}}
 \hfill	
  \subfloat[P5 -- spec-CSP spatial patterns.]{
	   \centering
	   \includegraphics[trim={0cm 4.7cm 0cm 4.2cm},clip,width=0.9\textwidth]{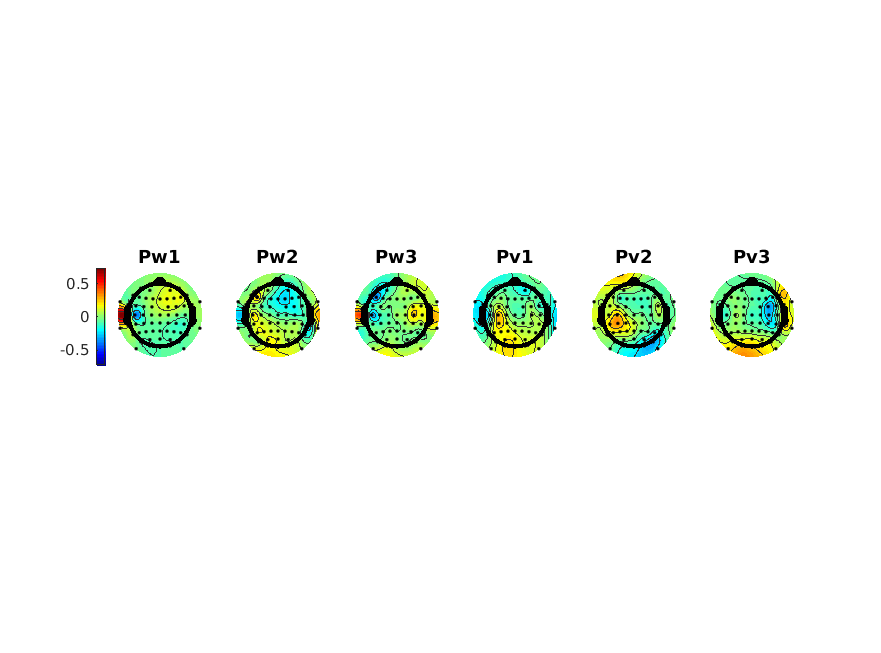}}
 \hfill 	
  \subfloat[P5 -- spec-CSP spectral filters.]{
	   \centering
	   \includegraphics[trim={1cm 0.5cm 1cm 13cm},clip,width=0.9\textwidth]{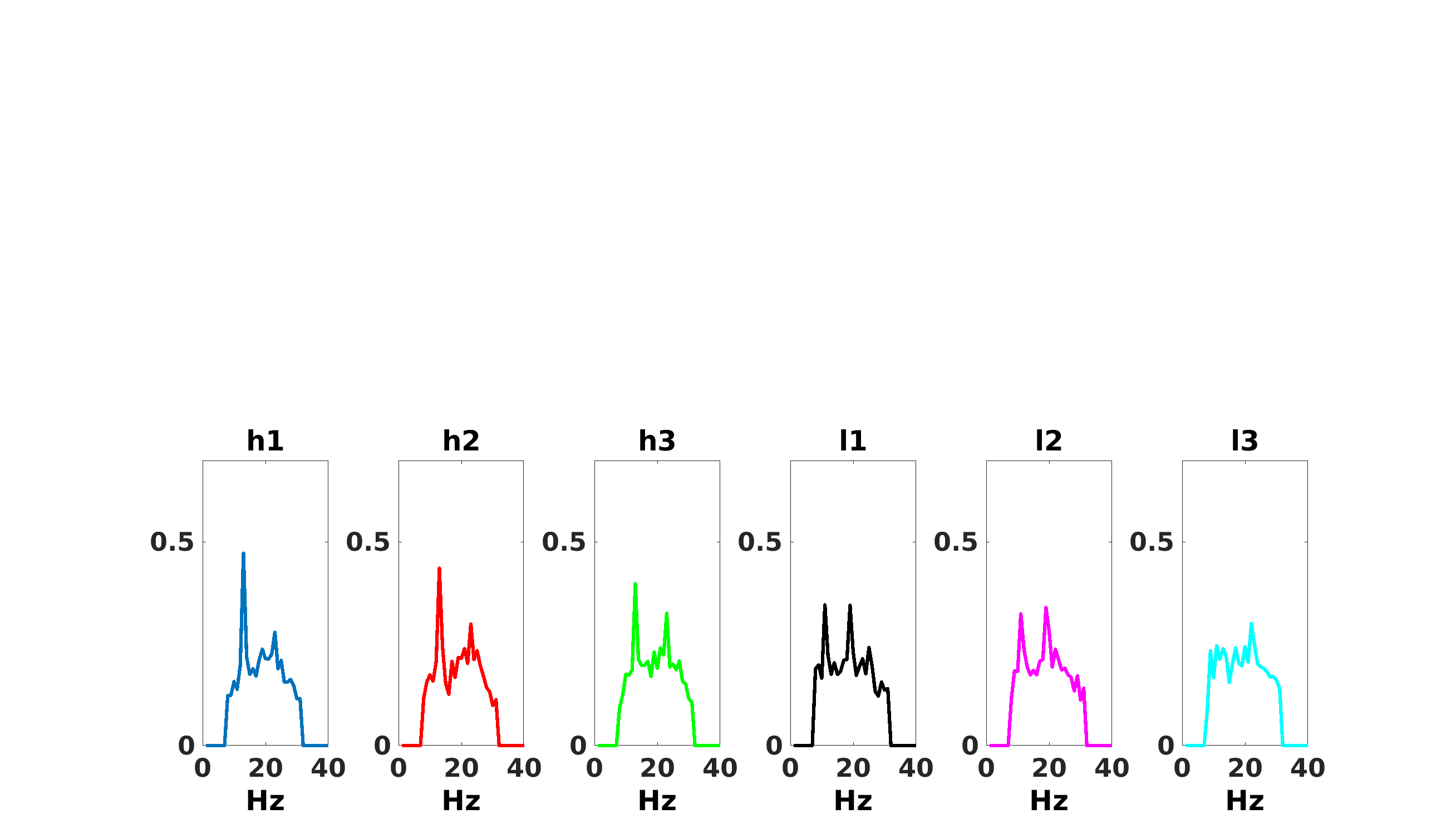}}
 \hfill	
 \subfloat[P5 -- SACSP spatial patterns.]{
	   \centering
	   \includegraphics[trim={0cm 4.7cm 0cm 4.2cm},clip,width=0.9\textwidth]{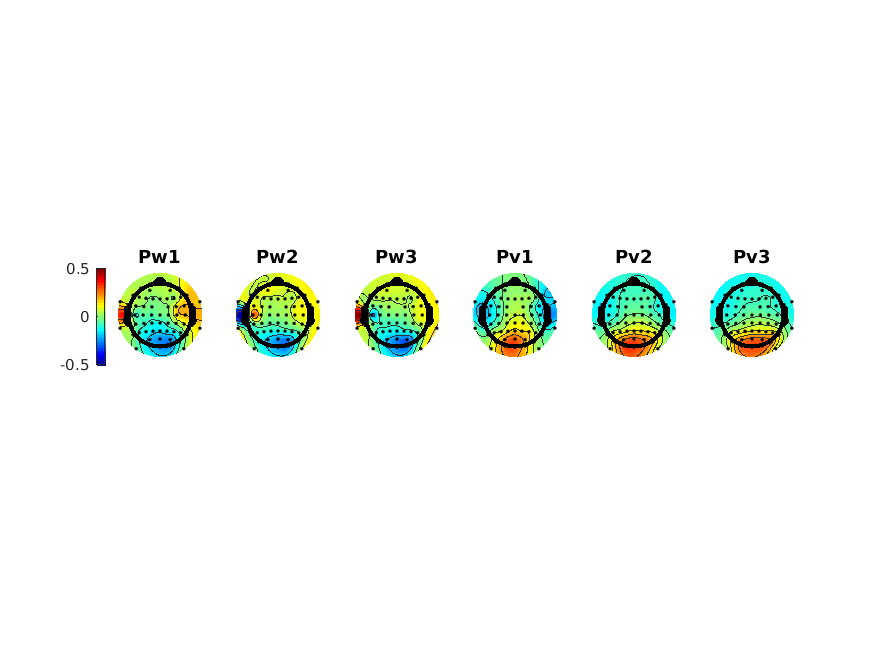}}
 \hfill 	
  \subfloat[P5 -- SACSP spectral filters.]{
	   \centering
	   \includegraphics[trim={1cm 0.5cm 1cm 13cm},clip,width=0.9\textwidth]{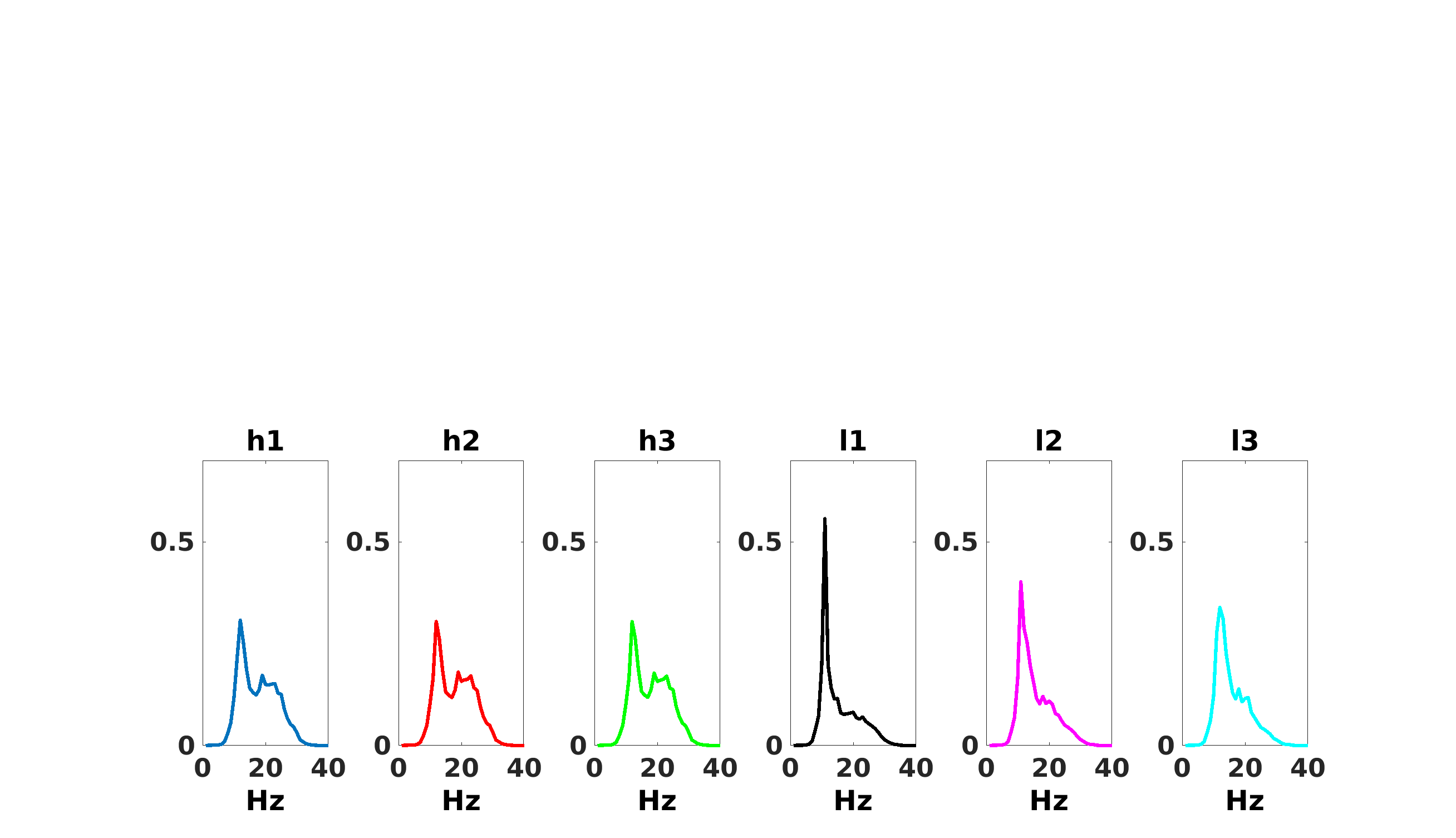}}
\caption{Spatial patterns and spectral filters trained on the calibration data for P5. }
\label{SACSP5}
\end{figure}

\begin{figure}
  \subfloat[P6 -- CSP spatial patterns.]{
	   \centering
	   \includegraphics[trim={0cm 4.7cm 0cm 4.2cm},clip,width=0.9\textwidth]{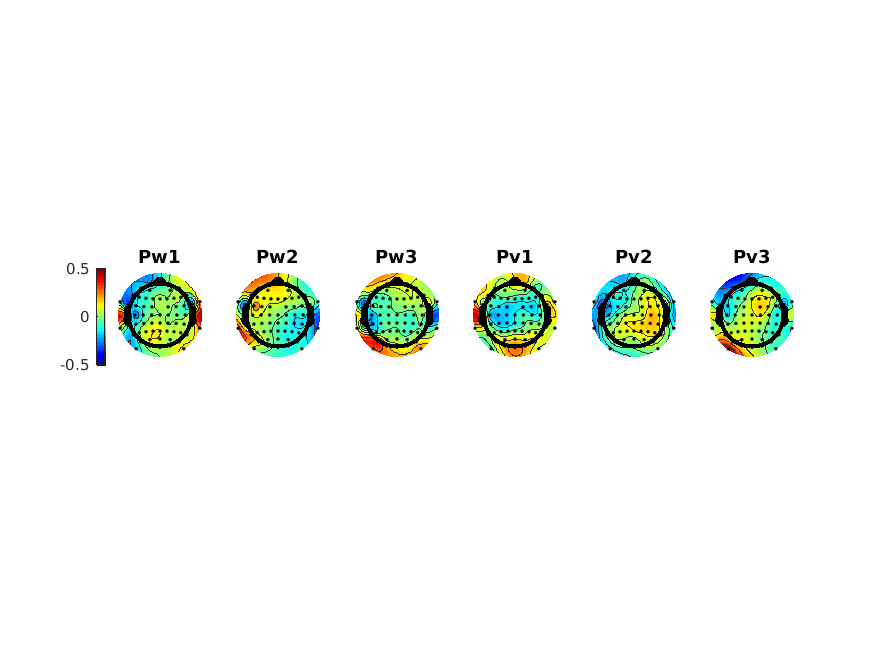}}
 \hfill 	
  \subfloat[P6 -- CCACSP spatial patterns.]{
	   \centering
	   \includegraphics[trim={0cm 4.7cm 0cm 4.2cm},clip,width=0.9\textwidth]{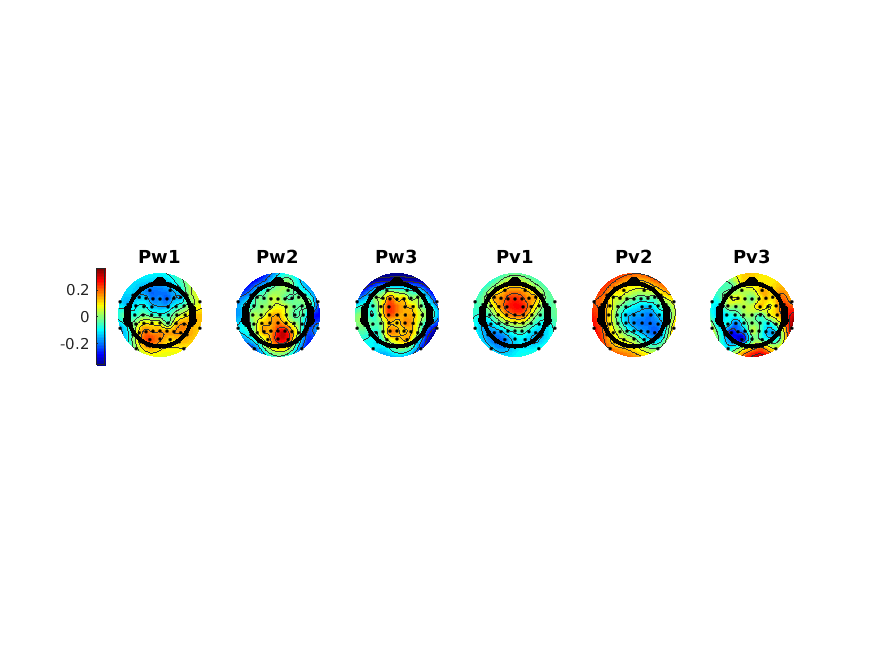}}
 \hfill	
  \subfloat[P6 -- spec-CSP spatial patterns.]{
	   \centering
	   \includegraphics[trim={0cm 4.7cm 0cm 4.2cm},clip,width=0.9\textwidth]{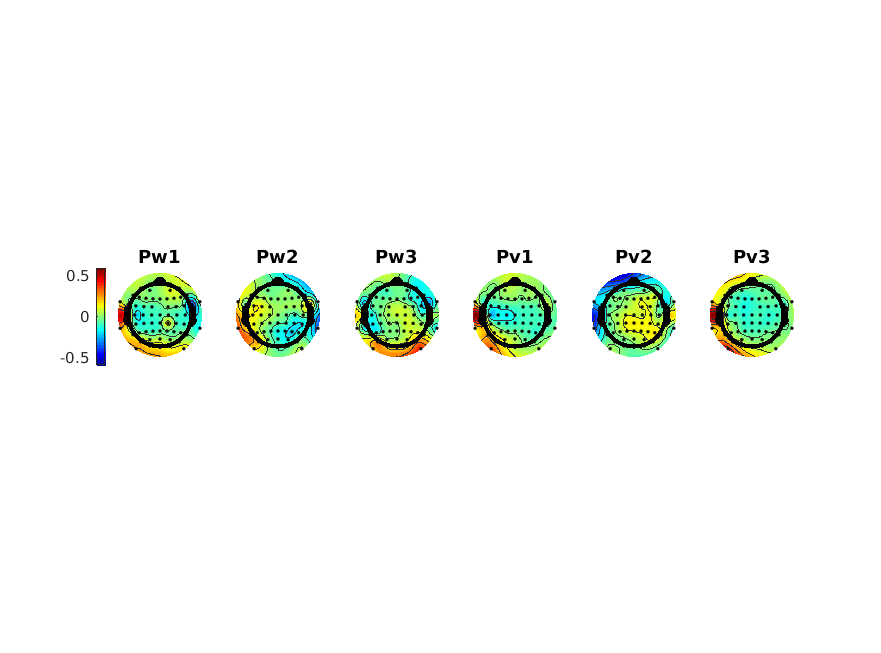}}
 \hfill 	
  \subfloat[P6 -- spec-CSP spectral filters.]{
	   \centering
	   \includegraphics[trim={1cm 0.5cm 1cm 13cm},clip,width=0.9\textwidth]{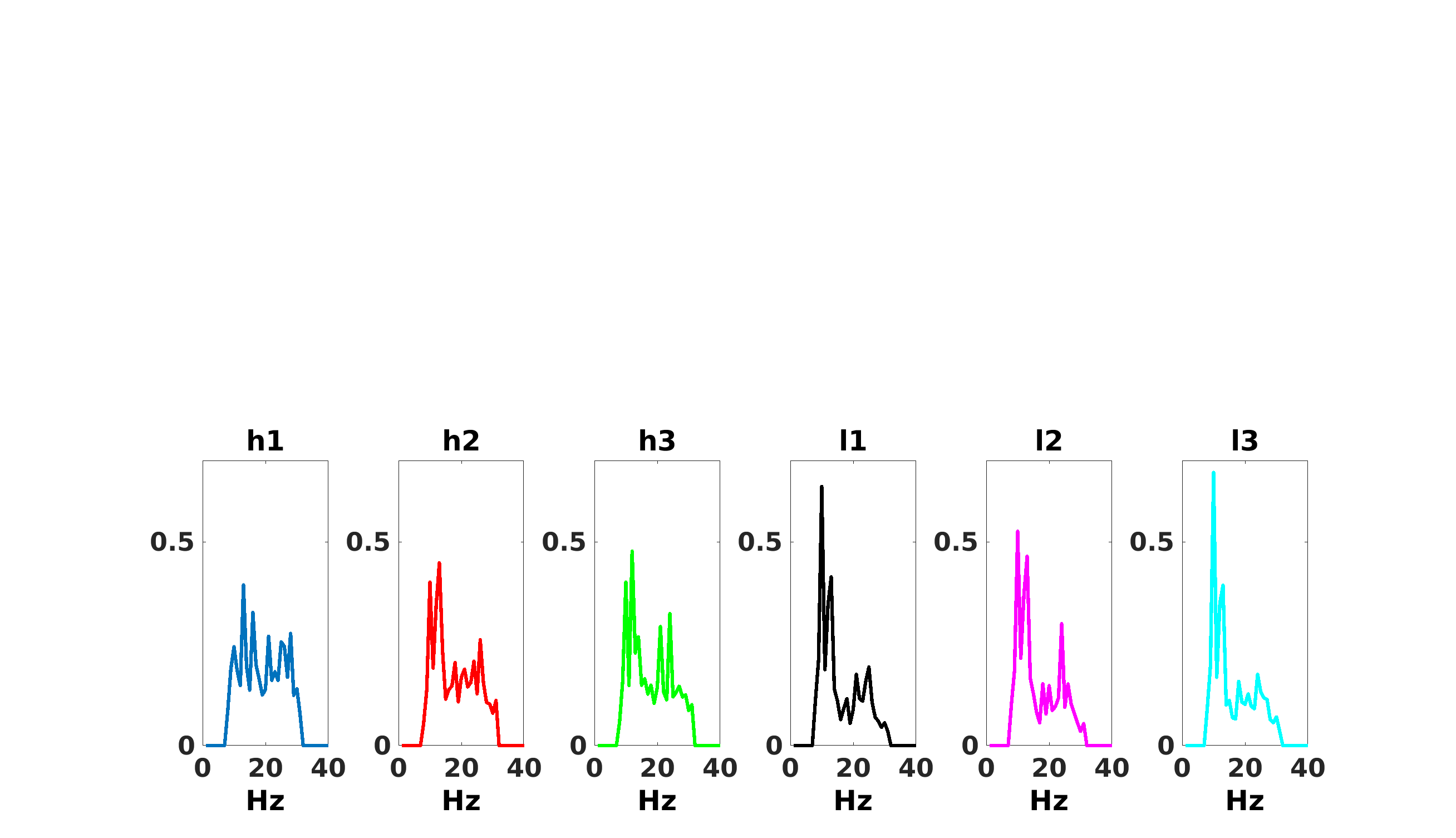}}
 \hfill	
 \subfloat[P6 -- SACSP spatial patterns.]{
	   \centering
	   \includegraphics[trim={0cm 4.7cm 0cm 4.2cm},clip,width=0.9\textwidth]{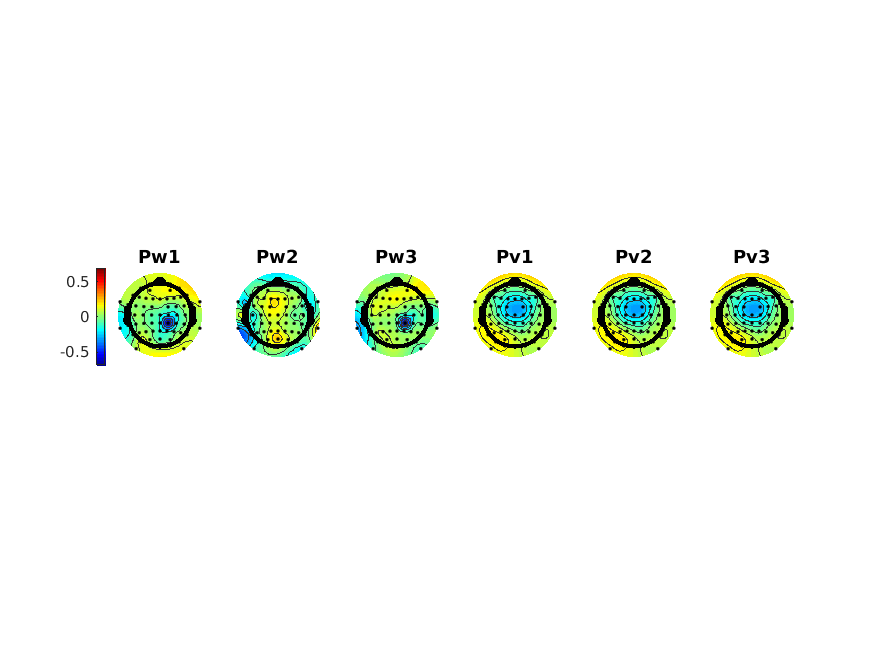}}
 \hfill 	
  \subfloat[P6 -- SACSP spectral filters.]{
	   \centering
	   \includegraphics[trim={1cm 0.5cm 1cm 13cm},clip,width=0.9\textwidth]{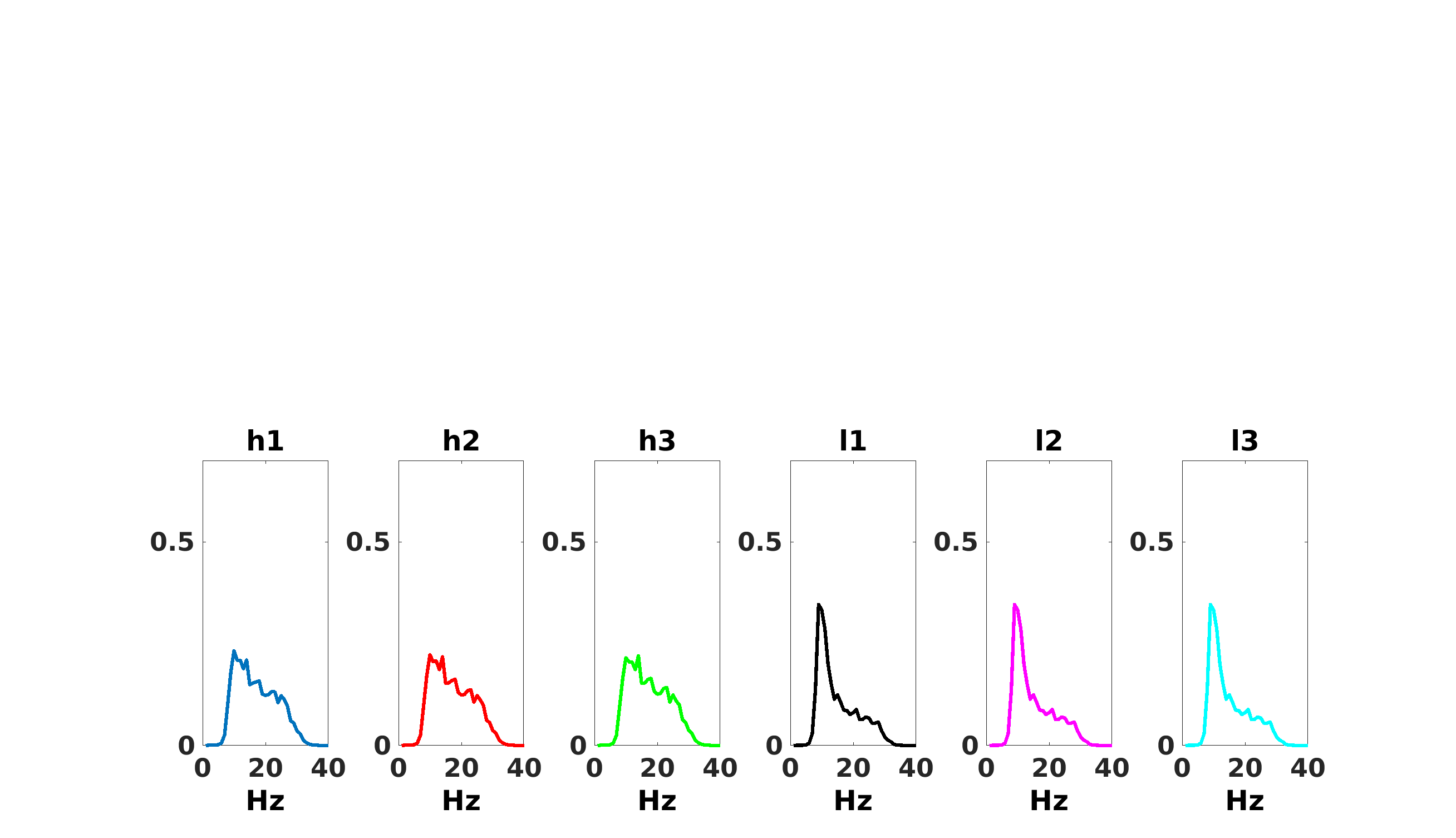}}
\caption{Spatial patterns and spectral filters trained on the calibration data for P6. }
\label{SACSP6}
\end{figure}

\begin{figure}
  \subfloat[P7 -- CSP spatial patterns.]{
	   \centering
	   \includegraphics[trim={0cm 4.7cm 0cm 4.2cm},clip,width=0.9\textwidth]{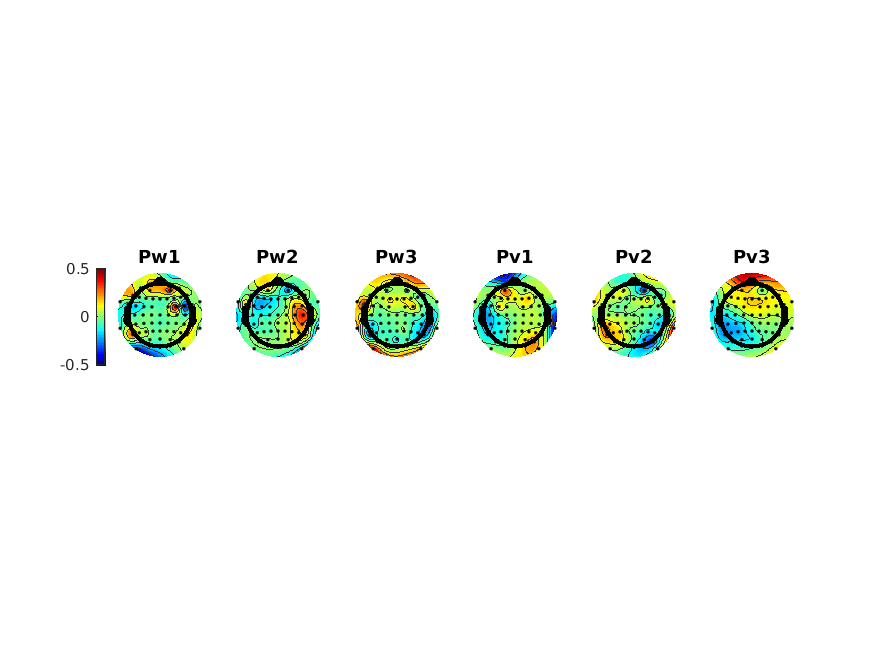}}
 \hfill 	
  \subfloat[P7 -- CCACSP spatial patterns.]{
	   \centering
	   \includegraphics[trim={0cm 4.7cm 0cm 4.2cm},clip,width=0.9\textwidth]{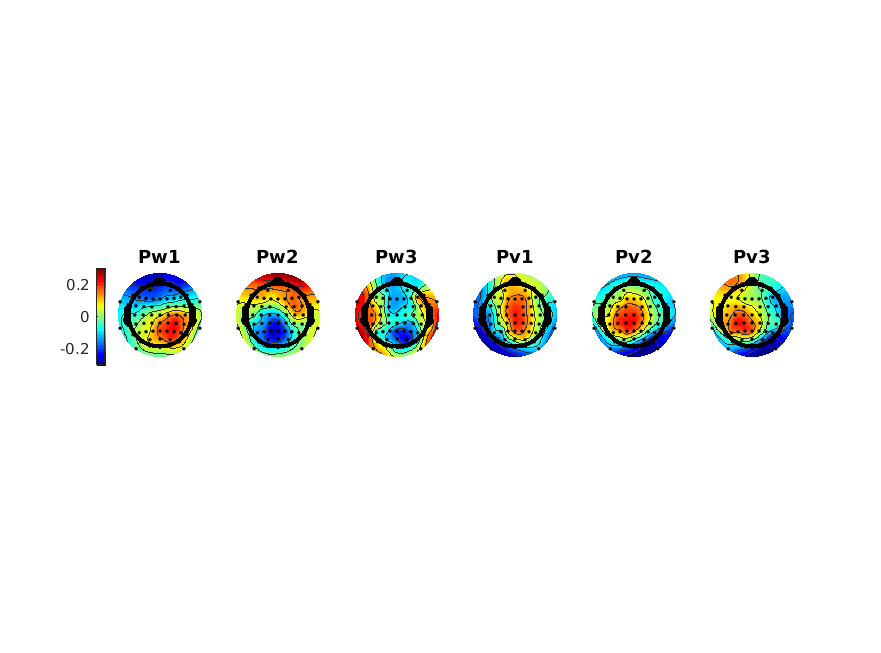}}
 \hfill	
  \subfloat[P7 -- spec-CSP spatial patterns.]{
	   \centering
	   \includegraphics[trim={0cm 4.7cm 0cm 4.2cm},clip,width=0.9\textwidth]{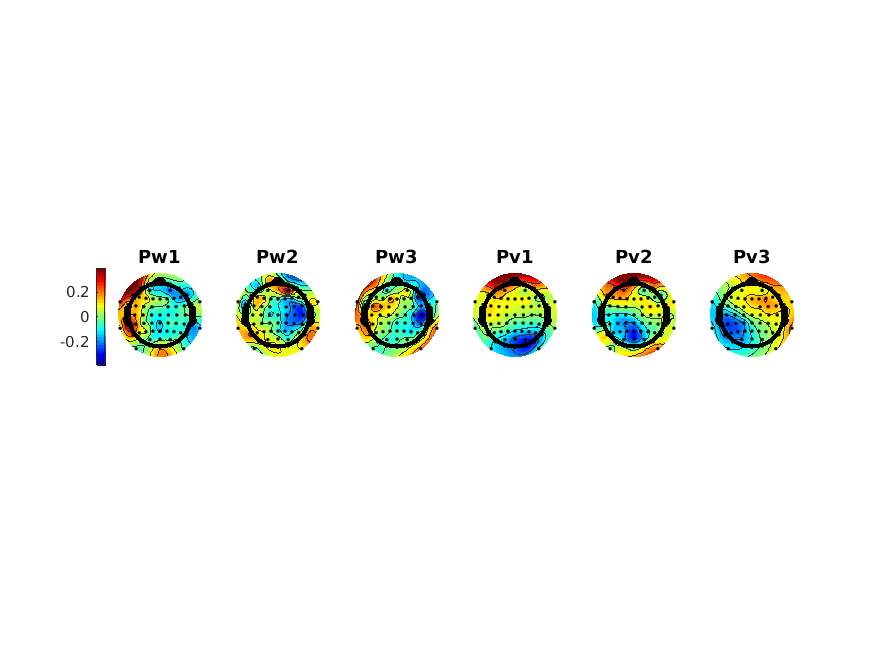}}
 \hfill 	
  \subfloat[P7 -- spec-CSP spectral filters.]{
	   \centering
	   \includegraphics[trim={1cm 0.5cm 1cm 13cm},clip,width=0.9\textwidth]{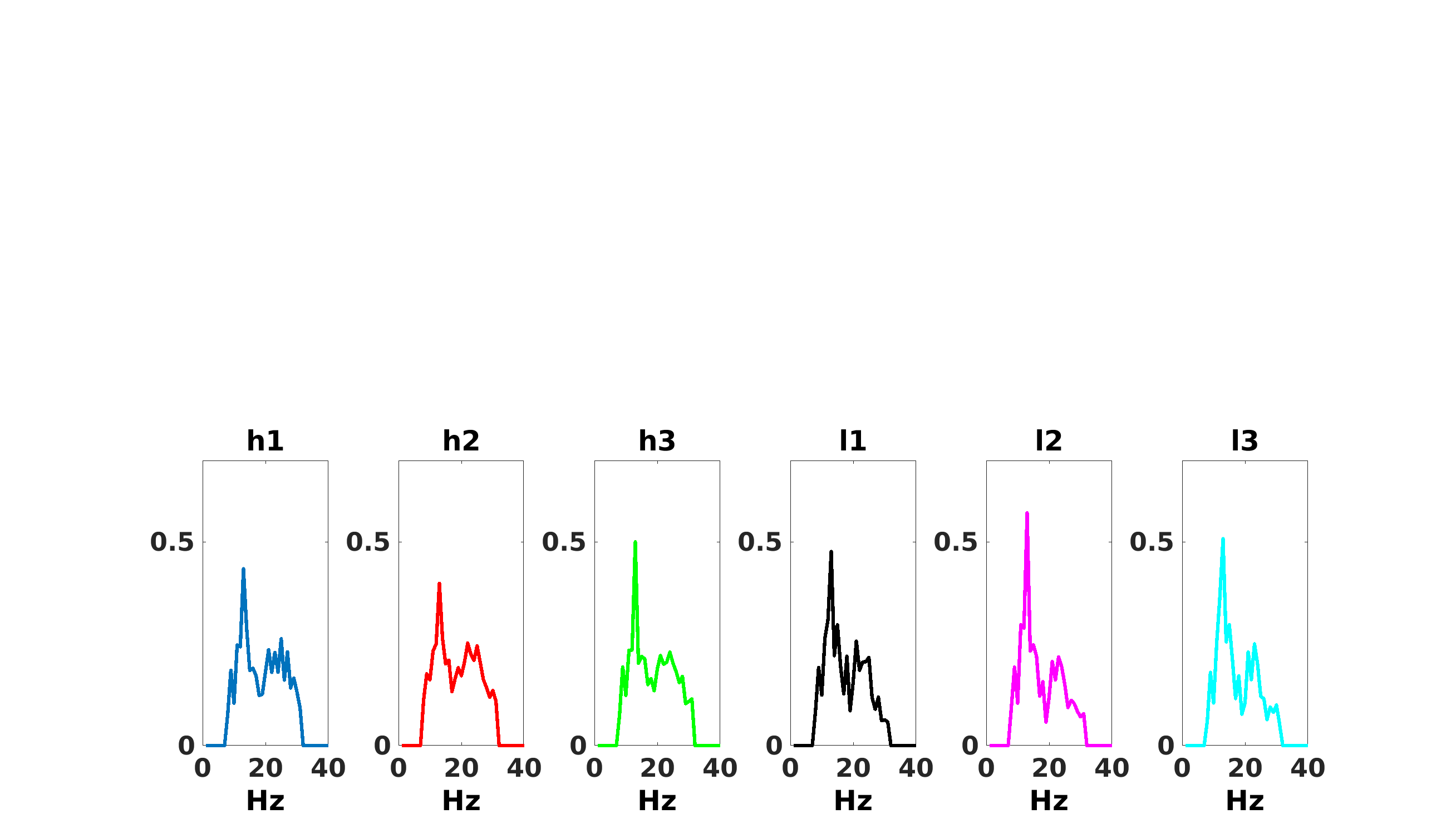}}
 \hfill	
 \subfloat[P7 -- SACSP spatial patterns.]{
	   \centering
	   \includegraphics[trim={0cm 4.7cm 0cm 4.2cm},clip,width=0.9\textwidth]{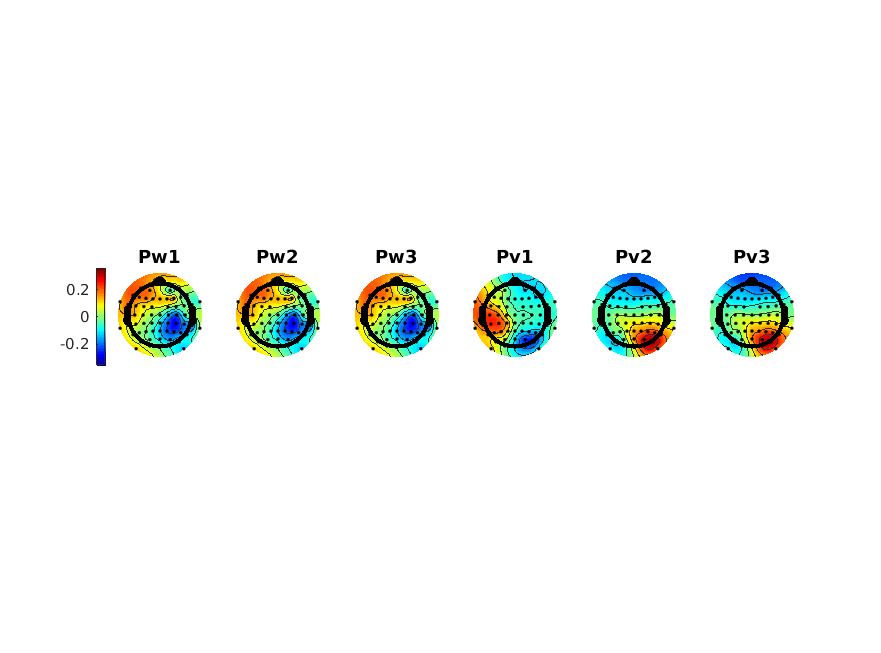}}
 \hfill 	
  \subfloat[P7 -- SACSP spectral filters.]{
	   \centering
	   \includegraphics[trim={1cm 0.5cm 1cm 13cm},clip,width=0.9\textwidth]{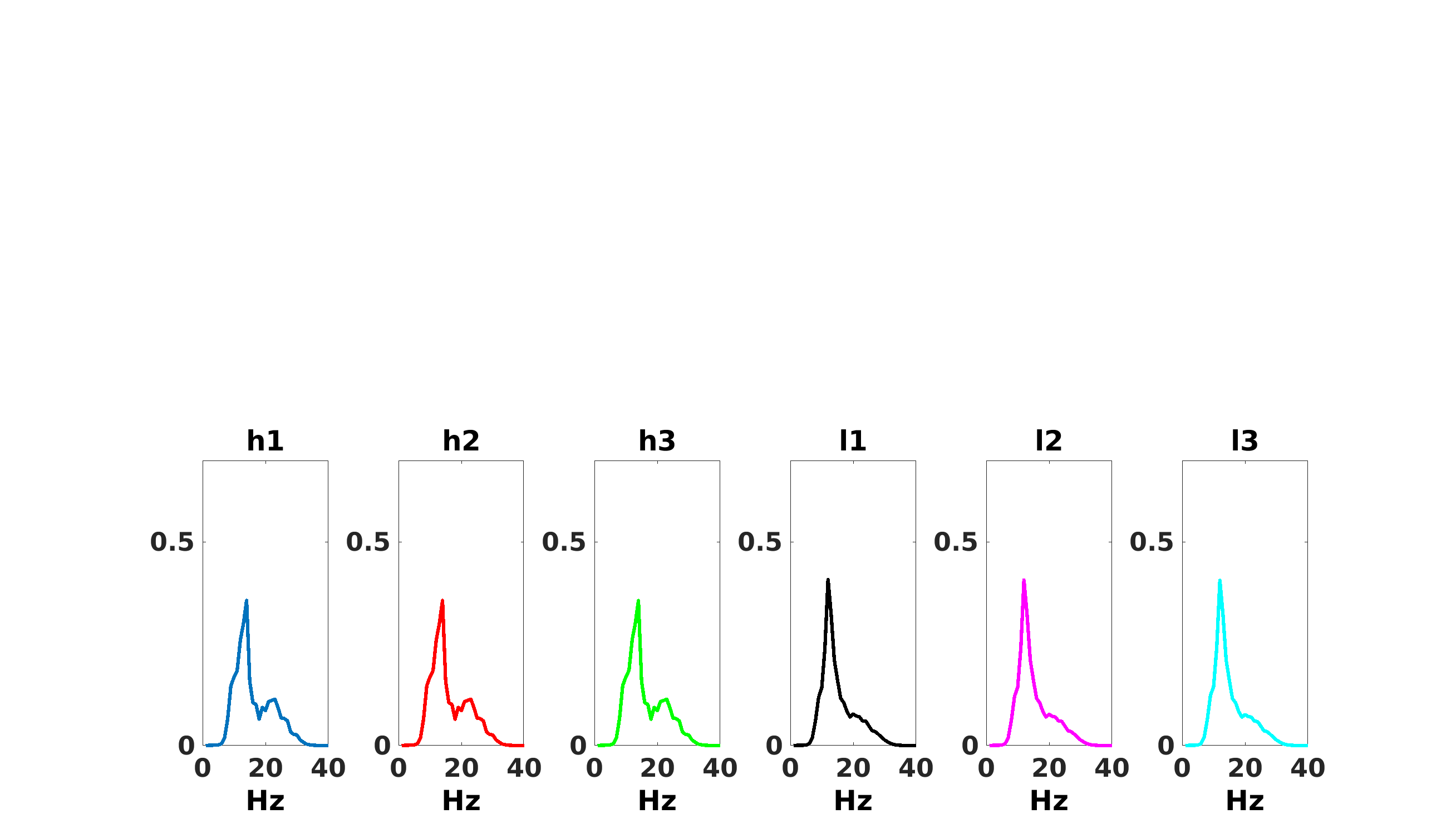}}
\caption{Spatial patterns and spectral filters trained on the calibration data for P7. }
\label{SACSP7}
\end{figure}

\begin{figure}
  \subfloat[P8 -- CSP spatial patterns.]{
	   \centering
	   \includegraphics[trim={0cm 4.7cm 0cm 4.2cm},clip,width=0.9\textwidth]{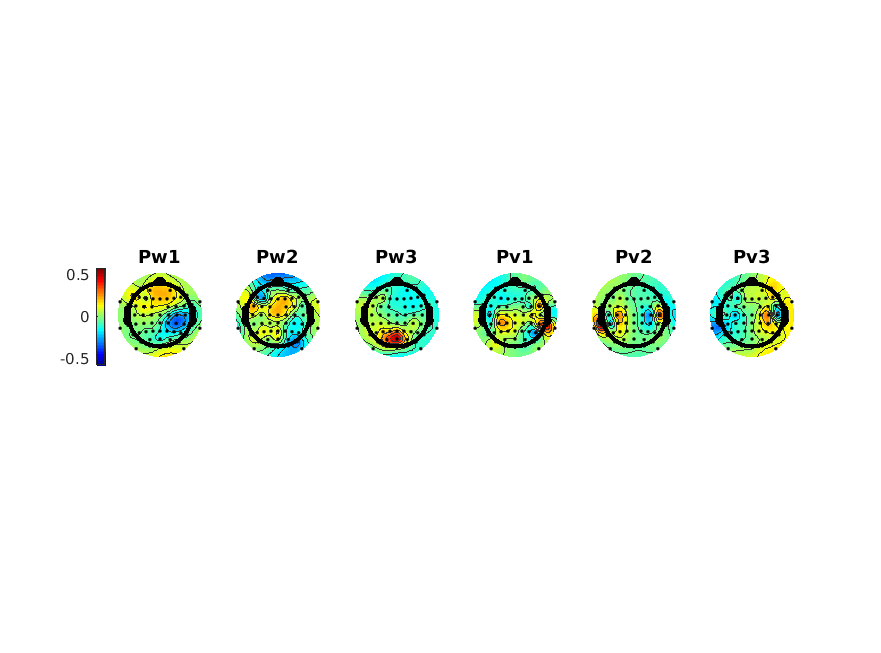}}
 \hfill 	
  \subfloat[P8 -- CCACSP spatial patterns.]{
	   \centering
	   \includegraphics[trim={0cm 4.7cm 0cm 4.2cm},clip,width=0.9\textwidth]{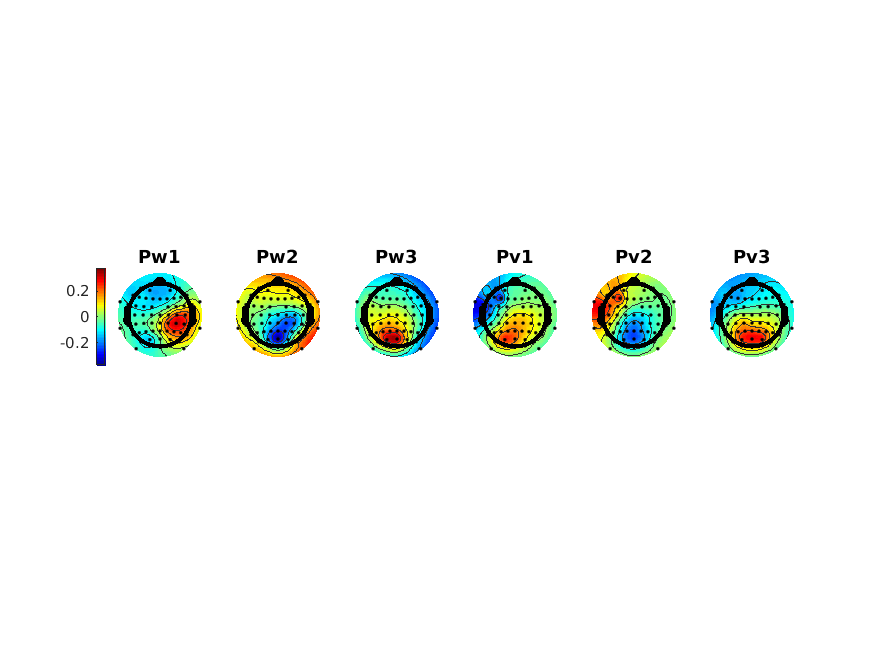}}
 \hfill	
  \subfloat[P8 -- spec-CSP spatial patterns.]{
	   \centering
	   \includegraphics[trim={0cm 4.7cm 0cm 4.2cm},clip,width=0.9\textwidth]{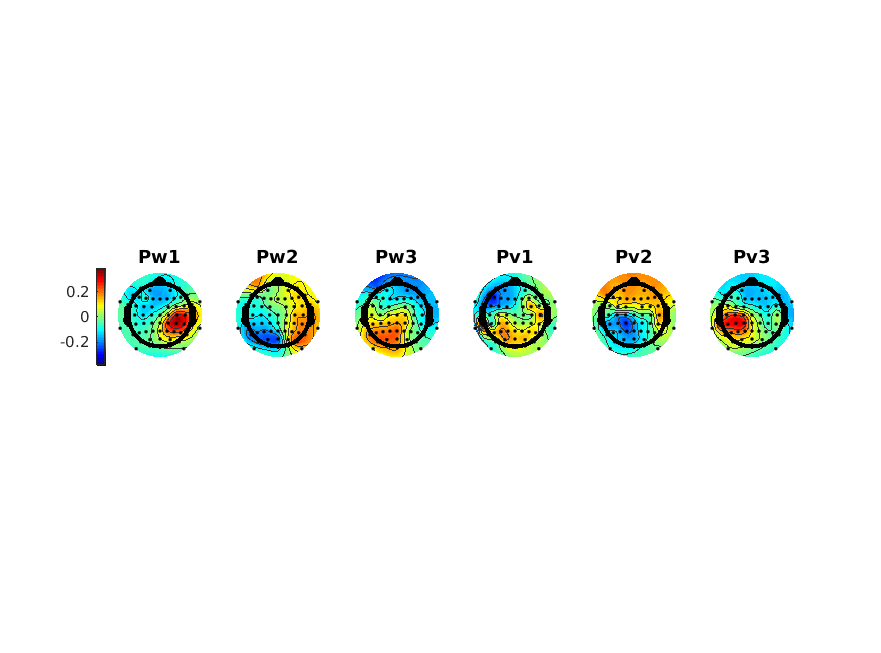}}
 \hfill 	
  \subfloat[P8 -- spec-CSP spectral filters.]{
	   \centering
	   \includegraphics[trim={1cm 0.5cm 1cm 13cm},clip,width=0.9\textwidth]{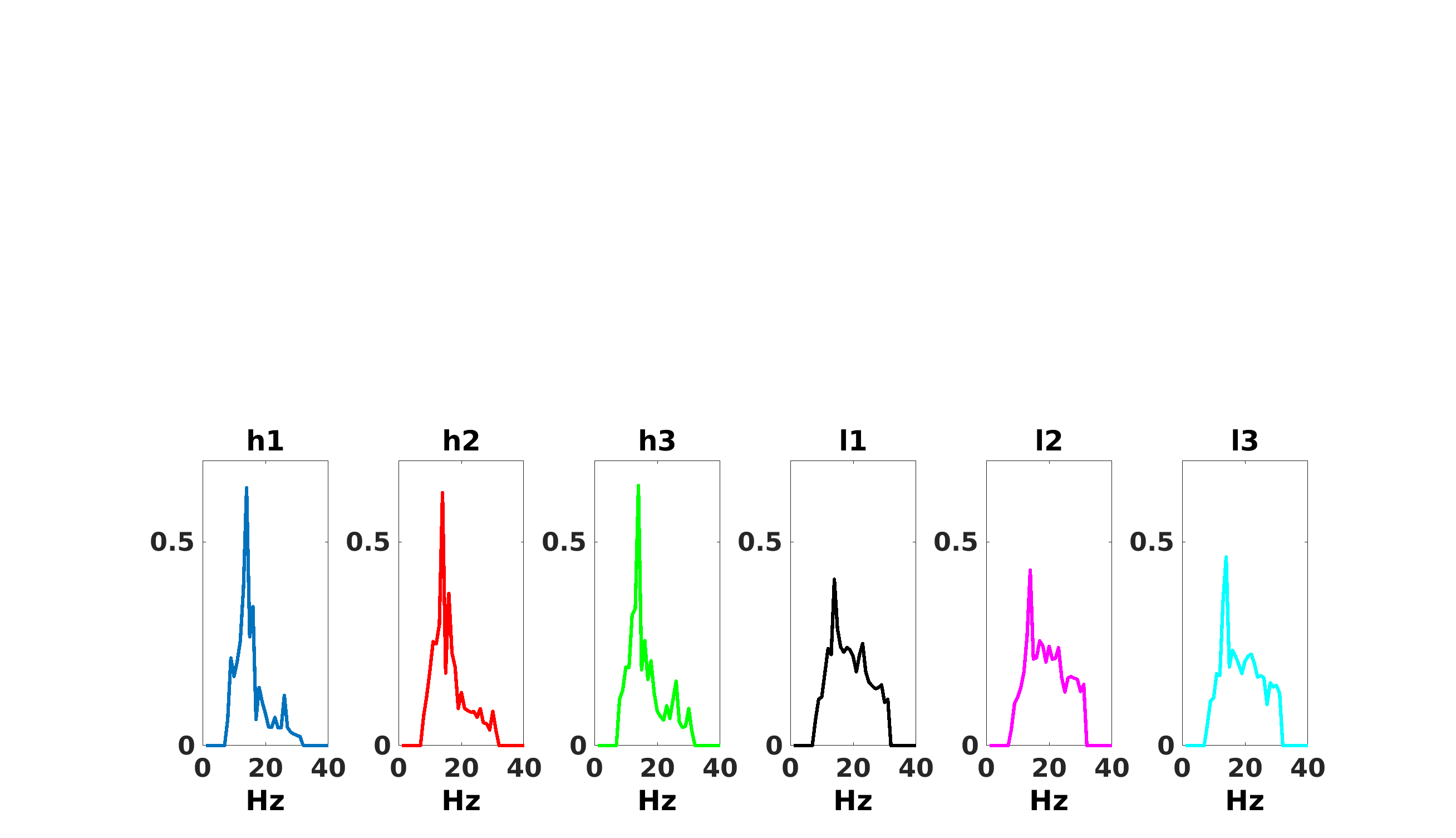}}
 \hfill	
 \subfloat[P8 -- SACSP spatial patterns.]{
	   \centering
	   \includegraphics[trim={0cm 4.7cm 0cm 4.2cm},clip,width=0.9\textwidth]{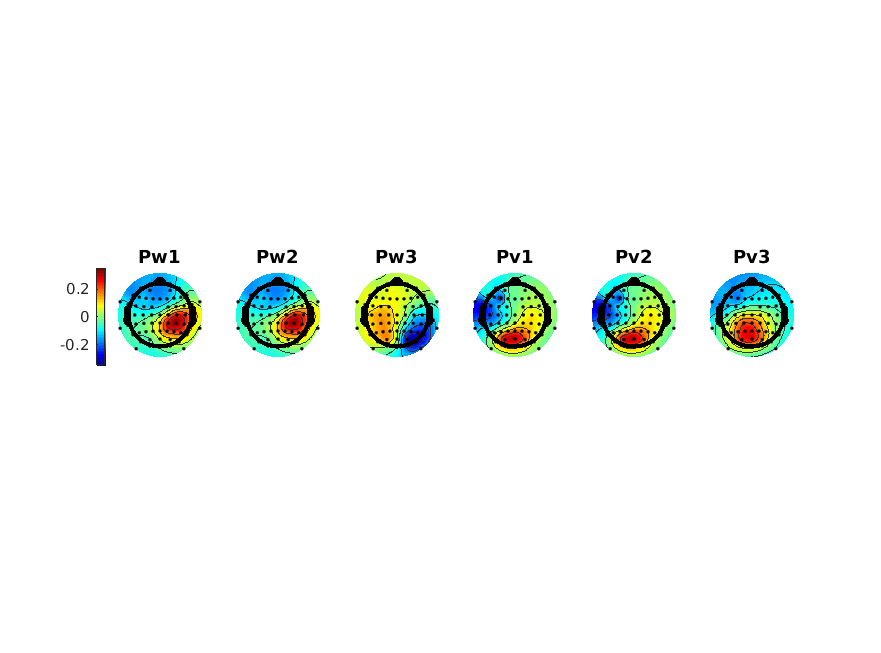}}
 \hfill 	
  \subfloat[P8 -- SACSP spectral filters.]{
	   \centering
	   \includegraphics[trim={1cm 0.5cm 1cm 13cm},clip,width=0.9\textwidth]{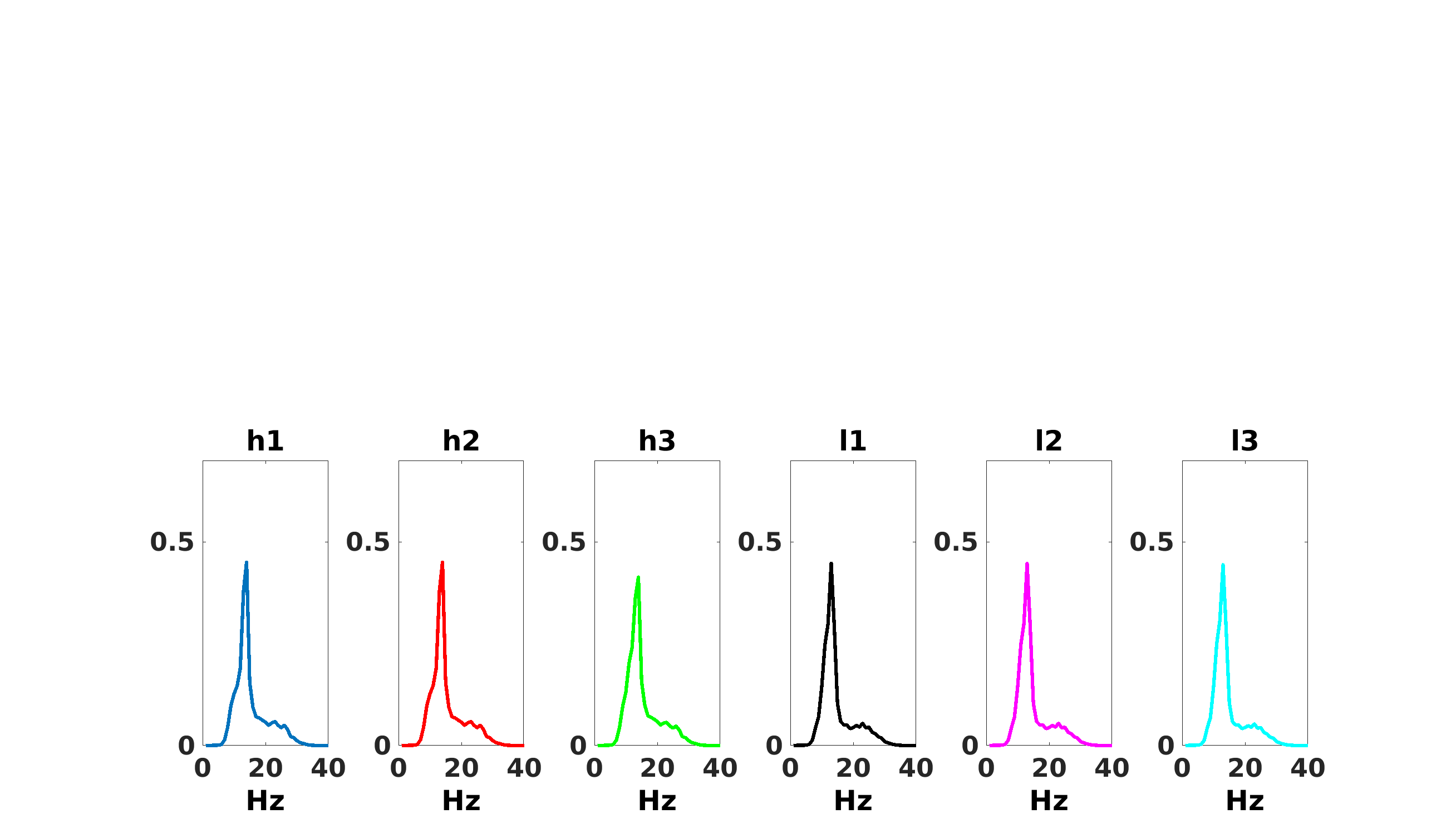}}
\caption{Spatial patterns and spectral filters trained on the calibration data for P8. }
\label{SACSP8}
\end{figure}

\begin{figure}
  \subfloat[P9 -- CSP spatial patterns.]{
	   \centering
	   \includegraphics[trim={0cm 4.7cm 0cm 4.2cm},clip,width=0.9\textwidth]{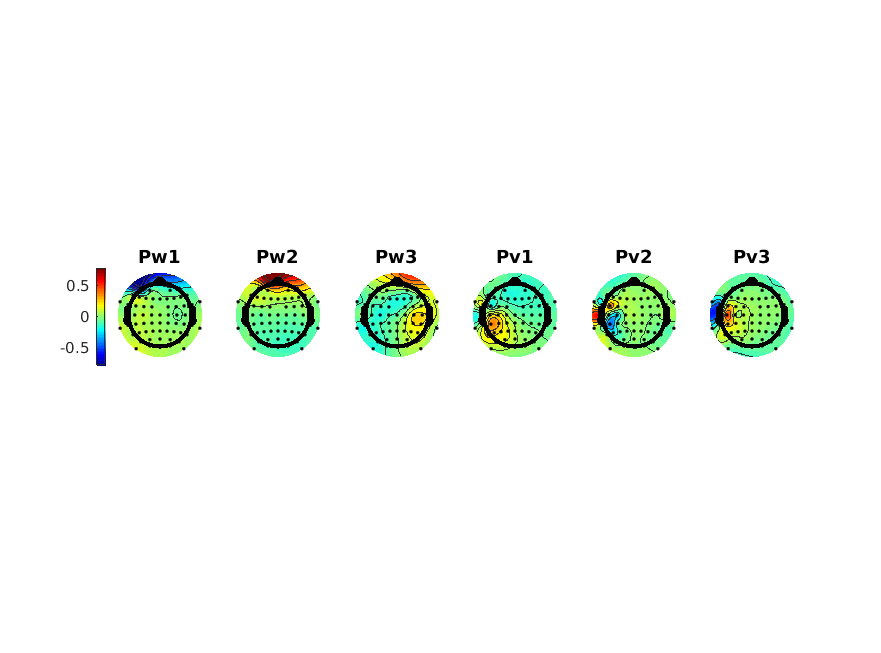}}
 \hfill 	
  \subfloat[P9 -- CCACSP spatial patterns.]{
	   \centering
	   \includegraphics[trim={0cm 4.7cm 0cm 4.2cm},clip,width=0.9\textwidth]{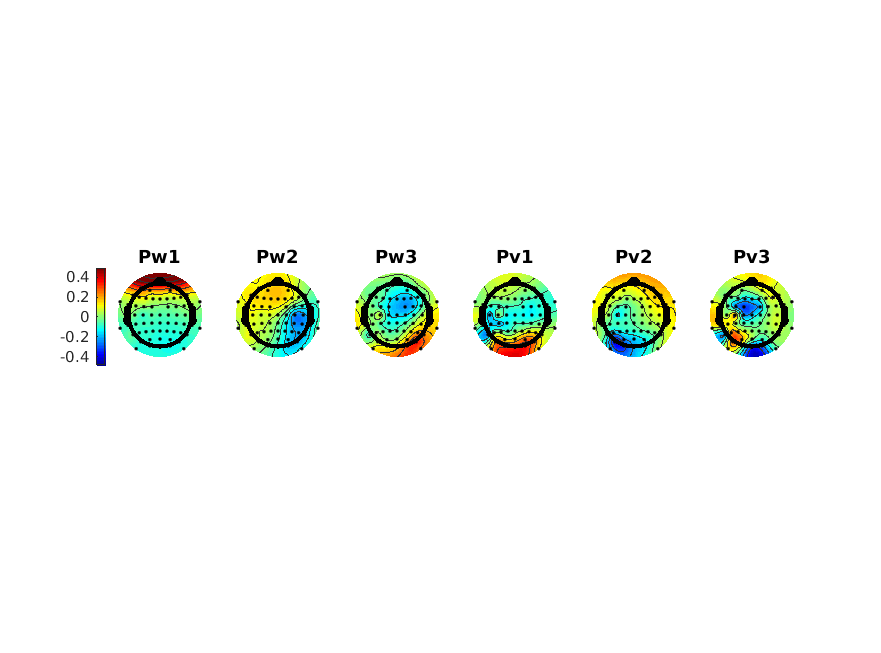}}
 \hfill	
  \subfloat[P9 -- spec-CSP spatial patterns.]{
	   \centering
	   \includegraphics[trim={0cm 4.7cm 0cm 4.2cm},clip,width=0.9\textwidth]{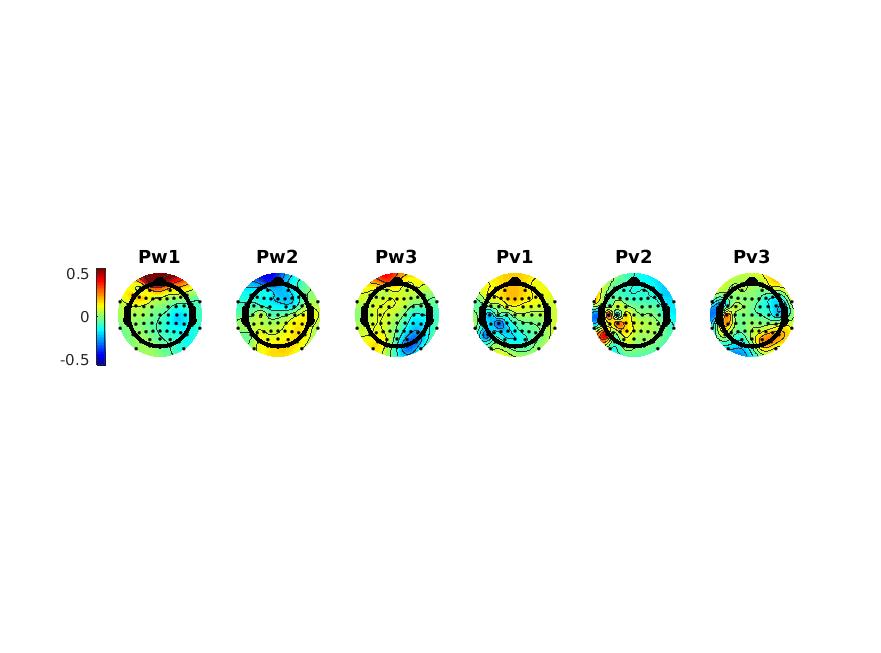}}
 \hfill 	
  \subfloat[P9 -- spec-CSP spectral filters.]{
	   \centering
	   \includegraphics[trim={1cm 0.5cm 1cm 13cm},clip,width=0.9\textwidth]{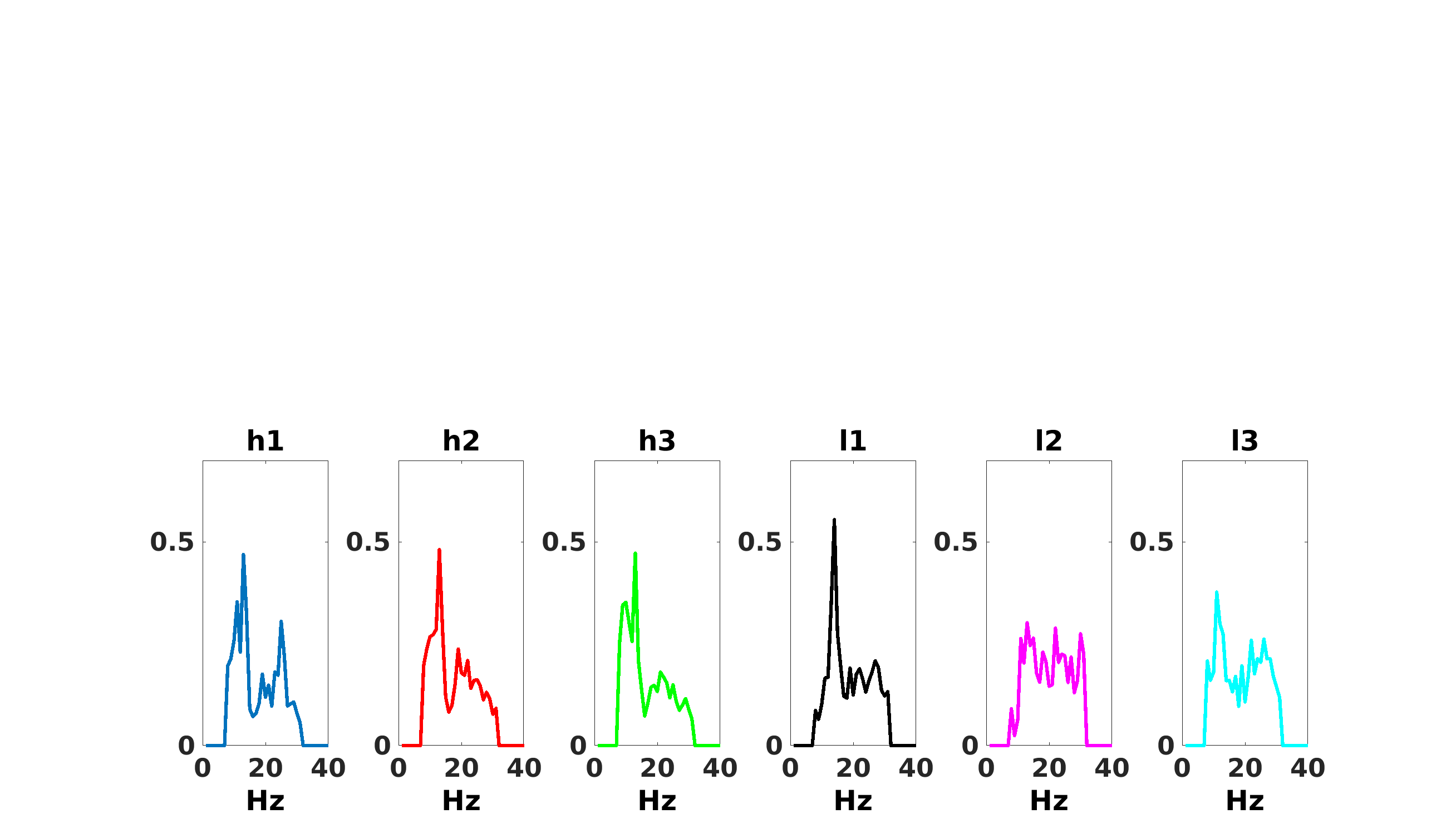}}
 \hfill	
 \subfloat[P9 -- SACSP spatial patterns.]{
	   \centering
	   \includegraphics[trim={0cm 4.7cm 0cm 4.2cm},clip,width=0.9\textwidth]{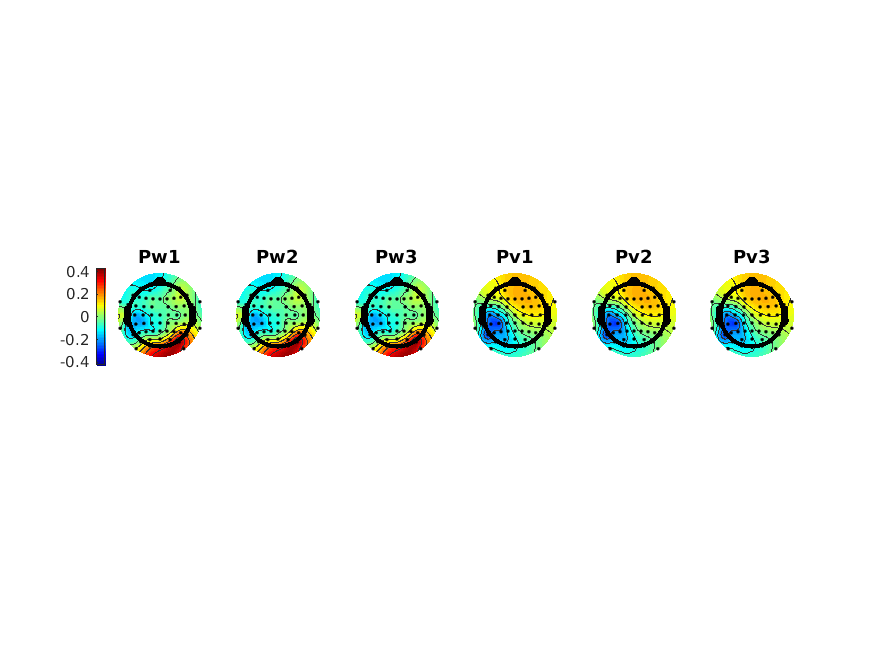}}
 \hfill 	
  \subfloat[P9 -- SACSP spectral filters.]{
	   \centering
	   \includegraphics[trim={1cm 0.5cm 1cm 13cm},clip,width=0.9\textwidth]{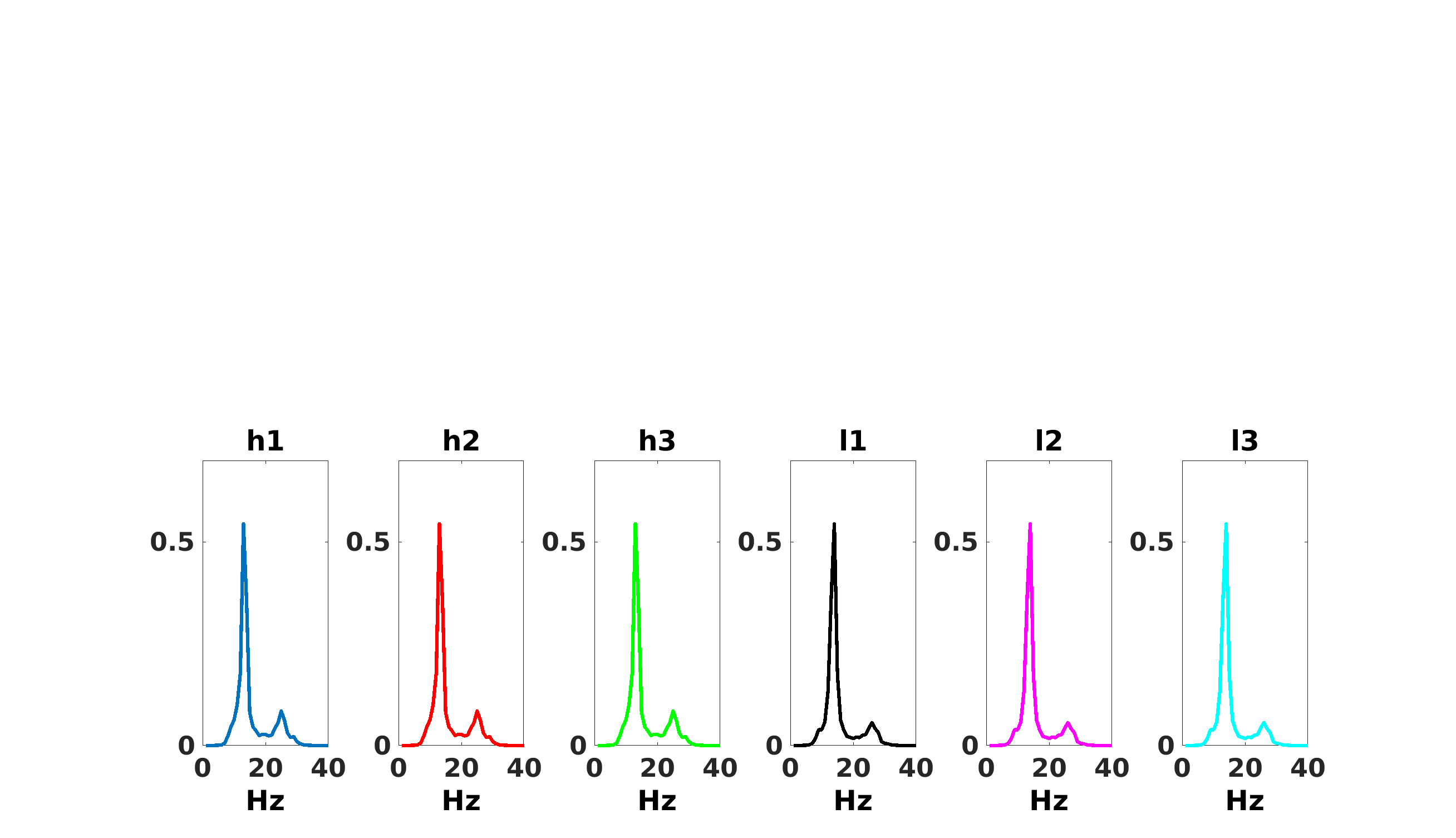}}
\caption{Spatial patterns and spectral filters trained on the calibration data for P9. }
\label{SACSP9}
\end{figure}

\begin{figure}
  \subfloat[P10 -- CSP spatial patterns.]{
	   \centering
	   \includegraphics[trim={0cm 4.7cm 0cm 4.2cm},clip,width=0.9\textwidth]{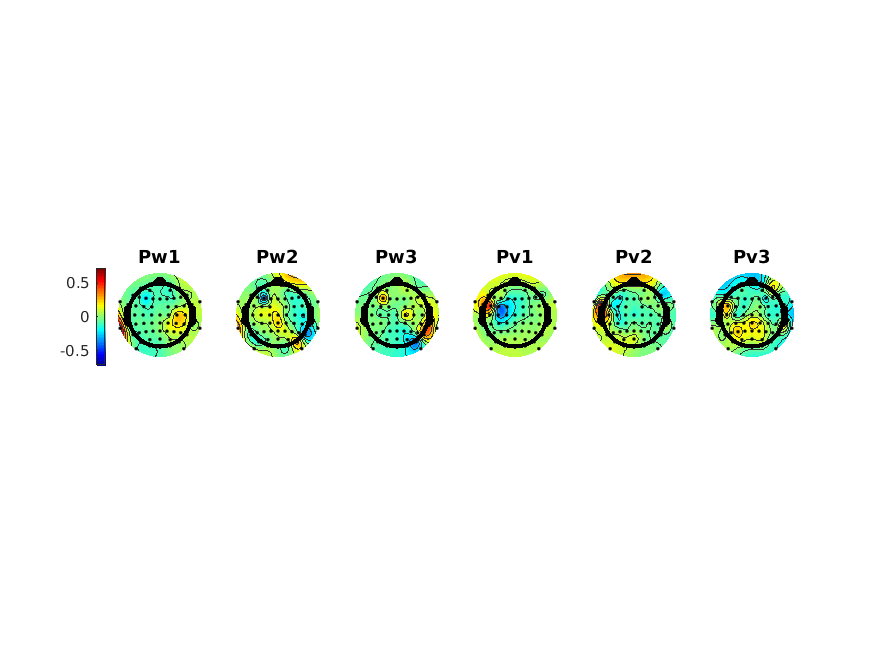}}
 \hfill 	
  \subfloat[P10 -- CCACSP spatial patterns.]{
	   \centering
	   \includegraphics[trim={0cm 4.7cm 0cm 4.2cm},clip,width=0.9\textwidth]{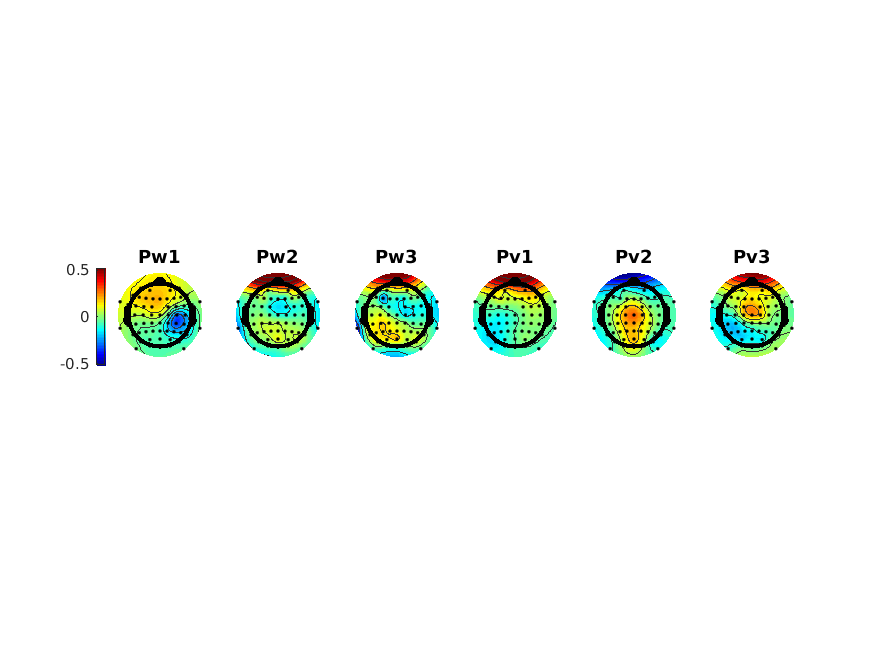}}
 \hfill	
  \subfloat[P10 -- spec-CSP spatial patterns.]{
	   \centering
	   \includegraphics[trim={0cm 4.7cm 0cm 4.2cm},clip,width=0.9\textwidth]{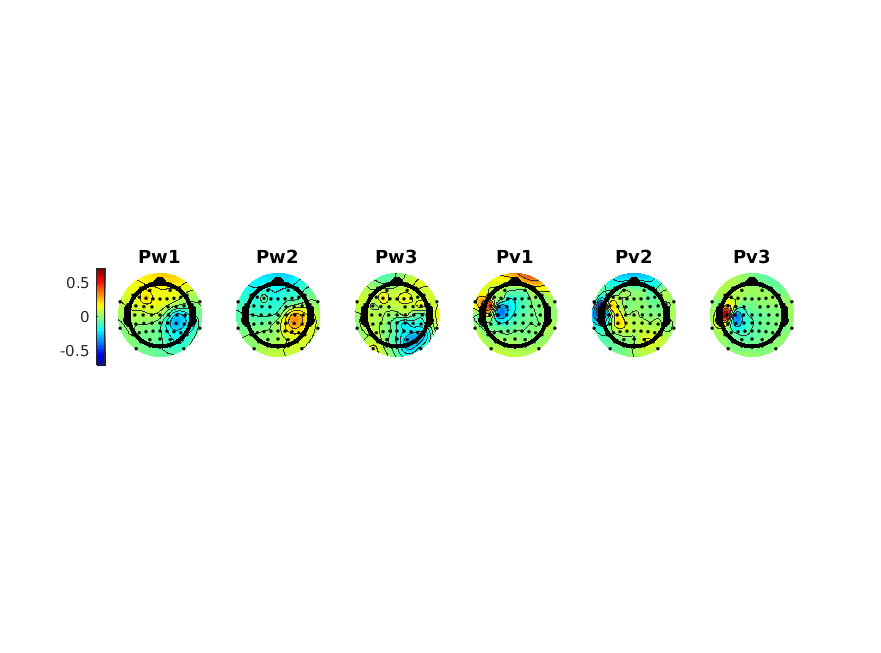}}
 \hfill 	
  \subfloat[P10 -- spec-CSP spectral filters.]{
	   \centering
	   \includegraphics[trim={1cm 0.5cm 1cm 13cm},clip,width=0.9\textwidth]{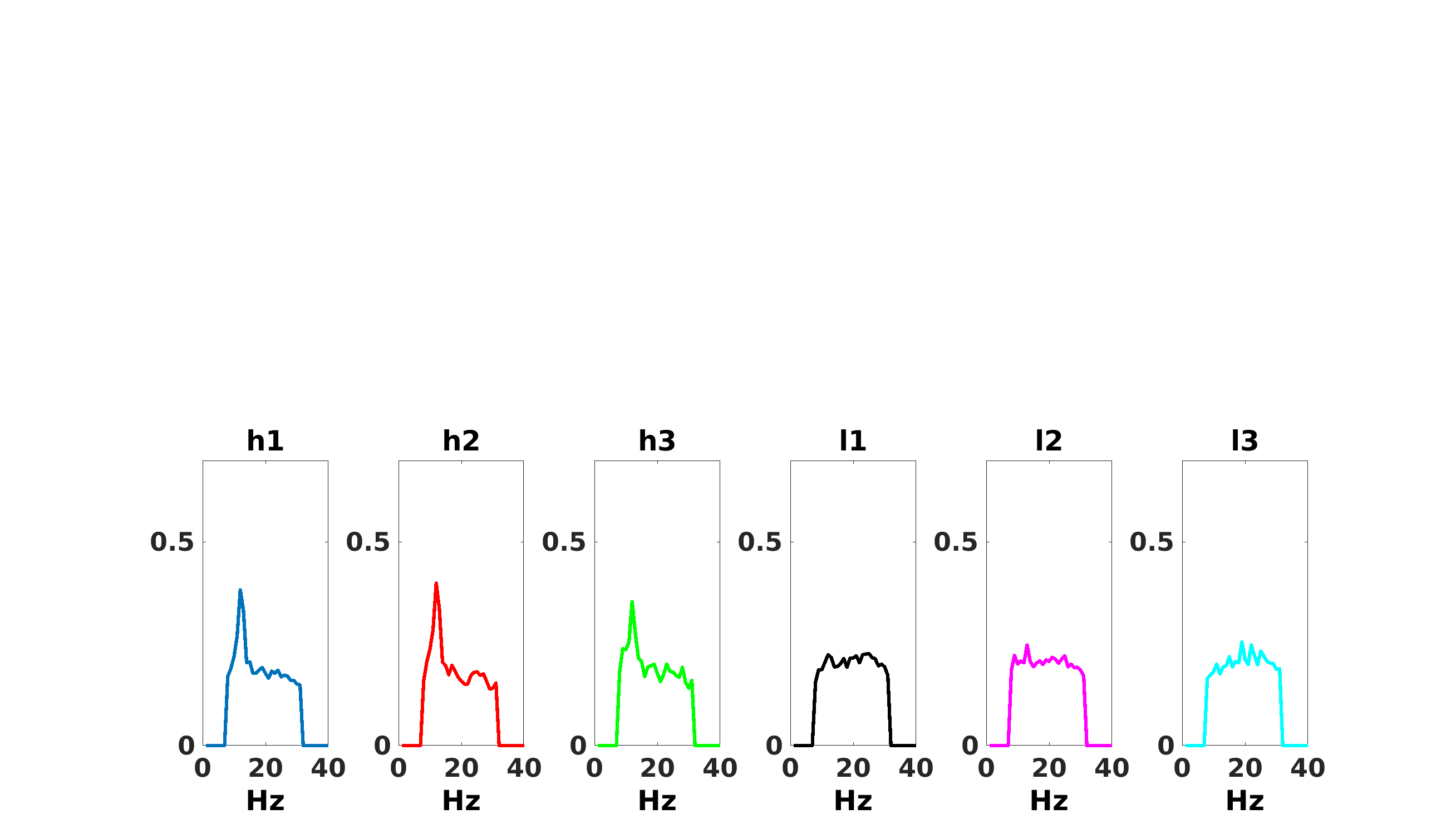}}
 \hfill	
 \subfloat[P10 -- SACSP spatial patterns.]{
	   \centering
	   \includegraphics[trim={0cm 4.7cm 0cm 4.2cm},clip,width=0.9\textwidth]{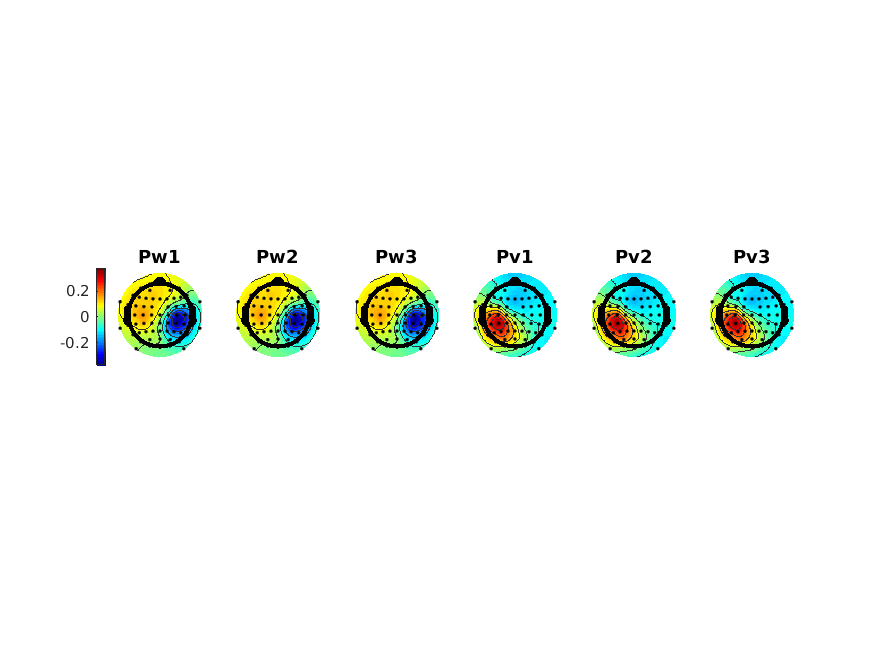}}
 \hfill 	
  \subfloat[P10 -- SACSP spectral filters.]{
	   \centering
	   \includegraphics[trim={1cm 0.5cm 1cm 13cm},clip,width=0.9\textwidth]{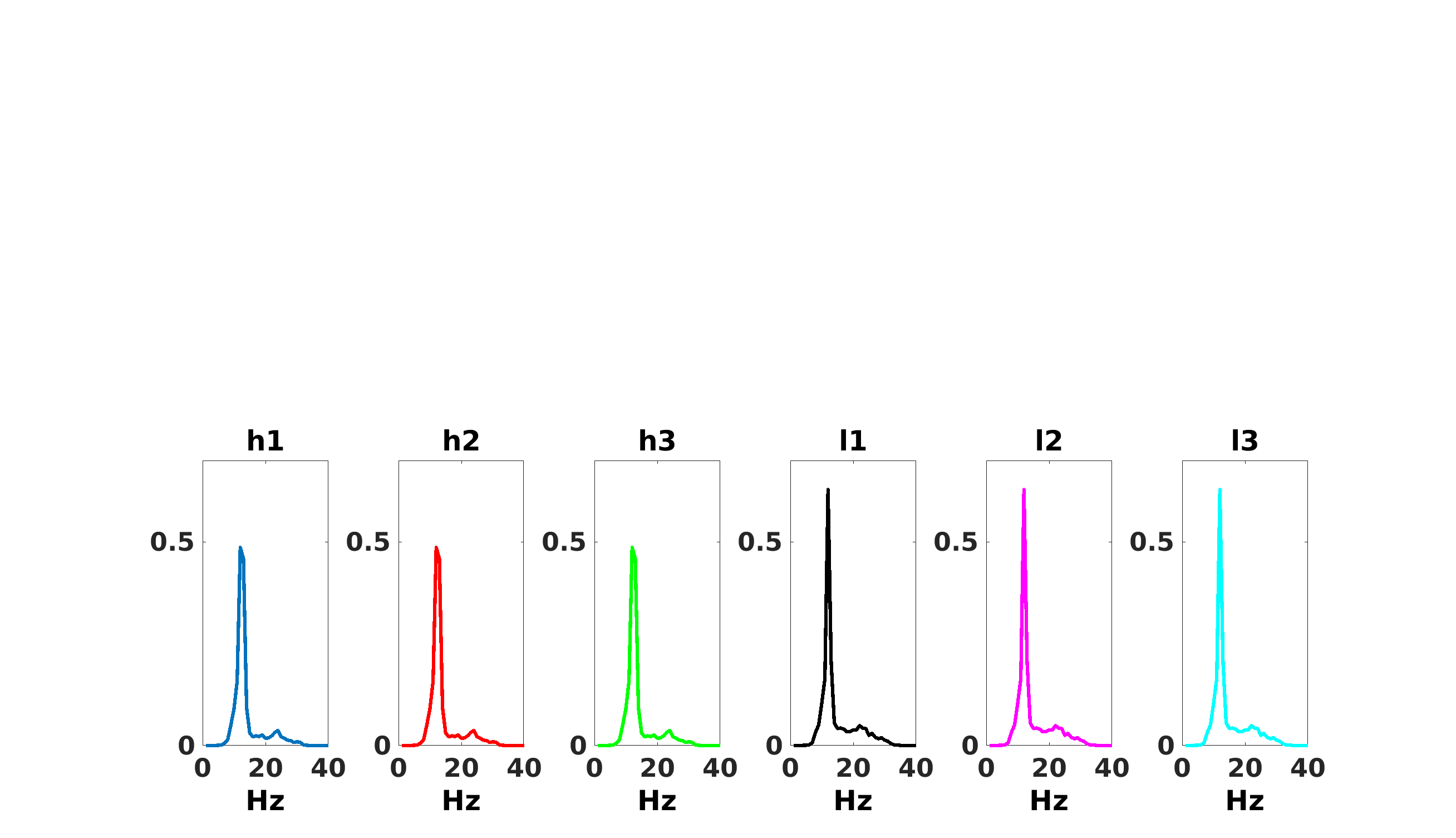}}
\caption{Spatial patterns and spectral filters trained on the calibration data for P10. }
\label{SACSP10}
\end{figure}

\begin{figure}
  \subfloat[P11 -- CSP spatial patterns.]{
	   \centering
	   \includegraphics[trim={0cm 4.7cm 0cm 4.2cm},clip,width=0.9\textwidth]{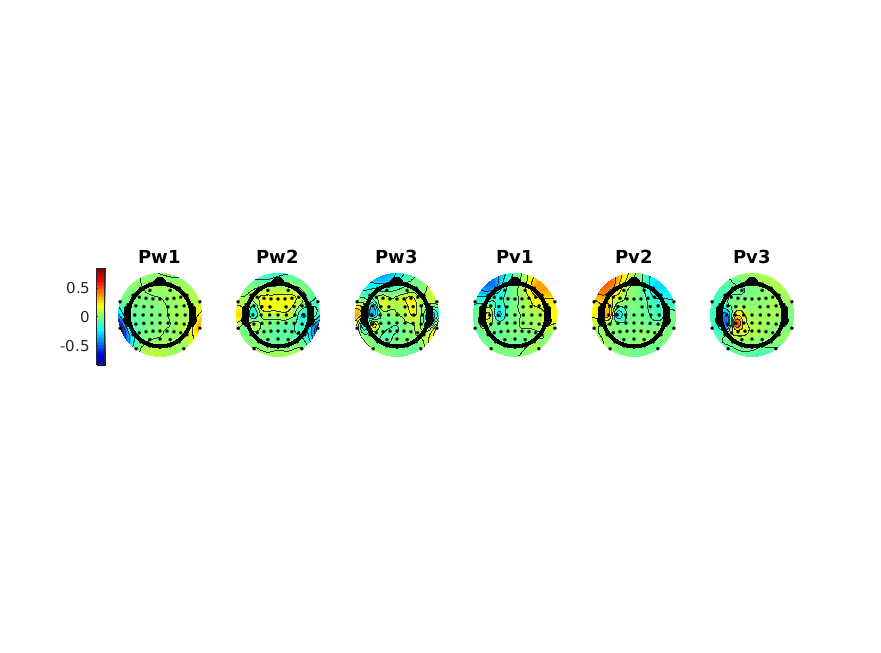}}
 \hfill 	
  \subfloat[P11 -- CCACSP spatial patterns.]{
	   \centering
	   \includegraphics[trim={0cm 4.7cm 0cm 4.2cm},clip,width=0.9\textwidth]{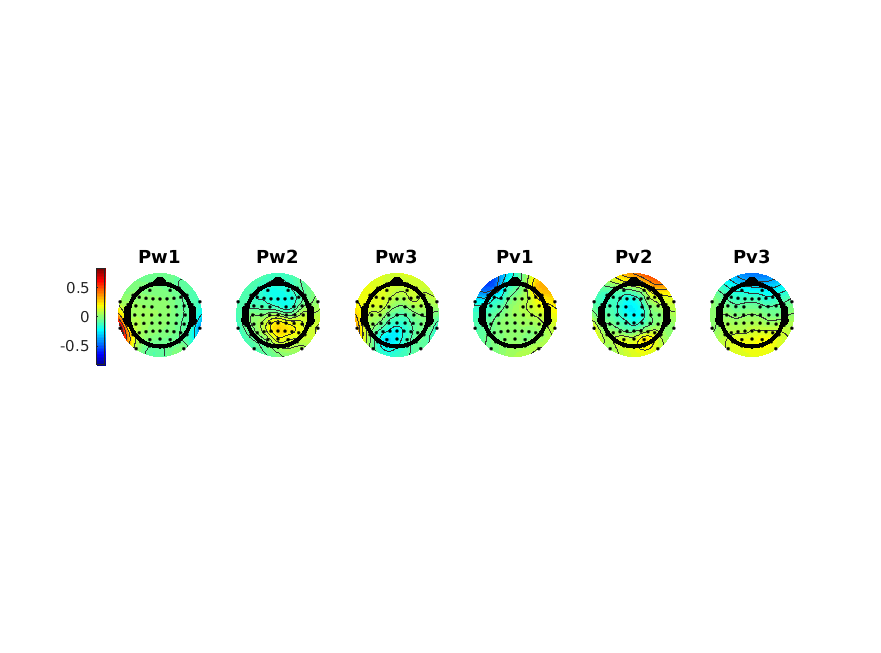}}
 \hfill	
  \subfloat[P11 -- spec-CSP spatial patterns.]{
	   \centering
	   \includegraphics[trim={0cm 4.7cm 0cm 4.2cm},clip,width=0.9\textwidth]{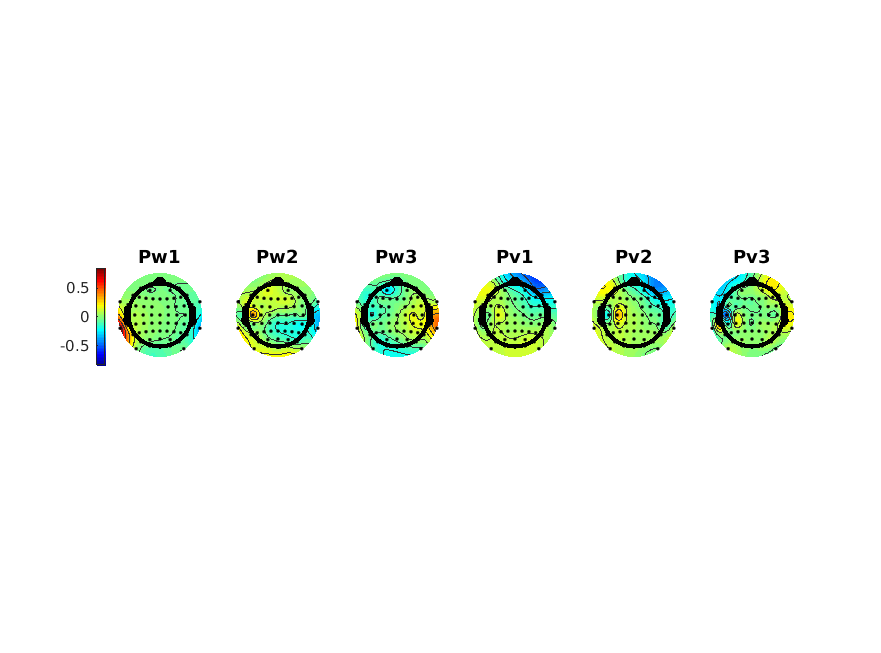}}
 \hfill 	
  \subfloat[P11 -- spec-CSP spectral filters.]{
	   \centering
	   \includegraphics[trim={1cm 0.5cm 1cm 13cm},clip,width=0.9\textwidth]{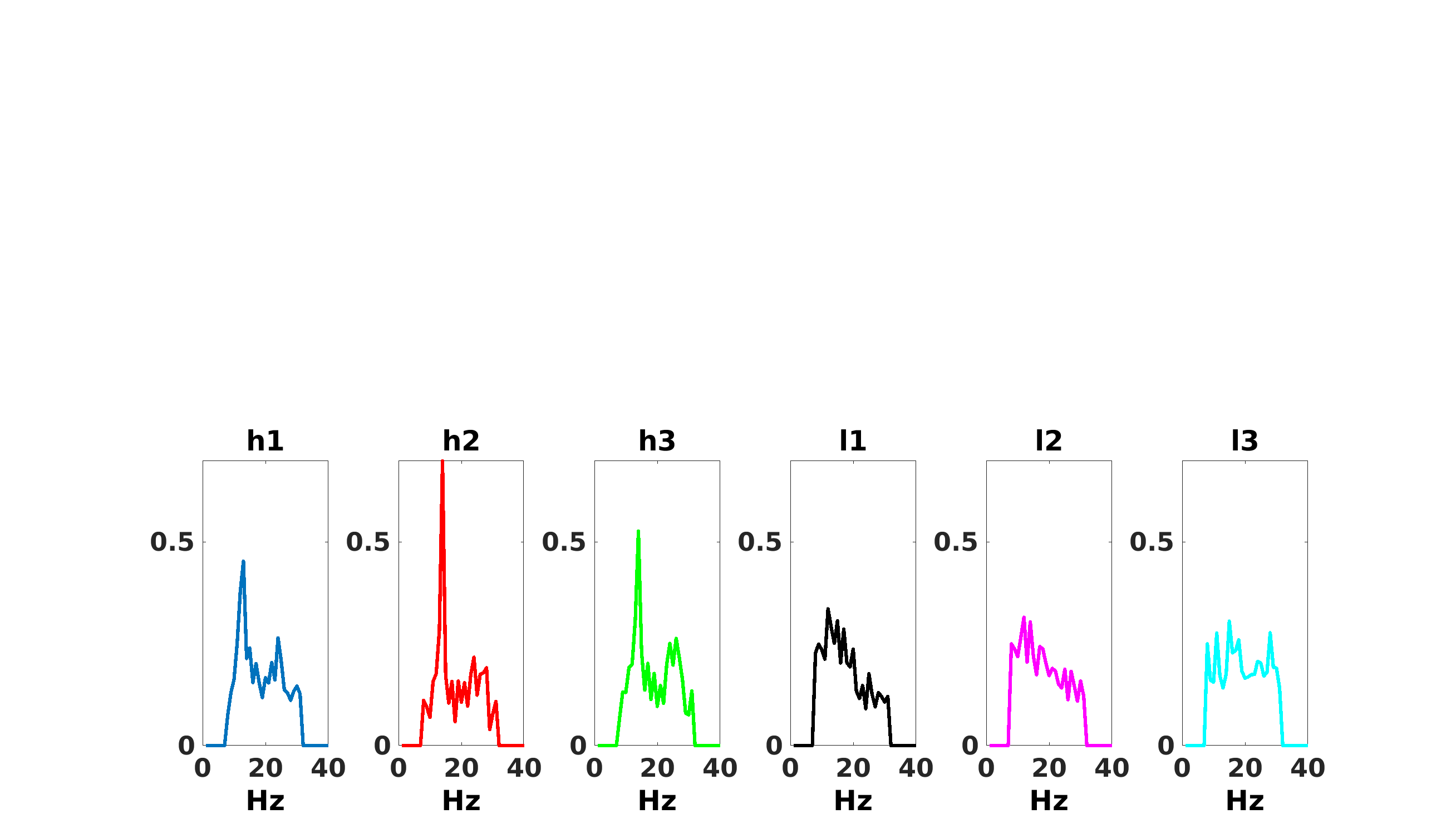}}
 \hfill	
 \subfloat[P11 -- SACSP spatial patterns.]{
	   \centering
	   \includegraphics[trim={0cm 4.7cm 0cm 4.2cm},clip,width=0.9\textwidth]{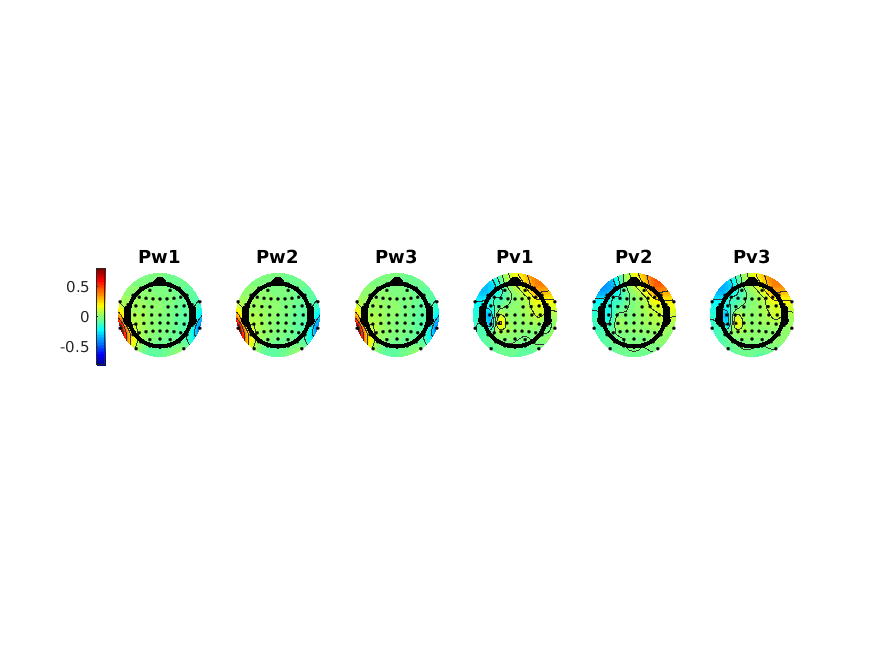}}
 \hfill 	
  \subfloat[P11 -- SACSP spectral filters.]{
	   \centering
	   \includegraphics[trim={1cm 0.5cm 1cm 13cm},clip,width=0.9\textwidth]{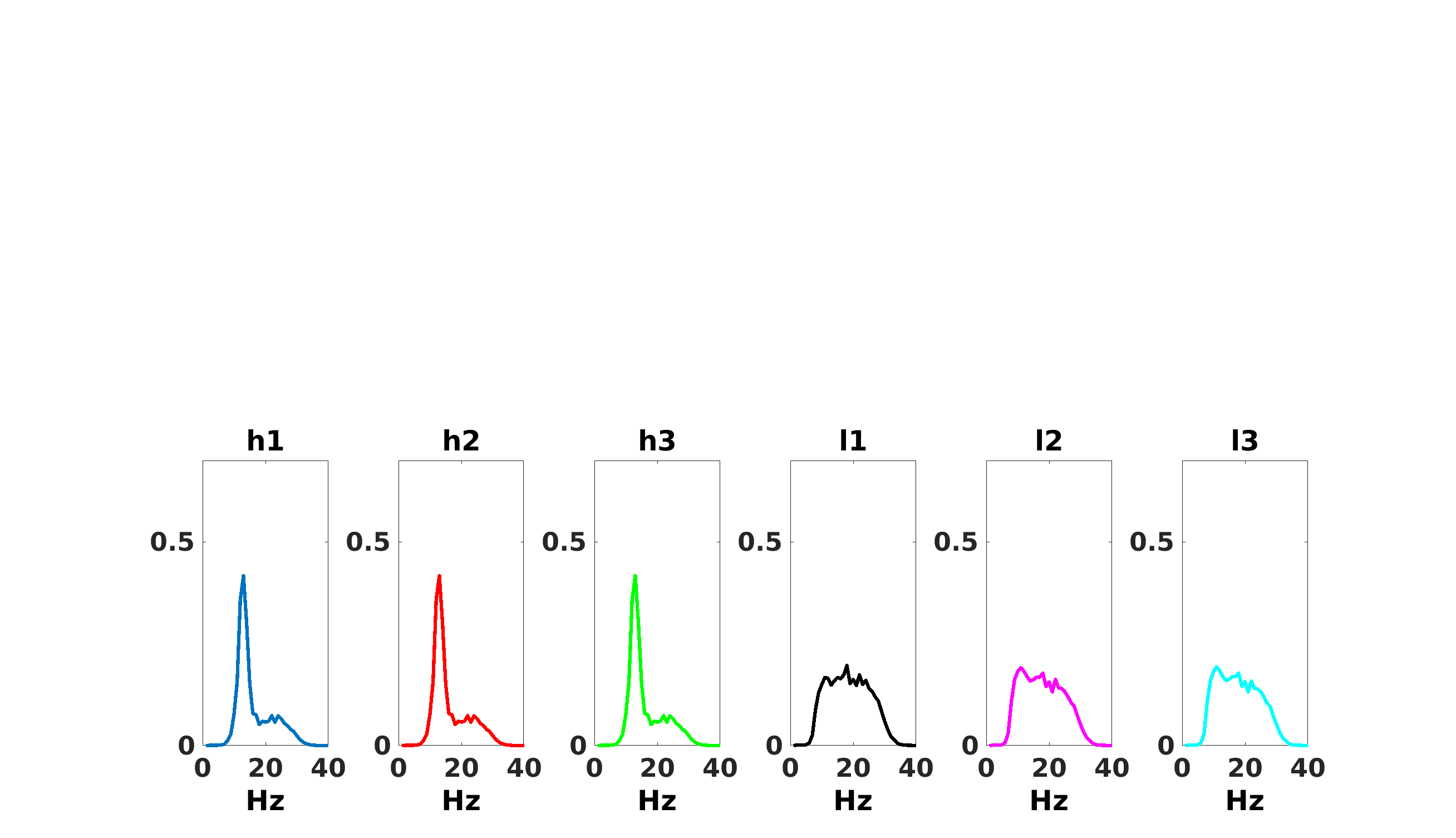}}
\caption{Spatial patterns and spectral filters trained on the calibration data for P11. }
\label{SACSP11}
\end{figure}

\begin{figure}
  \subfloat[P12 -- CSP spatial patterns.]{
	   \centering
	   \includegraphics[trim={0cm 4.7cm 0cm 4.2cm},clip,width=0.9\textwidth]{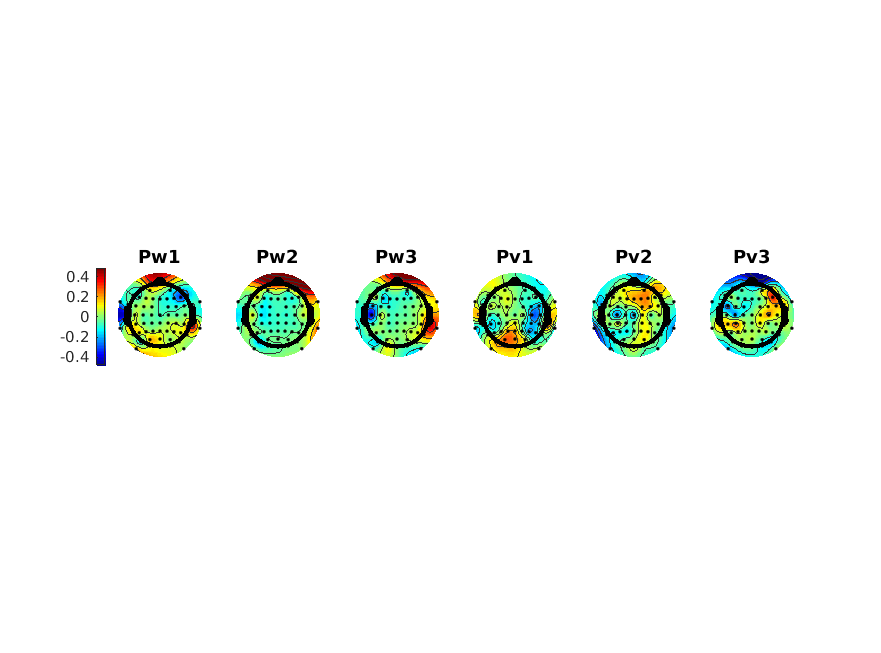}}
 \hfill 	
  \subfloat[P12 -- CCACSP spatial patterns.]{
	   \centering
	   \includegraphics[trim={0cm 4.7cm 0cm 4.2cm},clip,width=0.9\textwidth]{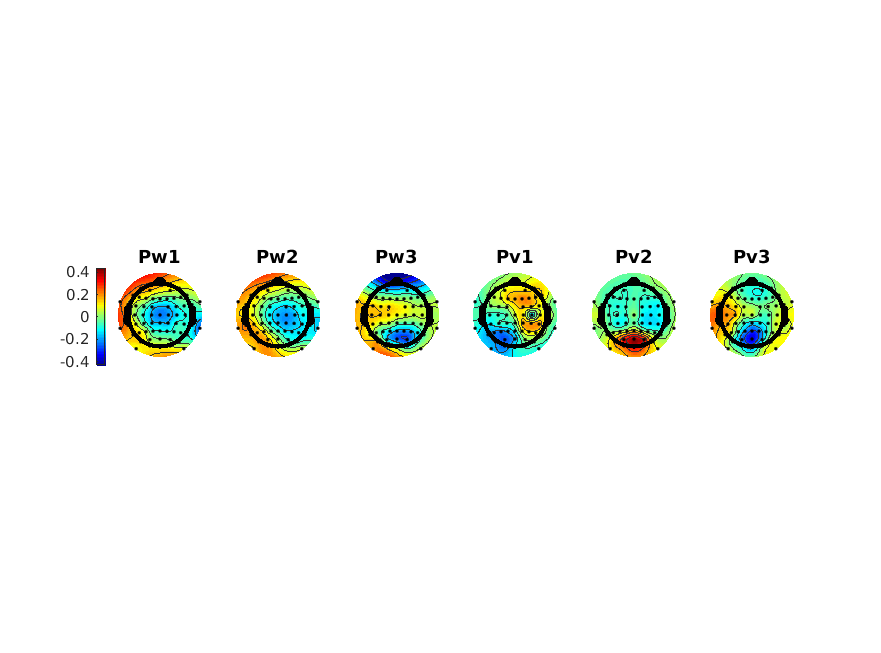}}
 \hfill	
  \subfloat[P12 -- spec-CSP spatial patterns.]{
	   \centering
	   \includegraphics[trim={0cm 4.7cm 0cm 4.2cm},clip,width=0.9\textwidth]{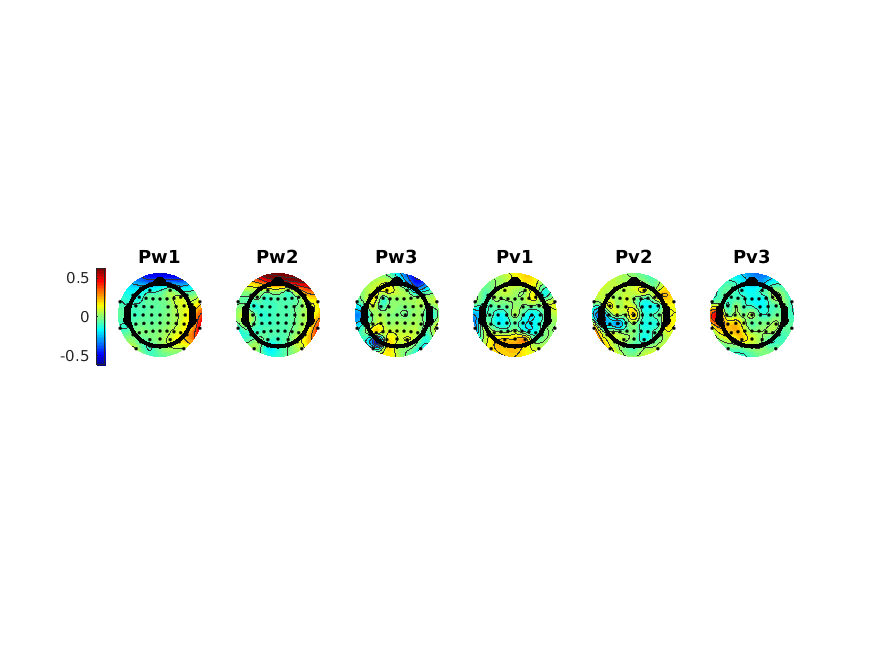}}
 \hfill 	
  \subfloat[P12 -- spec-CSP spectral filters.]{
	   \centering
	   \includegraphics[trim={1cm 0.5cm 1cm 13cm},clip,width=0.9\textwidth]{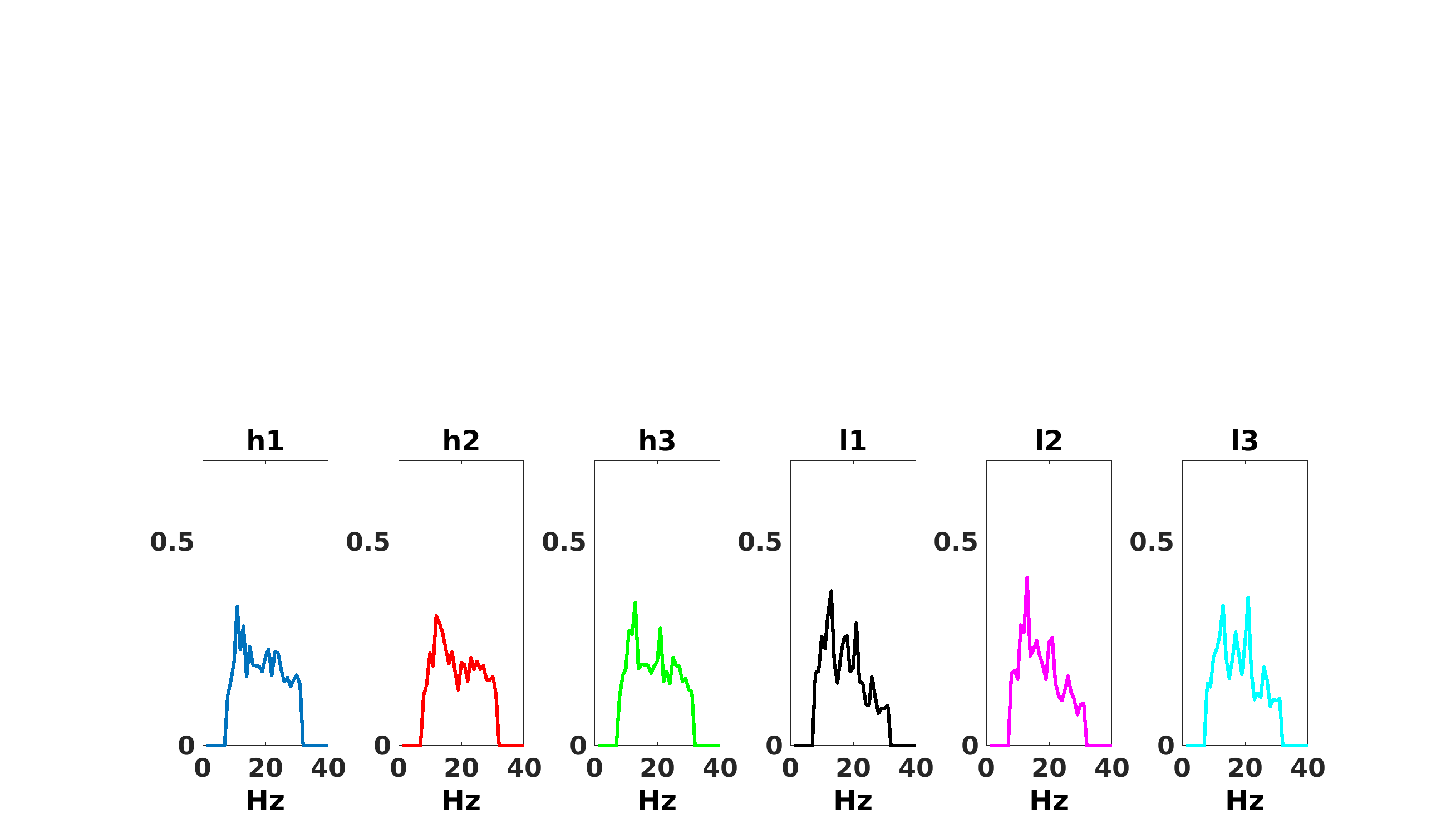}}
 \hfill	
 \subfloat[P12 -- SACSP spatial patterns.]{
	   \centering
	   \includegraphics[trim={0cm 4.7cm 0cm 4.2cm},clip,width=0.9\textwidth]{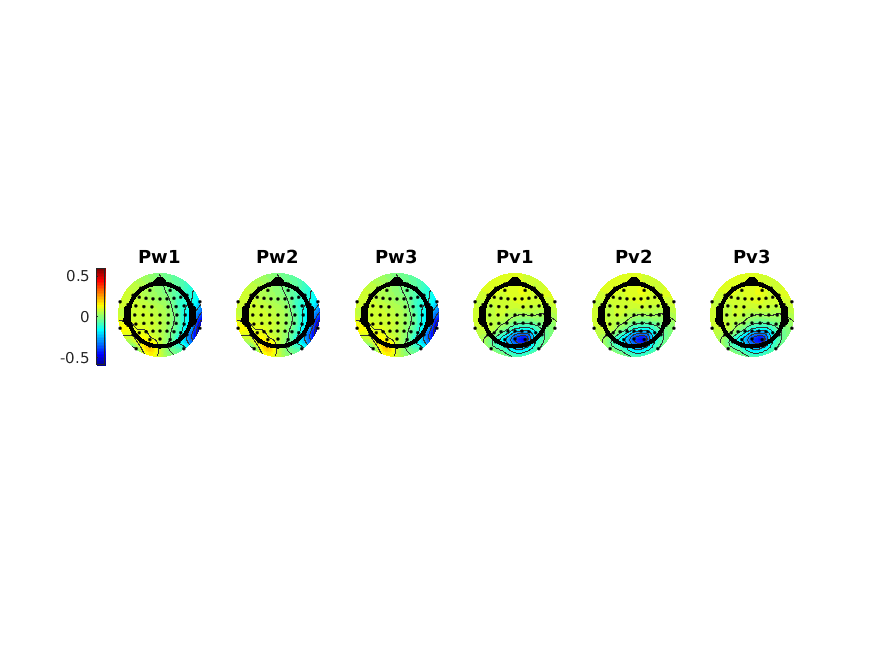}}
 \hfill 	
  \subfloat[P12 -- SACSP spectral filters.]{
	   \centering
	   \includegraphics[trim={1cm 0.5cm 1cm 13cm},clip,width=0.9\textwidth]{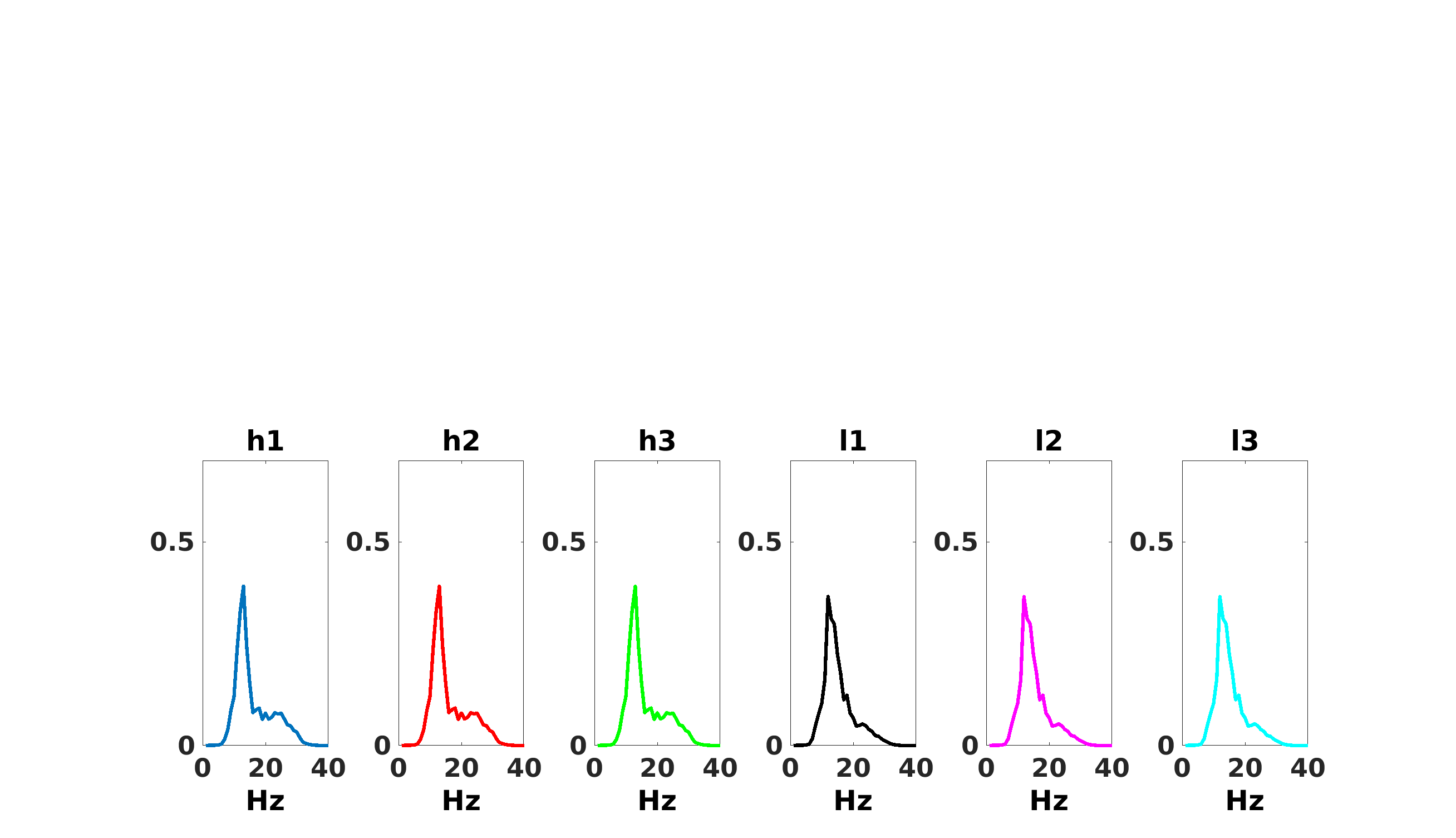}}
\caption{Spatial patterns and spectral filters trained on the calibration data for P12. }
\label{SACSP12}
\end{figure}

Figures \ref{SACSP1} to \ref{SACSP12} present the learnt spatial patterns and spectral filters for right and left motor imagery classes on calibration data for each participant (P1-P12) with CSP, CCACSP, spec-CSP and SACSP methods.

